\documentclass[12pt]{article}

\pdfoutput=1
\usepackage{color}
\usepackage{amsmath, amssymb}
\usepackage{epsfig, palatino}
\usepackage{pstricks,pst-node,pst-tree}
\usepackage{epic}

\usepackage{mathrsfs}

\usepackage{ae} 
\usepackage[T1]{fontenc}
\usepackage[ansinew]{inputenc}
\usepackage{amsmath}
\usepackage{amssymb}
\usepackage{graphicx}
\usepackage{color}
\definecolor{darkblue}{cmyk}{0.9,0.9,0,0}
\usepackage[colorlinks=true,linkcolor=darkblue,citecolor=darkblue,urlcolor=darkblue]{hyperref}
\usepackage{cite}
\usepackage{hyperref}
\usepackage{wasysym}

\newcommand{\comment}[1]{}

\newcommand{\beq}{\begin{equation}}
\newcommand{\eeq}{\end{equation}}
\newcommand{\beqq}{\begin{equation*}}
\newcommand{\eeqq}{\end{equation*}}
\newcommand\beqa{\begin{eqnarray}}
\newcommand\eeqa{\end{eqnarray}}
\newcommand\beqaa{\begin{eqnarray*}}
\newcommand\eeqaa{\end{eqnarray*}}
\newcommand\bea{\begin{array}}
\newcommand\eea{\end{array}}

\def\XXint#1#2#3{{\setbox0=\hbox{$#1{#2#3}{\int}$}
\vcenter{\hbox{$#2#3$}}\kern-.5\wd0}}

\newcommand{\nn}{\nonumber}

\newcommand{\neqa}{\nonumber\end{eqnarray}} 
\newcommand{\la}[1]{\label{#1}}

\def\tr{{\rm tr~}}

\renewcommand{\d}{\partial}

\newcommand{\<}{{\langle}}
\renewcommand{\>}{{\rangle}}

\newcommand{\re}{\relax{\rm I\kern-.18em R}}

\renewcommand{\sp}{p\hspace{-.40em}/}

\newcommand{\ft}[2]{{\textstyle\frac{#1}{#2}}}

\def\su2{{SU(2)}}

\def\[{\left[}
\def\]{\right]}

\def\({\left(}
\def\){\right)}
\def\[{\left[}
\def\]{\right]}

\def\<{\langle}
\def\>{\rangle}

\def\i2{\frac{i}{2}}

\def\spi{\relax{\rm \pi\kern-0.5em /}}
\def\sA{\relax{\rm A\kern-0.5em /}}
\def\sp{\relax{\rm p\kern-0.5em /}}
\def\sd{\relax{\rm \d\kern-0.5em /}}
\def\sk{\relax{\rm k\kern-0.5em /}}
\def\sn{\relax{\rm n\kern-0.5em /}}
\def\sl{\relax{\rm l\kern-0.5em /}}
\def\sP{\relax{\rm P\kern-0.7em /}}
\def\sBethe{\relax{\rm \Bethe\kern-0.5em /}}
\def\cN{{\cal N}}

\def\One{1\hskip-.16cm1}

\def\cR{{\cal R}}
\def\cO{{\cal O}}

\def\cN{{\cal N}}

\def\cW{{\cal W}}

\def\2F1{\,_2{\rm F}_1}

 \newcommand{\Blue}[1]{{\color{blue}#1\color{black}}}
\newcommand{\Red}[1]{{\color{red}#1\color{black}}}

\usepackage{varioref}
\usepackage{makeidx}
\makeindex

\usepackage[english]{babel}
\begin{document}

\thispagestyle{empty}

\renewcommand{\thefootnote}{\fnsymbol{footnote}}
\setcounter{page}{1}
\setcounter{footnote}{0}
\setcounter{figure}{0}
\begin{center}
$$$$
{\Large\textbf{\mathversion{bold}
Space-time S-matrix and Flux-tube S-matrix II.\\ 
Extracting and Matching Data
}\par}

\vspace{1.0cm}

\textrm{Benjamin Basso$^{\displaystyle\pentagon}$, Amit Sever$^{{\displaystyle\pentagon},{\displaystyle\Box}}$ and Pedro Vieira$^{\displaystyle\pentagon}$}
\\ \vspace{1.2cm}
\footnotesize{
\textit{$^{\displaystyle\pentagon}$Perimeter Institute for Theoretical Physics,
Waterloo, Ontario N2L 2Y5, Canada\\
$^{\displaystyle\Box}$School of Natural Sciences, Institute for Advanced Study, Princeton, NJ 08540, USA
}  
\vspace{4mm}
}

\par\vspace{1.5cm}

\textbf{Abstract}\vspace{2mm}
\end{center}
We elaborate on a non-perturbative formulation of scattering amplitudes/null polygonal Wilson loops in planar ${\cal N}=4$ Super-Yang-Mills theory. The construction is based on a decomposition of the Wilson loop into elementary building blocks named {\it pentagon transitions}. Our discussion expands on a previous letter of the authors where these transitions were introduced and analyzed for the so-called gluonic excitations. In this paper we revisit these transitions and extend the analysis to the sector of scalar excitations. We restrict ourselves to the single particle transitions and bootstrap their finite coupling expressions using a set of axioms. Besides these considerations, the main focus of the paper is on the extraction of perturbative data from scattering amplitudes at weak coupling and its comparison against the proposed pentagon transitions. We present several tests for both the hexagon and heptagon (MHV and NMHV) amplitudes up to two- and three-loop orders. In attached notebooks we provide explicit higher-loop predictions obtained from our method.

\noindent

\setcounter{page}{1}
\renewcommand{\thefootnote}{\arabic{footnote}}
\setcounter{footnote}{0}

 \def\nref#1{{(\ref{#1})}}

\newpage

\tableofcontents

    \parskip 5pt plus 1pt   \jot = 1.5ex
\section{Introduction}
We often sum over states to compute physical quantities. We do it when computing partition functions in quantum or statistical physics. Also, when studying correlation functions in a conformal field theory, we can successively apply the operator product expansion (OPE) to write a general $n$-point function as multiple sums over the states generated by the fusion of the local operators.
What is less widely known is that a similar strategy can be applied for computing the vacuum expectation values of null polygonal Wilson loops $\cal W$ in conformal gauge theories~\cite{OPEpaper}.

This method, which parallels the one for correlators, goes under the name of OPE as well. It entails however summing over a rather different class of states: namely, the complete set of excitations $\psi$ of the flux tube supported by two null Wilson lines~\cite{AldayMaldacena}. For a generic polygon we have to perform this sum as many times as needed to fully decompose the evolution of the flux-tube state along the loop. This has to be done $n-5$ times for an $n$-edged Wilson loop, in accord with the counting of conformal invariants of the loop. In a previous communication \cite{short} we pushed ahead with this idea and proposed that these multiple sums should be organized into the sequence
\beq\la{decompositionIntro}
\mathcal{W}=\sum_{\psi_i}  \[\prod\limits_{i=1}^{n-5} e^{-E_i\tau_i+ip_i\sigma_i+im_i\phi_i} \]P(0|\psi_1)\ P({\psi}_1|\psi_2)\ \dots\  P({\psi}_{n-6}|\psi_{n-5})\ P({\psi}_{n-5}|0) \,,
\eeq
which reflects the decomposition~(\ref{OPEvsOPE}) of the WL into a sequence of overlapping squares and pentagons. The geometrical data of the loop, or equivalently the set of (4D) cross ratios $\{\tau_i, \sigma_i, \phi_i\}$, appears in the first factor only and couples directly to the energy, momentum, and angular momentum $\{E_i, p_i, m_i\}$ of the flux-tube state $\psi_i$ defined on the $i$-th square. The other elementary building blocks, which arise from the unions of two consecutive squares, are the \textit{pentagon transitions} $P(\psi_i|\psi_j)$ between the states $\psi_i$ and $\psi_j$. They are independent of the global geometry and fully determined by the flux-tube dynamics. They are the analogues of the structure constants for local operators: 
\vspace{-2mm}
\beq
\centering
\def\svgwidth{11cm}
\begingroup%
  \makeatletter%
  \providecommand\color[2][]{%
    \errmessage{(Inkscape) Color is used for the text in Inkscape, but the package 'color.sty' is not loaded}%
    \renewcommand\color[2][]{}%
  }%
  \providecommand\transparent[1]{%
    \errmessage{(Inkscape) Transparency is used (non-zero) for the text in Inkscape, but the package 'transparent.sty' is not loaded}%
    \renewcommand\transparent[1]{}%
  }%
  \providecommand\rotatebox[2]{#2}%
  \ifx\svgwidth\undefined%
    \setlength{\unitlength}{916.42021484bp}%
    \ifx\svgscale\undefined%
      \relax%
    \else%
      \setlength{\unitlength}{\unitlength * \real{\svgscale}}%
    \fi%
  \else%
    \setlength{\unitlength}{\svgwidth}%
  \fi%
  \global\let\svgwidth\undefined%
  \global\let\svgscale\undefined%
  \makeatother%
  \begin{picture}(1,0.45449334)%
    \put(0,0){\includegraphics[width=\unitlength]{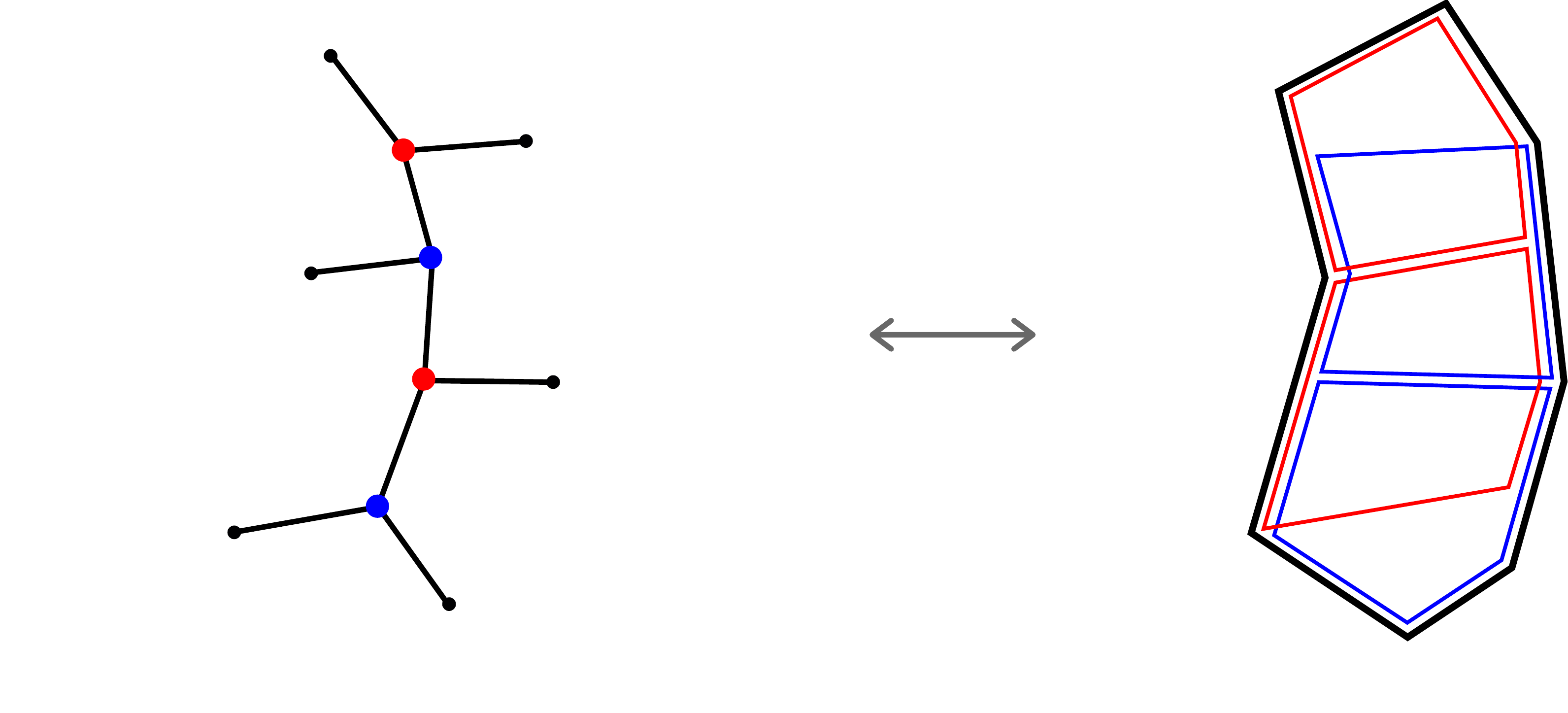}}%
    \put(0.88446196,0.1620079){\color[rgb]{0,0,0}\makebox(0,0)[lb]{\smash{$\psi_1$}}}%
    \put(0.90715897,0.24057448){\color[rgb]{0,0,0}\makebox(0,0)[lb]{\smash{$\psi_2$}}}%
    \put(0.9001266,0.31449082){\color[rgb]{0,0,0}\rotatebox{15.43987883}{\makebox(0,0)[lb]{\smash{$\psi_3$}}}}%
    \put(0.26465896,0.15502421){\color[rgb]{0,0,0}\makebox(0,0)[lb]{\smash{$\psi_1$}}}%
    \put(0.2821182,0.24581225){\color[rgb]{0,0,0}\makebox(0,0)[lb]{\smash{$\psi_2$}}}%
    \put(0.27338858,0.32088698){\color[rgb]{0,0,0}\makebox(0,0)[lb]{\smash{$\psi_3$}}}%
    \put(0.29421808,0.05099856){\color[rgb]{0,0,0}\makebox(0,0)[lb]{\smash{$\cO$}}}%
    \put(0.10767062,0.09888104){\color[rgb]{0,0,0}\makebox(0,0)[lb]{\smash{$\cO$}}}%
    \put(0.36105039,0.20024005){\color[rgb]{0,0,0}\makebox(0,0)[lb]{\smash{$\cO$}}}%
    \put(0.15904494,0.26042722){\color[rgb]{0,0,0}\makebox(0,0)[lb]{\smash{$\cO$}}}%
    \put(0.34315169,0.35726105){\color[rgb]{0,0,0}\makebox(0,0)[lb]{\smash{$\cO$}}}%
    \put(0.19102991,0.43200943){\color[rgb]{0,0,0}\makebox(0,0)[lb]{\smash{$\cO$}}}%
    \put(-0.00084307,0.00211264){\color[rgb]{0,0,0}\makebox(0,0)[lb]{\smash{$\text{OPE for Correlation Functions}$}}}%
    \put(0.71323982,0.00211264){\color[rgb]{0,0,0}\makebox(0,0)[lb]{\smash{$\text{OPE for Wilson Loops}$}}}%
  \end{picture}%
\endgroup%
 
\la{OPEvsOPE}
\eeq

Whenever it applies, the pentagon decomposition provide us with the complete information on the Wilson loop $\cal W$. This remains true when we have a marginal coupling $\lambda$ in the theory: in this case all the OPE data become coupling dependent and the decomposition holds regardless of the specific value of the coupling as long as the spectrum of flux-tube excitations remains gapped. 
This is the situation in  planar $\mathcal{N}=4$ Super-Yang-Mills theory for generic values of the 't Hooft coupling $\lambda = g^2_{YM}N$. What is even more interesting is that in this theory null polygonal Wilson loops are believed to be the same as color-ordered scattering amplitudes~\cite{AM,AmplitudeWilson}. By means of this duality the pentagon decomposition should then stand as a non-perturbative definition of scattering amplitudes in this theory.

For both practical and theoretical reasons, the most relevant pentagon transitions are those associated to single-particle states. These are the lightest states of the flux tube at generic coupling. They form the seed for building up the contributions of heavier states that display more intricate transitions. In this paper we will expand on~\cite{short} and study two specific, but prominent, examples of single-particle transitions: the ones corresponding to the scalar and the gluonic excitations of the flux tube, respectively. They form the almost complete list of the fundamental, i.e., twist one, excitations of the flux tube~\cite{BenDispPaper} -- the only ones missing being the fermions, which will be analyzed in a future publication.

From a practical view point, the transitions for the lightest excitations are the easiest ones to extract as they dominate at large $\tau_i$ -- meaning that they decay the slowest in the multi-collinear limit $\tau_i\to \infty$. Their contributions can thus be neatly separated from the rest, especially at weak coupling.
These transitions also occupy a distinctive position numerically. Clearly, if we manage to compute exactly the pentagon transitions for all the lightest states of the flux tube, we will already have a good approximation to $\mathcal{W}$ for generic kinematics. The situation is similar to the conventional OPE for correlators. If we have at our disposal the OPE structure constants for the first few operators with lowest scaling dimensions, summing over them already gives an excellent estimate of an higher-point correlation function. This is so unless we are close to the radius of convergency of the OPE, in which case we might need a lot more operators to get a good approximation. Of course, in all cases, if we sum over all the states then we get the full exact result without any approximation.

This way of organizing our expansion -- by the importance of the states that are \textit{flowing}~-- should be contrasted with more conventional approaches based on perturbation theory. In perturbation theory we power expand around $\lambda=0$. The term proportional to $\lambda^0$ is the tree level result and the terms suppressed as $\lambda^l$ are called the $l$-th loops contributions. If we keep all terms in the perturbative expansion we can in principle re-sum the series and obtain the exact result without any approximation.  The two expansions are clearly different.   At any loop order, we typically need to sum over all the states to get the complete result from the OPE. Reciprocally, to get the exact contribution of any given state we need to re-sum all orders in perturbation theory. Still, the data from one approach provides valuable constraints and checks on the other one.

In this paper we will explain how to make use of this interesting interplay {between the OPE expansion and perturbation theory}  and illustrate how to efficiently extract information about the pentagon transitions from the abundant knowledge and literature on scattering amplitudes/Wilson loops at weak coupling. We shall make extensive use, in particular, of the work \cite{Bourjaily:2013mma} which builds on the recent developments about the all-loop integrand \cite{ArkaniHamed:2010kv,ArkaniHamed:2009dn} and where all the one-loop amplitudes were given in manifestly conformally invariant notation. Another source of invaluable data comes from higher-loop results based on the Q-bar equation \cite{Qbar} or its predecessor \cite{simonHep}, as well as on the symbol technology \cite{VerguPaper,Lance,Lance-Drummond,Dixon} which extend results based on more traditional techniques \cite{Remainder,DelDuca:2010zg}.
Our discussion here will be centered around the six and seven points amplitudes corresponding to the hexagon and heptagon (super) Wilson loops. These are the simplest possible ones and by far the most well studied both in the OPE framework and in perturbation theory. They play a pivotal role in our approach since they give us the most direct access to the pentagon transitions of interest.

From a theoretical perspective the single-particle pentagon transitions are obviously the most fundamental ones. For them, a set of axioms or \textit{bootstrap} has been proposed in~\cite{short}. It lays the ground for the application of the powerful integrability technology (see~\cite{review} for a recent review) and allows one to come up with an educated guess for the pentagon transitions at finite coupling. This was illustrated in~\cite{short} for the gluonic transitions. In this paper we elaborate on this conjecture and generalize it to the scalar transitions. We shall provide further evidence for our finite coupling expressions by matching them against data up to the available orders in perturbation theory. We will also present several predictions for the OPE behaviour of Wilson loops that should be very helpful for constraining the expectation values of these loops at higher-loop orders.

Certainly the most remarkable property of the pentagon transitions is their relation to the flux tube S-matrix. This is what we called the \textit{fundamental relation} in \cite{short}. Precisely,  
it states that the pentagon transition $P(\psi_1|\psi_2)$ between two single-particle states is related to the transition $P(\psi_2|\psi_1)$ by a very simple factor: the flux-tube S-matrix amplitude between states $\psi_1$ and $\psi_2$. A single-particle state can always be parametrized by its momentum such that we should observe that
\beq
\centering\def\svgwidth{9cm}
\begingroup%
  \makeatletter%
  \providecommand\color[2][]{%
    \errmessage{(Inkscape) Color is used for the text in Inkscape, but the package 'color.sty' is not loaded}%
    \renewcommand\color[2][]{}%
  }%
  \providecommand\transparent[1]{%
    \errmessage{(Inkscape) Transparency is used (non-zero) for the text in Inkscape, but the package 'transparent.sty' is not loaded}%
    \renewcommand\transparent[1]{}%
  }%
  \providecommand\rotatebox[2]{#2}%
  \ifx\svgwidth\undefined%
    \setlength{\unitlength}{497.62695312bp}%
    \ifx\svgscale\undefined%
      \relax%
    \else%
      \setlength{\unitlength}{\unitlength * \real{\svgscale}}%
    \fi%
  \else%
    \setlength{\unitlength}{\svgwidth}%
  \fi%
  \global\let\svgwidth\undefined%
  \global\let\svgscale\undefined%
  \makeatother%
  \begin{picture}(1,0.45798404)%
    \put(0,0){\includegraphics[width=\unitlength]{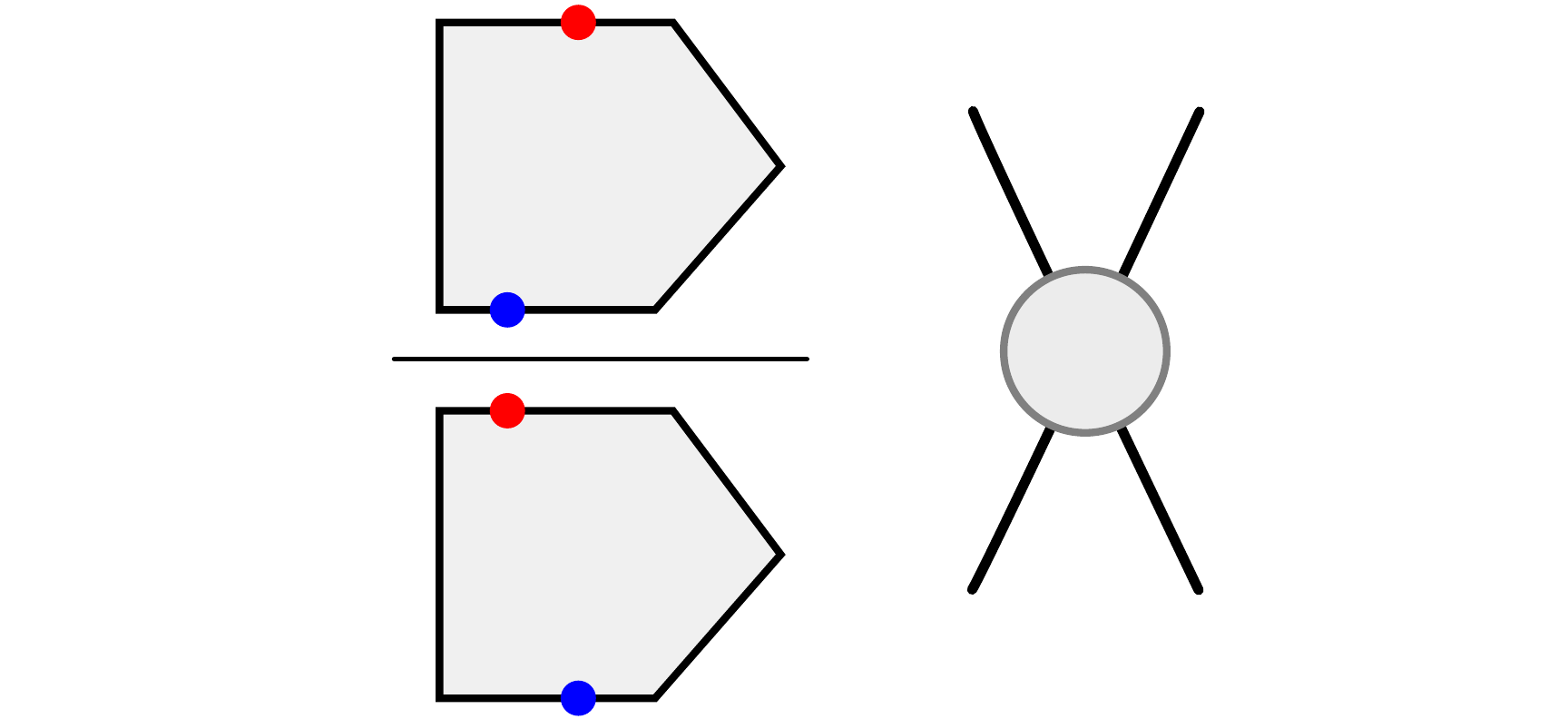}}%
    \put(-0.02129325,0.21813102){\color[rgb]{0,0,0}\makebox(0,0)[lb]{\smash{$\displaystyle \frac{P(p_1|p_2)}{P(p_2|p_1)}\ =$}}}%
    \put(0.35740864,0.39927517){\color[rgb]{0,0,0}\makebox(0,0)[lb]{\smash{$p_2$}}}%
    \put(0.30450014,0.28807425){\color[rgb]{0,0,0}\makebox(0,0)[lb]{\smash{$p_1$}}}%
    \put(0.55102303,0.21813102){\color[rgb]{0,0,0}\makebox(0,0)[lb]{\smash{$=$}}}%
    \put(0.78895227,0.21813102){\color[rgb]{0,0,0}\makebox(0,0)[lb]{\smash{$\displaystyle = \ S(p_1,p_2)$}}}%
    \put(0.76250654,0.05003542){\color[rgb]{0,0,0}\makebox(0,0)[lb]{\smash{$p_2$}}}%
    \put(0.60023035,0.04994358){\color[rgb]{0,0,0}\makebox(0,0)[lb]{\smash{$p_1$}}}%
    \put(0.67641817,0.21813102){\color[rgb]{0,0,0}\makebox(0,0)[lb]{\smash{$\displaystyle S$}}}%
    \put(0.31093066,0.15303334){\color[rgb]{0,0,0}\makebox(0,0)[lb]{\smash{$p_1$}}}%
    \put(0.3445476,0.03916606){\color[rgb]{0,0,0}\makebox(0,0)[lb]{\smash{$p_2$}}}%
  \end{picture}%
\endgroup%
 \la{funrel}
\eeq
This relation neatly relates the space-time S-matrix which is built out of the pentagon transitions $P(p_1|p_2)$ to the flux-tube S-matrix $S(p_1, p_2)$. As an illustration of how this relation works we will now consider the simplest possible amplitude and to the lowest possible order in perturbation theory: a tree level NMHV amplitude. By choosing carefully a component of this amplitude we shall see how the equality~(\ref{funrel}) comes out for a complex scalar excitation $Z$ of the flux tube.

\subsection{The Flux-tube S-matrix...} 

The flux-tube S-matrix can be obtained in many different ways, reflecting the many equivalent pictures for the flux tube itself~\cite{AldayMaldacena}.
This flux can be viewed as a twist-two operator at large spin, or as the dual of the so-called Gubser-Klebanov-Polyakov (GKP) string~\cite{GKP}, or again as a $\Pi$-shaped null Wilson line~\cite{Belitsky,Wang}. The first picture, which relates directly to the integrable spin-chain description, is by far the best one computationally. Nonetheless it is the last one that carries the clearest physical interpretation. In this case, the flux tube excitations are represented by insertions of adjoint fields at positions $x(\sigma_1),x(\sigma_2),$ etc., along the Wilson line, here parameterized with the null coordinate $x=x^-$. For example, a complex scalar field $Z$ can be inserted (conformally along the line) by 
\beq\la{scalarinsertion}
\star\,{\cal Z}(\sigma)\,\star\equiv \star|\d_\sigma x(\sigma)|^s Z\(x(\sigma)\)\star
\eeq
 where the stars stand for the Wilson line and where $s=1/2$ is the $SL(2)$ conformal spin of the scalar field.  More generally a state with $N$ scalars is written as
\beq
\int\limits_{-\infty < \sigma_1 < \dots < \sigma_N < +\infty}\!\!\!\!\!\!\!\!\!\!\!\!\!\!d\sigma_1\dots d\sigma_N\ \psi(\sigma_1,\dots,\sigma_N)\,\,\star{\cal Z}(\sigma_1)\star\dots\star{\cal Z}(\sigma_N)\,\star \,, \la{insertionWL}
\eeq
with $\psi(\sigma_1,\dots,\sigma_N)$ the wave function in position space. The eigenstates of the flux tube can be fixed by diagonalizing the (euclidean) time evolution $-\partial_{\tau} = N +\mathcal{H}$ with the light-ray Hamiltonian
\beq\label{lightrayH}
{\cal H}\cdot \psi=2g^2\sum_{j=1}^{N}  \int\limits_{\sigma_{j-1}-\sigma_j}^{\sigma_{j+1}-\sigma_j}\!\!\!\!{dt\over\sinh|t|}\[e^{-|t|}\psi(\dots)-\psi(\dots,\sigma_{j-1},\sigma_j+t,\sigma_{j+1},\dots)\] \, ,
\eeq
here at one loop with $g^2 \equiv \lambda/(4\pi)^2$ and where $\sigma_{N+1} = -\sigma_0 = \infty$.\footnote{For excitations (i.e, fields) with conformal spin $s\ne\ft{1}{2}$ the Hamiltonian takes a slightly more complicated form~\cite{toappear}.}
It is of the familiar type~\cite{Braun,Belitsky,Wang} and acts by displacing local insertions between their nearest neighbours. It commutes with the momentum $p = -i\sum^N_i\partial_{\sigma_i}$ which then solves the one-particle problem immediately. The plane wave $\psi(\sigma)= e^{ip\sigma}$ diagonalizes $\mathcal{H}$ with the one-loop energy 
\beq
\gamma(p) = 2g^2(\psi(s+i\tfrac{p}{2})+\psi(s-i\tfrac{p}{2})-2\psi(1))\, ,
\eeq
and $\psi(z) = \partial_z\log\Gamma(z)$, in agreement with the spin-chain prediction~\cite{BGK,BenDispPaper,Straps}.

To read out the S-matrix $S(p_1, p_2)$ one should solve the two-particle problem $\mathcal{H}\cdot \psi=(\gamma(p_1)+\gamma(p_2))\psi$ whose solution is of the generic type
\beq
\psi(\sigma_1,\sigma_2) = e^{ip_1 \sigma_1+ip_2 \sigma_2} f_{p_1,p_2}({\sigma_1-\sigma_2}) + e^{ip_2 \sigma_1+ip_1\sigma_2}S(p_1 ,p_2)f_{p_2,p_1}({\sigma_1-\sigma_2})\, , \la{2ptWave}
\eeq
with some partial wave $f_{p_1,p_2}(\sigma)$ normalized to $1$ at large separation $\sigma \rightarrow -\infty$. It is a difficult task to get $f$ by using the integral operator $\mathcal{H}$ directly. A shortcut is found by using integrability: the crucial point is that $\mathcal{H}$ commutes with the  diagonal component of a transfer matrix, whose construction follows the habitual procedure (see~\cite{Wang} for the case at hand). This turns the problem into an hypergeometric differential equation, whose solution is given by~(\ref{2ptWave}) with
\beq
f_{p_1,p_2}(\sigma) = {}_2F_1\!\left(1-s+\tfrac{ip_1}{2}, 1-s-\tfrac{ip_2}{2}, 1+\tfrac{ip_1}{2}-\tfrac{ip_2}{2} |e^{2\sigma}\right)\, .
\eeq
The regularity of the wave function at $\sigma_1=\sigma_2$ fixes then the S-matrix,
\beq\la{scalarSmatrix}
S(p_1, p_2) = \frac{\Gamma(s-\frac{ip_1}{2})\Gamma(s+\frac{ip_2}{2})\Gamma(\frac{ip_1}{2}-\frac{ip_2}{2})}{\Gamma(s+\frac{ip_1}{2})\Gamma(s-\frac{ip_2}{2})\Gamma(\frac{ip_2}{2}-\frac{ip_1}{2})}\, ,
\eeq
in accord again with the spin-chain expression (which could be extracted from~\cite{BGK}). This route to the flux-tube S-matrix makes its extraction physically more transparent. It shares also many similarities with the analysis of the Balitsky-Fadin-Kuraev-Lipatov (BFKL) Hamiltonian performed in~\cite{Bartels}, hence promising an interesting interplay between the two approaches. At higher loops, the integrable spin-chain description takes over and predicts the flux-tube S-matrix at any value of the coupling, as shown in~\cite{toappearAdam,Peng} for scalar excitations.

\subsection{... and the Spacetime S-matrix}\la{Intropart2}
The usual Wilson loop computes maximal helicity violating (MHV) amplitudes~\cite{AM,AmplitudeWilson}. The (Next-to)$^k$MHV amplitudes are given by the super Wilson loop \cite{superloopskinner,superloopsimon} up to a tree-level MHV prefactor. The various components of the N$^k$MHV amplitudes correspond to different field insertions along the various edges and cusps of the Wilson loop. Some of these components have a simple OPE interpretation. This is particularly true for those where we just insert at tree level a scalar $Z$ and its conjugate $\bar Z$ at the bottom and top cusps of the polygon, see~\cite{SuperOPE, Wang}. For them we expect the same kind of decomposition as in (\ref{decompositionIntro}) to apply. The only obvious difference is that we should sum over states with the given overall R-charge. This implies in particular that the first pentagon transition $P(0|\psi_1)$ no longer stands for the creation of the state $\psi_1$ from the vacuum, but instead represents the creation of the state $\psi_1$ in the background of a scalar insertion.\footnote{The operatorial definition of the various components of the super Wilson loop might receive loop corrections. In this paper we will refer to these components by their tree level form. E.g., a \textit{scalar component} is a component that corresponds at tree level to a scalar insertion at a cusp. At higher loops this component also includes for instance fermions integrated on the neighboring edges \cite{superloopsimon,Andrei}. Similarly, throughout the paper, when we refer to an excitation being inserted at a cusp it should be understood that this is a tree level concept and that at loop level the excitation will typically be more delocalized. \la{ftDisclaimer}
}
To signal this  difference we add a star to this transition and write it as $P_*(0|\psi_1)$ -- similarly for the last transition that lies at the top of the loop. 

Consider now an heptagon Wilson loop with one complex scalar $Z$ inserted at the bottom cusp $x$ and a conjugate scalar $\bar Z$ at the top cusp $y$. These scalars are coming with certain kinematical factors made out of the usual spinor helicity brackets ($1/\<7\,1\>$ and $1/\<4\,5\>$ respectively) and with an overall $1/g$ each \cite{superloopsimon}. Those are conformal factors that insure the conformal invariance of the final result. 
To evaluate the Wilson loop at tree level we simply connect the two scalars by the free scalar propagator $g^2/(x-y)^2$. Hence \vspace{-.2mm} 
\beq
\centering
\def\svgwidth{8.5cm}
\begingroup%
  \makeatletter%
  \providecommand\color[2][]{%
    \errmessage{(Inkscape) Color is used for the text in Inkscape, but the package 'color.sty' is not loaded}%
    \renewcommand\color[2][]{}%
  }%
  \providecommand\transparent[1]{%
    \errmessage{(Inkscape) Transparency is used (non-zero) for the text in Inkscape, but the package 'transparent.sty' is not loaded}%
    \renewcommand\transparent[1]{}%
  }%
  \providecommand\rotatebox[2]{#2}%
  \ifx\svgwidth\undefined%
    \setlength{\unitlength}{544.13125bp}%
    \ifx\svgscale\undefined%
      \relax%
    \else%
      \setlength{\unitlength}{\unitlength * \real{\svgscale}}%
    \fi%
  \else%
    \setlength{\unitlength}{\svgwidth}%
  \fi%
  \global\let\svgwidth\undefined%
  \global\let\svgscale\undefined%
  \makeatother%
  \begin{picture}(1,0.71471807)%
    \put(0,0){\includegraphics[width=\unitlength]{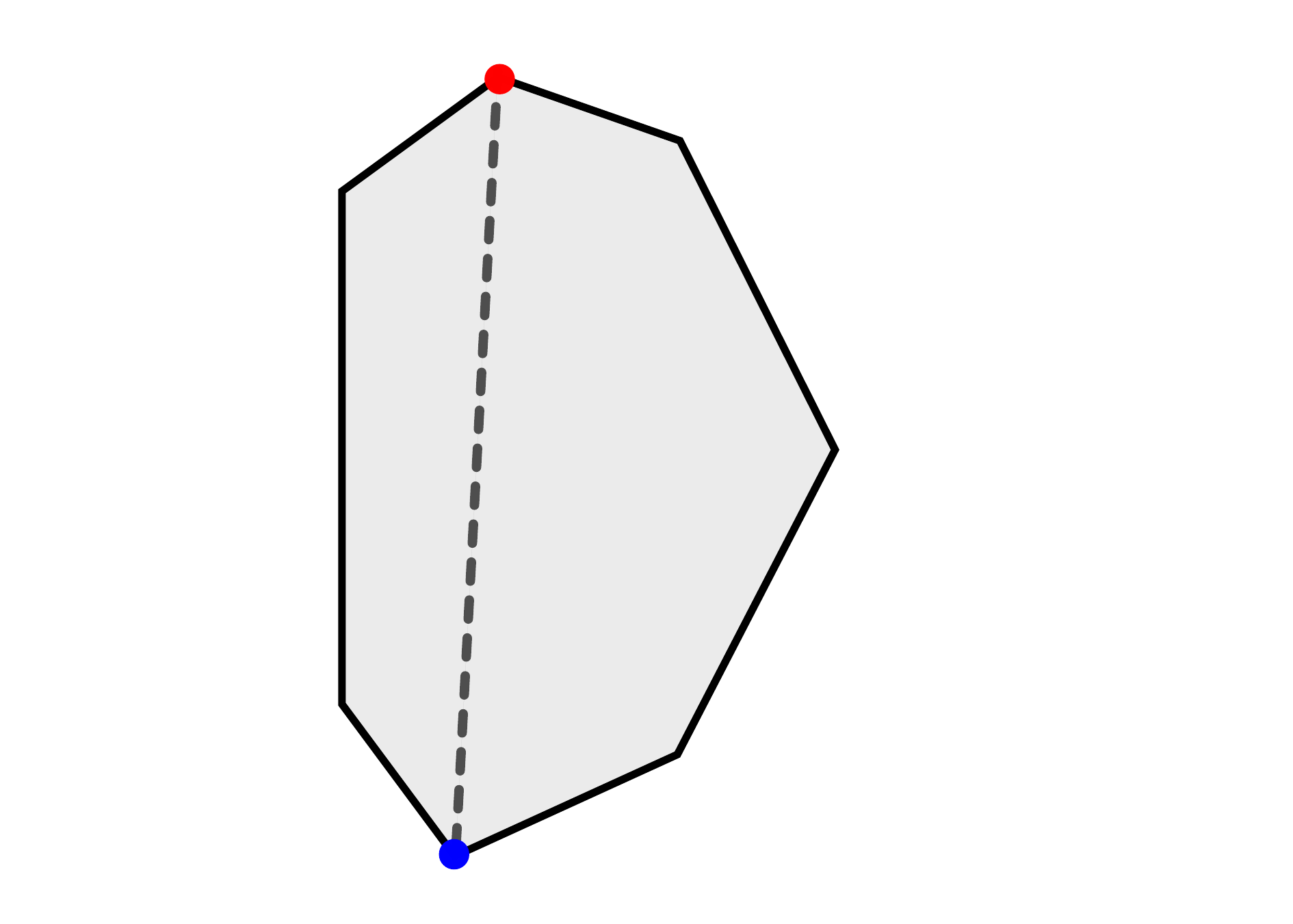}}%
    \put(0.26771593,0.0869464){\color[rgb]{0.4,0.4,0.4}\makebox(0,0)[lb]{\smash{7}}}%
    \put(0.45296535,0.05754173){\color[rgb]{0.4,0.4,0.4}\makebox(0,0)[lb]{\smash{1}}}%
    \put(0.59704815,0.22220789){\color[rgb]{0.4,0.4,0.4}\makebox(0,0)[lb]{\smash{2}}}%
    \put(0.59998861,0.49273085){\color[rgb]{0.4,0.4,0.4}\makebox(0,0)[lb]{\smash{3}}}%
    \put(0.4588462,0.64269467){\color[rgb]{0.4,0.4,0.4}\makebox(0,0)[lb]{\smash{4}}}%
    \put(0.29123958,0.61623047){\color[rgb]{0.4,0.4,0.4}\makebox(0,0)[lb]{\smash{5}}}%
    \put(0.2265493,0.3545289){\color[rgb]{0.4,0.4,0.4}\makebox(0,0)[lb]{\smash{6}}}%
    \put(0.39269439,0.37071978){\color[rgb]{0,0,0}\makebox(0,0)[lb]{\smash{${\displaystyle{g^2\over(x-y)^2}}$}}}%
    \put(0.28241826,0.00671369){\color[rgb]{0,0,0}\makebox(0,0)[lb]{\smash{$\Blue{Z(x)}$}}}%
    \put(0.31728381,0.68344116){\color[rgb]{0,0,0}\makebox(0,0)[lb]{\smash{$\Red{\bar Z(y)}$}}}%
    \put(0.37759928,0.12834132){\color[rgb]{0,0,0}\makebox(0,0)[lb]{\smash{$\Blue{1\over g\<7\,1\>}$}}}%
    \put(0.39860261,0.57781272){\color[rgb]{0,0,0}\makebox(0,0)[lb]{\smash{$\Red{1\over g\<4\,5\>}$}}}%
    \put(0.7036311,0.35503322){\color[rgb]{0,0,0}\makebox(0,0)[lb]{\smash{${\displaystyle=
{1\over\Blue{\<7\,1\>}
(x-y)^2
\Red{\<4\,5\>}}}$}}}%
    \put(-0.00208098,0.35503322){\color[rgb]{0,0,0}\makebox(0,0)[lb]{\smash{${\displaystyle{\cal R}^{(7145)}_\text{tree}=}$}}}%
  \end{picture}%
\endgroup%

\label{TreeLevelScalar}
\eeq
where we use blue and red colors to indicate the two conjugate field insertions. It could hardly be simpler. The final expression can be written in terms of four-component momentum twistors, which are briefly reviewed in appendix~\ref{geomeryappendix}. We get
\beq\la{treelevelheptagon}
{\cal R}_\text{tree}^{(7145)}={1\over\<7,1,4,5\>} \,,
\eeq
where the bracket in the denominator stands for the determinant of the four momentum twistors. The result is now manifestly conformally invariant, since conformal transformations act linearly as $SL(4)$ transformations on the twistors.

To make contact with the pentagon decomposition (\ref{decompositionIntro}), we should express the result (\ref{treelevelheptagon}) in terms of the OPE variables $\{\sigma_i,\tau_i,\phi_i\}$. The procedure, which was previously explained in~\cite{heptagonPaper,SuperOPE,short}, goes as follows. The heptagon is composed from a sequence of four squares: 
\beq
\centering
\def\svgwidth{5cm}
\begingroup%
  \makeatletter%
  \providecommand\color[2][]{%
    \errmessage{(Inkscape) Color is used for the text in Inkscape, but the package 'color.sty' is not loaded}%
    \renewcommand\color[2][]{}%
  }%
  \providecommand\transparent[1]{%
    \errmessage{(Inkscape) Transparency is used (non-zero) for the text in Inkscape, but the package 'transparent.sty' is not loaded}%
    \renewcommand\transparent[1]{}%
  }%
  \providecommand\rotatebox[2]{#2}%
  \ifx\svgwidth\undefined%
    \setlength{\unitlength}{344.20854492bp}%
    \ifx\svgscale\undefined%
      \relax%
    \else%
      \setlength{\unitlength}{\unitlength * \real{\svgscale}}%
    \fi%
  \else%
    \setlength{\unitlength}{\svgwidth}%
  \fi%
  \global\let\svgwidth\undefined%
  \global\let\svgscale\undefined%
  \makeatother%
  \begin{picture}(1,1.07377793)%
    \put(0,0){\includegraphics[width=\unitlength]{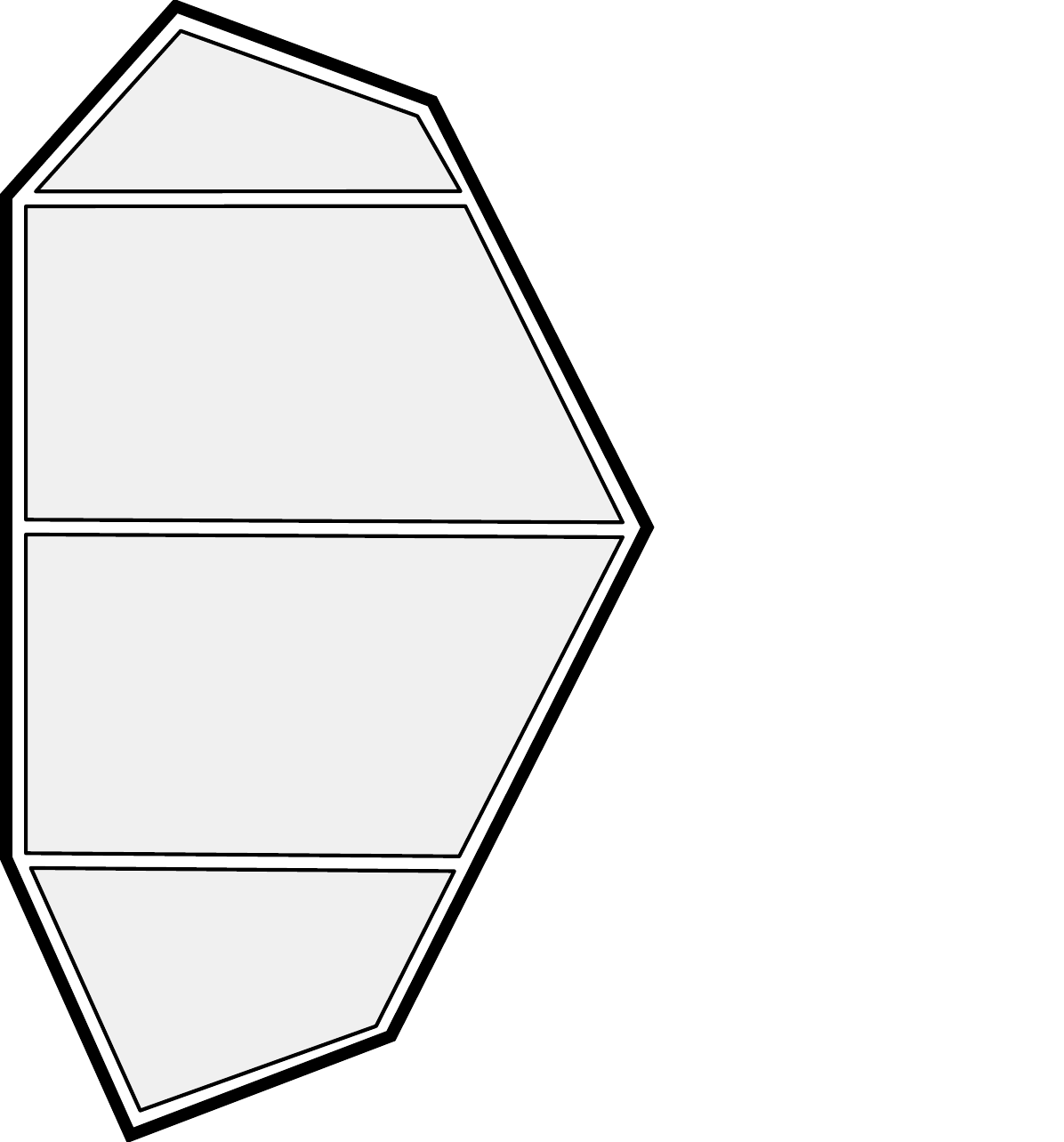}}%
    \put(0.05677398,0.40274849){\color[rgb]{0,0,0}\makebox(0,0)[lb]{\smash{$\{\tau_1,\sigma_1,\phi_1\}$}}}%
    \put(0.05677398,0.72813263){\color[rgb]{0,0,0}\makebox(0,0)[lb]{\smash{$\{\tau_2,\sigma_2,\phi_2\}$}}}%
    \put(0.46113005,0.94805432){\color[rgb]{0,0,0}\makebox(0,0)[lb]{\smash{$\text{Top square}$}}}%
    \put(0.46113005,0.15783571){\color[rgb]{0,0,0}\makebox(0,0)[lb]{\smash{$\text{Bottom square}$}}}%
    \put(0.57269032,0.39025295){\color[rgb]{0,0,0}\makebox(0,0)[lb]{\smash{$\text{1st middle square}$}}}%
    \put(0.57269032,0.73423046){\color[rgb]{0,0,0}\makebox(0,0)[lb]{\smash{$\text{2nd middle square}$}}}%
  \end{picture}%
\endgroup%

\eeq
It has in particular two middle squares, each of which has three conformal symmetries. These play an important role in the OPE approach since they allow us to parametrize all conformally inequivalent heptagons by acting with the symmetries of the middle squares on the twistors (i.e., on the cusps) located to their bottom.  
The corresponding six parameters, $\{\tau_1,\sigma_1,\phi_1\}$ and $\{\tau_2,\sigma_2,\phi_2\}$, are the six independent conformal cross ratios of an heptagon. The whole construction can be made very explicit by using a particular choice of momentum twistors, as done in (\ref{Heptagon2}) in appendix \ref{gluingappendix}. 
Using these twistors, the determinant in (\ref{treelevelheptagon}) is now expressed in terms of our OPE parameters. To single out the single-particle contribution in (\ref{decompositionIntro}), which is the one of interest here, we then expand at large $\tau_1$ and $\tau_2$. This step has a clear geometrical interpretation and corresponds to the collinear limit where the bottom and top squares are being flattened. We find
\beq
\begin{array}{c}
\displaystyle {\cal R}^{(7145)}_\text{tree} =e^{-\tau_1-\tau_2}\times {1\over e^{\sigma_1-\sigma_2}+e^{\sigma_2-\sigma_1}+e^{\sigma_1+\sigma_2}}+\dots= \,\,\, \\ \\ \\   \\ \\ \\ \\ \\ \\  \\ \\ 
\end{array}  
\centering
\def\svgwidth{4cm}
\begingroup%
  \makeatletter%
  \providecommand\color[2][]{%
    \errmessage{(Inkscape) Color is used for the text in Inkscape, but the package 'color.sty' is not loaded}%
    \renewcommand\color[2][]{}%
  }%
  \providecommand\transparent[1]{%
    \errmessage{(Inkscape) Transparency is used (non-zero) for the text in Inkscape, but the package 'transparent.sty' is not loaded}%
    \renewcommand\transparent[1]{}%
  }%
  \providecommand\rotatebox[2]{#2}%
  \ifx\svgwidth\undefined%
    \setlength{\unitlength}{261.75742188bp}%
    \ifx\svgscale\undefined%
      \relax%
    \else%
      \setlength{\unitlength}{\unitlength * \real{\svgscale}}%
    \fi%
  \else%
    \setlength{\unitlength}{\svgwidth}%
  \fi%
  \global\let\svgwidth\undefined%
  \global\let\svgscale\undefined%
  \makeatother%
  \begin{picture}(1,1.29585293)%
    \put(0,0){\includegraphics[width=\unitlength]{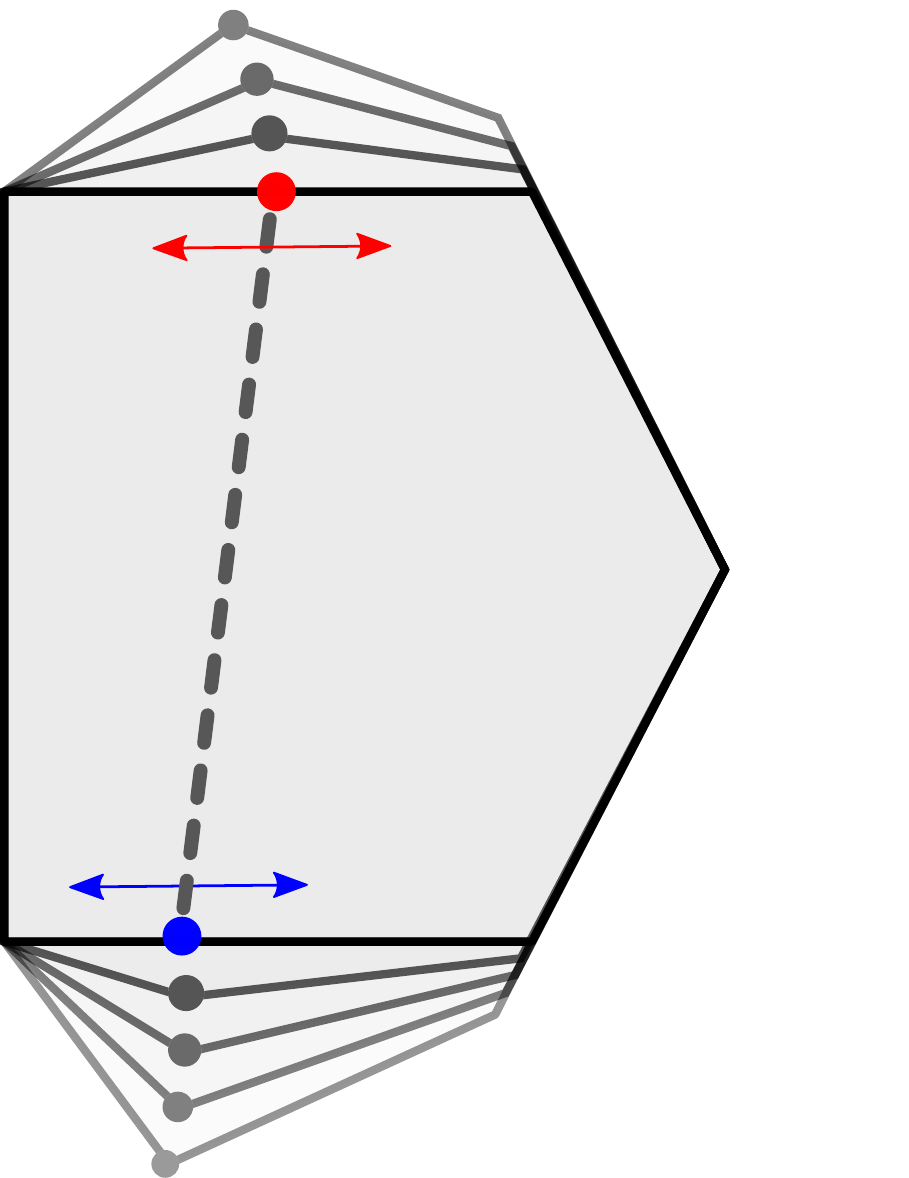}}%
    \put(0.47024812,0.05012647){\color[rgb]{0,0,0}\makebox(0,0)[lb]{\smash{${\color{gray}\tau_1\to \infty}$}}}%
    \put(0.50692345,1.25429487){\color[rgb]{0,0,0}\makebox(0,0)[lb]{\smash{${\color{gray}\tau_2\to \infty}$}}}%
    \put(0.2685348,0.36186555){\color[rgb]{0,0,0}\makebox(0,0)[lb]{\smash{${\color{blue} \sigma_1}$}}}%
    \put(0.35411021,0.94866839){\color[rgb]{0,0,0}\makebox(0,0)[lb]{\smash{${\color{red} \sigma_2}$}}}%
  \end{picture}%
\endgroup%
 \vspace{-2.5cm} \la{leadingscalarposition}
\eeq
Finally, to decompose (\ref{leadingscalarposition}) into energy eigenstates, we go to Fourier space
\beq
{\cal R}^{(7145)}_{\textrm{tree}}=e^{-\tau_1-\tau_2}\int \frac{dp_1\,dp_2}{(2\pi)^2}\,e^{ip_1\sigma_1+ip_2\sigma_2} f(p_1,p_2) +\dots
\eeq
where the integration contours are slightly shifted in the upper-half planes, and read that
\beq
f(p_1,p_2) = \frac{1}{4} \Gamma\(\frac{1}{2}+\frac{ip_1}{2}\)\Gamma\(-\frac{ip_1}{2}-\frac{ip_2}{2}\)\Gamma\(\frac{1}{2}+\frac{ip_2}{2}\)\, . \la{integrand}
\eeq
As explained in \cite{short} -- and in much detail in this paper -- this integrand should be identified with the product 
\beq
f(p_1,p_2)=P_*(0|p_1)\,\mu(p_1)\,P(-p_1|p_2)\,\mu(p_2)\, P_*(-p_2|0)  \la{fP}
\eeq
of pentagon transitions $P$ and square measures $\mu$. Most of them drop out from the ratio $f(-p_1, p_2)/f(-p_2, p_1)$ leaving
\beq
\frac{P(p_1|p_2)}{P(p_2|p_1)}= \frac{\Gamma(\frac{1}{2}-\frac{ip_1}{2})\Gamma(\frac{1}{2}+\frac{ip_2}{2})\Gamma(\frac{ip_1}{2}-\frac{ip_2}{2})}{\Gamma(\frac{1}{2}+\frac{ip_1}{2})\Gamma(\frac{1}{2}-\frac{ip_2}{2})\Gamma(\frac{ip_2}{2}-\frac{ip_1}{2})}\, , \la{ratioP}
\eeq
and revealing the sought-after relation with the scalar flux-tube S-matrix~(\ref{scalarSmatrix}). 

This concludes our illustration of the fundamental relation (\ref{funrel}) for the scalar pentagon transition. We stress again that the computations of the left and right hand sides in (\ref{funrel}) are very different. The left hand side in (\ref{ratioP}) is obtained by Fourier transforming a \textit{tree-level} propagator while the right hand side comes from diagonalizing a \textit{one-loop} Hamiltonian. Still, and remarkably enough, these two distinct computations yield the same result at the end. When writing (\ref{funrel}) we explicitly assume that the phenomenon will persist at any loop order, such that at $l$ loops, for instance, the same NMHV component as above will compute the scalar flux-tube S-matrix coming from the $l+1$ loops light-ray Hamiltonian. The main goal of the paper is to provide evidence for the validity of the fundamental relation at higher loops for both scalar and gluonic transitions.

\section{Geometry and Decomposition} \la{Geometry}

\subsection{Square and Pentagon Transitions}

In this section we comment on the pentagon decomposition (\ref{decompositionIntro}), expanding on the discussion in~\cite{short}. 
A small but significant generalization of this decomposition is to allow for insertions of local fields at both the bottom and top cusps and/or edges of the polygon.
Taking this step enables us to encompass several components of the super Wilson loop~\cite{superloopskinner, superloopsimon} within the OPE framework~\cite{SuperOPE}, besides the usual bosonic loop.
One such example is the component (\ref{TreeLevelScalar}) encountered in the introduction. 
What they all have in common is that they can be computed by means of the sequence of transitions and propagations
\footnote{Null polygonal Wilson loops have well understood (cusp and collinear) UV divergences~\cite{Polyakov:1980ca,KK}. Here $\cal W$ is a (conformally invariant) ratio of polygonal Wilson loops \cite{short} which is free from these divergences. Its precise definition is recalled in section \ref{combiningSec}.}
\beq
\mathcal{W}={\< \text{vac}| \widehat{\mathcal{P}}_*}\, e^{-\tau_{n-5}\hat H+i\sigma_{n-5}{\hat{P}}+i\phi_{n-5}\hat J}  \,\widehat{\mathcal{P}} \dots \widehat{\mathcal{P}}\, e^{-\tau_{1}\hat H+i\sigma_{1}\hat P+i\phi_{1}\hat J}\,{\widehat{\mathcal{P}}_* |\text{vac}\>}\, , \la{decFormal}
\eeq
whose building blocks we now discuss in details.

The representation~(\ref{decFormal}) is based on a tessellation of the $n$-gon WL into a sequence of $n-3$ null squares, as depicted in (\ref{OPEvsOPE}). The propagation over anyone of these squares is simply given by $e^{-\tau_i \hat H+i\sigma_i {\hat{P}}+i\phi_i \hat J}$, where $\hat H$, ${\hat{P}}$ and $\hat J$ are the generators of the three conformal symmetries of the square. Since we always start and end with the vacuum state, the propagations in the first and last squares are trivial. Hence we only have $n-5$ non-trivial exponentials in (\ref{decFormal}), one for each \textit{middle square}. This leads to $3(n-5)$ coordinates $\{\tau_i,\sigma_i,\phi_i\}$ that parametrize all conformally inequivalent polygons (for explicit definitions see (\ref{ccr}) in appendix \ref{gluingappendix}). This matches with the number of independent cross-ratios for a polygon with $n$ null edges.

Each two consecutive squares in the decomposition (\ref{OPEvsOPE}) form a pentagon. Accordingly, the transitions between two neighboring squares is represented by a \textit{pentagon} operator $\widehat{\mathcal{P}}$ in (\ref{decFormal}). There are $n-4$ pentagon transitions in total and thus $n-4$ operators $\widehat{\mathcal{P}}$ in~(\ref{decFormal}). {For those components of the super loop where we have insertions at the bottom and top of the polygon, the first and the last pentagon operators are slightly more general: they may have additional fields insertions at their cusps and edges carrying some charges. To allow for that possibility, we added a star $*$ to the first and last transitions in (\ref{decFormal}).} 
 
To make contact with the decomposition (\ref{decompositionIntro}) we insert the resolution of the identity at every middle square of the Wilson loop. For example, for a bosonic heptagon WL we have 
 \begin{equation}
\centering
\def\svgwidth{6cm}
\begingroup%
  \makeatletter%
  \providecommand\color[2][]{%
    \errmessage{(Inkscape) Color is used for the text in Inkscape, but the package 'color.sty' is not loaded}%
    \renewcommand\color[2][]{}%
  }%
  \providecommand\transparent[1]{%
    \errmessage{(Inkscape) Transparency is used (non-zero) for the text in Inkscape, but the package 'transparent.sty' is not loaded}%
    \renewcommand\transparent[1]{}%
  }%
  \providecommand\rotatebox[2]{#2}%
  \ifx\svgwidth\undefined%
    \setlength{\unitlength}{403.71721191bp}%
    \ifx\svgscale\undefined%
      \relax%
    \else%
      \setlength{\unitlength}{\unitlength * \real{\svgscale}}%
    \fi%
  \else%
    \setlength{\unitlength}{\svgwidth}%
  \fi%
  \global\let\svgwidth\undefined%
  \global\let\svgscale\undefined%
  \makeatother%
  \begin{picture}(1,0.91550107)%
    \put(0,0){\includegraphics[width=\unitlength]{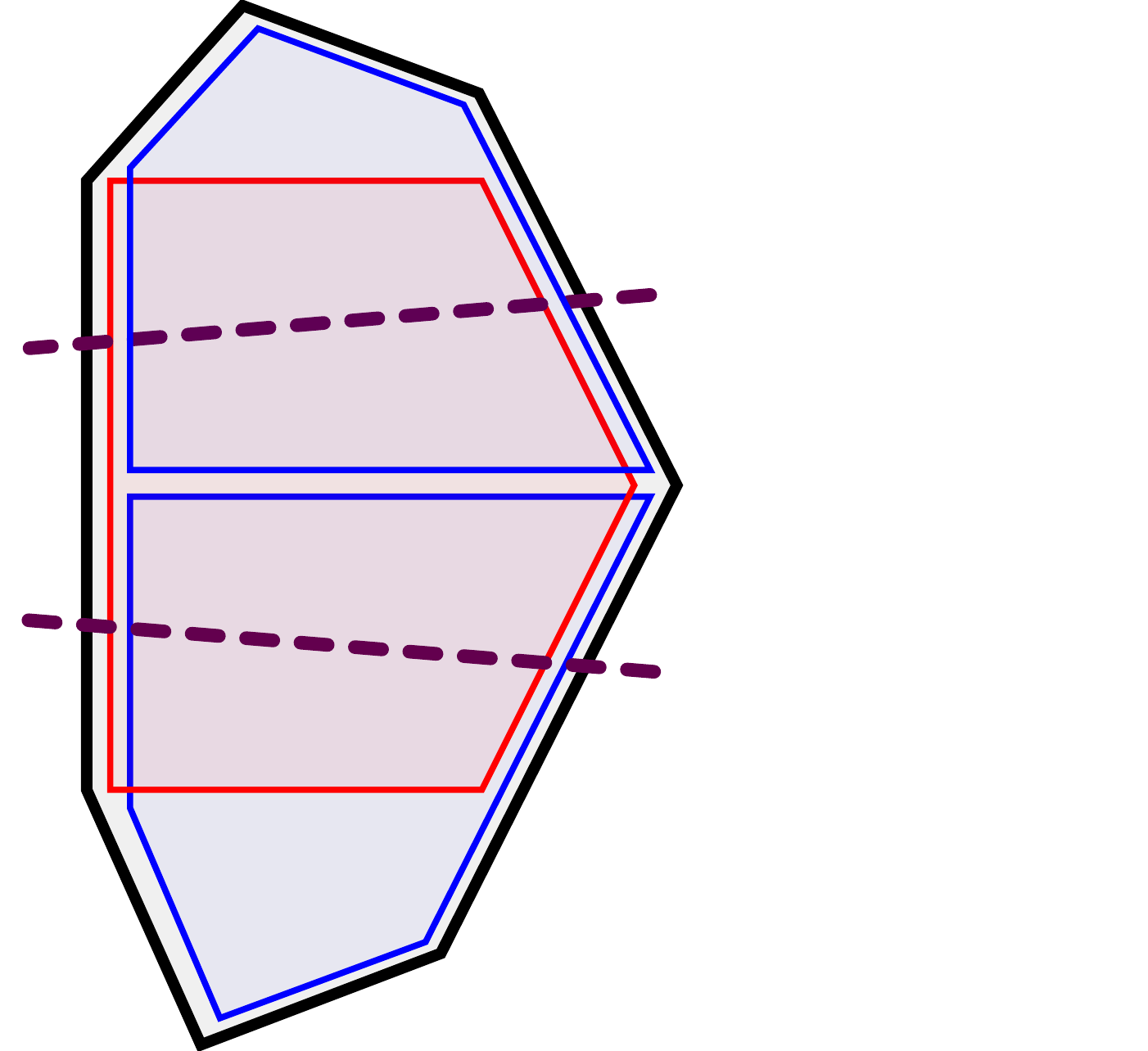}}%
    \put(0.59408455,0.31108402){\color[rgb]{0,0,0}\rotatebox{-3.95407721}{\makebox(0,0)[lb]{\smash{$\hat\One=\sum\limits_{\psi_1}|\psi_1\>\<\psi_1|$}}}}%
    \put(0.58695418,0.63898662){\color[rgb]{0,0,0}\rotatebox{4.80986875}{\makebox(0,0)[lb]{\smash{$\hat\One=\sum\limits_{\psi_2}|\psi_2\>\<\psi_2|$}}}}%
  \end{picture}%
\endgroup%
\nn
\eeq
while for the NMHV scalar component we have the same picture but with scalars inserted at the top and bottom cusps, see (\ref{TreeLevelScalar}). 

The states $\psi_i$ are eigenstates of the flux tube Hamiltonian $\hat H$ in the $i$-th middle square. As alluded to in the introduction, there are at least three equivalent ways of thinking about them:
\begin{itemize}
\item {\bf Null Wilson Line with Insertions.} A convenient way of thinking of the flux tube is as a square null Wilson loop~\cite{AldayMaldacena}. It is the most natural one here and will pervade our analysis. By further sending the top edge of this loop to infinity we get the $\Pi$-shaped Wilson loop~\cite{Belitsky,Wang} alluded to in the introduction. This step is useful when the focus is on the study of the excitations of the flux tube for instance. Excited states then correspond to insertions of local operators along the bottom line of the loop parameterized by the null direction $x^-$, as in (\ref{insertionWL}). 
When we flatten Wilson loops in the OPE approach, this is what we usually end up with, as illustrated in~(\ref{leadingscalarposition}). 

In (\ref{leadingscalarposition}) we dealt with the Born level expression at weak coupling. 
Starting at the next loop order, the operators acquire anomalous dimensions that are controlled by a light-ray Hamiltonian~\cite{Braun,Belitsky,Wang}, as the one introduced in~(\ref{lightrayH}) in a particular subsector of states. The mixing captured by this Hamiltonian originates from the UV divergences that accompany the insertions and are handled in the usual way: by regularizing/renormalizing the insertions. The eigenvalues of this Hamiltonian are nothing else than the anomalous component of the energies of the flux tube excitations.

The description of the GKP states in terms of Wilson lines with insertions is well suited for geometrical considerations. For example, as we will see below, the connection between the square and the pentagon transitions becomes quite transparent in this picture. 

\item {\bf Large Spin Operators.} Wilson lines are extended operators. They can be obtained from local operators with a large number of derivatives that delocalize them along a null direction. That is, the flux tube can be described by an operator with a very large spin. Excited states are obtained by sprinkling excitations in the middle of this `sea' of derivatives. For example, in this language, we can have two gluonic excitations $\Blue{F}=F_{z-}$ on top of the flux tube by means of a linear combination of local operators of the form\footnote{In (\ref{localOp}) we have two gluonic excitations ${\color{blue}F}$ plus two scalars $Z$. These scalars are already present for the vacuum (i.e., twist two) state $\cO_\text{vac}=\tr(Z\,DDDD\dots DDDD \, Z)+\dots$ since the derivatives need something to act on. They are not dynamical, however, and can be thought as being part of the background.}
\beq
\cO=\tr(Z\,DDDD\dots DDDD \,\Blue{F}\,  DDDD\dots DDDD \,\Blue{F} \,DDDD\dots DDDD \, Z) \la{localOp}
\eeq
where $D=D_-$ is the covariant derivative along the null direction $x^-$.

This picture is computationally appealing since a lot is known about single-trace operators in planar $\mathcal{N}=4$ SYM using the technology based on integrability. Thanks to this mapping to the integrable spin chain, the complete spectrum of flux-tube excitations and their associated dispersion relations were found at any coupling~\cite{BenDispPaper}. We can also derive, from the underlying spin-chain description, the way these excitations scatter -- i.e. their S-matrices -- at any coupling \cite{toappearAdam,toappear}. The energies enter directly the decomposition (\ref{decompositionIntro}) while the S-matrices are the fundamental objects governing the pentagon transitions. The use of integrability is then essential to our approach since it allows us to compute these objects at finite coupling.
\item  {\bf Excited GKP String.} Finally, it is sometimes convenient to think of the flux tube as the (dual of the) GKP string \cite{GKP}. Indeed, the string that ends on the null square at the boundary of AdS is dual to the two-point function of the large spin operators discussed above \cite{martin,OPEpaper}.\footnote{Strictly speaking, Gubser-Klebanov-Polyakov studied a folded string rotating in the middle of AdS \cite{GKP}. This description is related to the one invoked here by analytic continuation \cite{martin,arkady,OPEpaper}.} Excitations of the flux tube are dual to ripples on this string. For example, (\ref{localOp}) is dual to a folded string in $AdS_5$ with two bumps that are dual to the gluonic excitations, while (\ref{insertionWL}) involves fluctuations in the sphere $S^5$, dual to the scalar excitations. 

The string point of view is also quite instructive. Since it is based on a two dimensional quantum field theory,  non-trivial transformations such as mirror or crossing symmetries are conceptually simpler to grasp in this dual language. 
\end{itemize}
As we see, all these descriptions are complementary  and depending on the context we might find convenient to use one or the other. Let us now focus on some features that are common to all these descriptions.
\begin{itemize}
\item Since the flux is infinite and its excitations are gapped, the number of excitations $N$ is a conserved charge. These excitations can be of different kinds: there are fermions, gluons, scalars and also bound states of these more fundamental fields \cite{BenDispPaper}. We use a vector of indices ${\bf a}=\{a_1,\dots, a_N\}$ to indicate what kind of particles we are considering. For example, in (\ref{insertionWL}) we have ${\bf a}=\{Z,\dots,Z\}$ while for (\ref{localOp}) we get ${\bf a}=\{F,F\}$. Since it is typically clear which excitations are being discussed we will often omit the dependence on ${\bf a}$ in most formulae. 
\item The $N$ excitations have momenta $\{p_1,\dots,p_N\}$. These momenta are conjugate to a non-compact direction labelled (in each square) by $\sigma$ and as such they can take any real values. The sums in (\ref{decompositionIntro}) therefore include integrations over these momenta. The energy of the state is the sum of the energies of the constituents. 
As suggested by the integrability-based description, it appears convenient to introduce a so-called Bethe rapidity $u$. The single particle energy $E(u)$, the momentum $p(u)$, as well as the S-matrix between two excitations $S(u,v)$, are most naturally parametrized in terms of such rapidities. To leading order in perturbation theory, the rapidity and the momentum are roughly the same, $p=2u+O(g^2)$. We shall use ${\bf u}=\{u_1,\dots,u_N\}$ instead of momenta to parameterize the state. 
\end{itemize}

For now on we stick to the first description of the flux tube states, that is, in terms of Wilson lines with insertions. The pentagon transitions, which arise through the gluing of two squares in (\ref{OPEvsOPE}), are then depicted as
\beq 
\centering
\def\svgwidth{12cm}
\begingroup%
  \makeatletter%
  \providecommand\color[2][]{%
    \errmessage{(Inkscape) Color is used for the text in Inkscape, but the package 'color.sty' is not loaded}%
    \renewcommand\color[2][]{}%
  }%
  \providecommand\transparent[1]{%
    \errmessage{(Inkscape) Transparency is used (non-zero) for the text in Inkscape, but the package 'transparent.sty' is not loaded}%
    \renewcommand\transparent[1]{}%
  }%
  \providecommand\rotatebox[2]{#2}%
  \ifx\svgwidth\undefined%
    \setlength{\unitlength}{893.25107422bp}%
    \ifx\svgscale\undefined%
      \relax%
    \else%
      \setlength{\unitlength}{\unitlength * \real{\svgscale}}%
    \fi%
  \else%
    \setlength{\unitlength}{\svgwidth}%
  \fi%
  \global\let\svgwidth\undefined%
  \global\let\svgscale\undefined%
  \makeatother%
  \begin{picture}(1,0.15852206)%
    \put(0,0){\includegraphics[width=\unitlength]{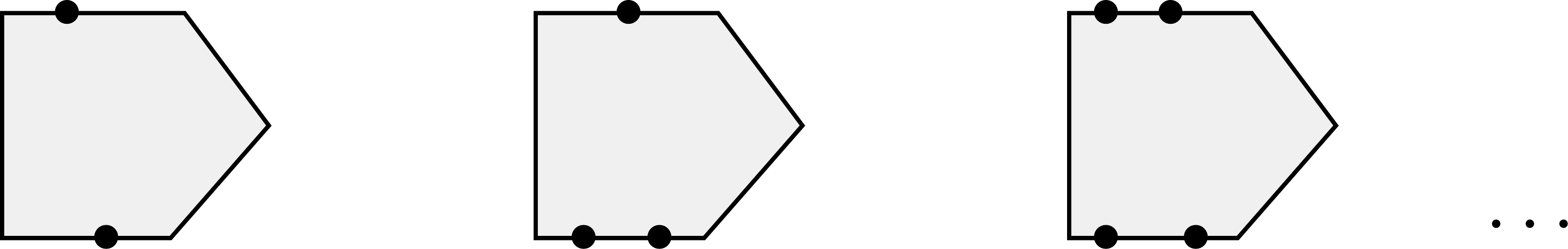}}%
    \put(0.05677989,0.02424558){\color[rgb]{0,0,0}\makebox(0,0)[lb]{\smash{$u$}}}%
    \put(0.03254595,0.12281329){\color[rgb]{0,0,0}\makebox(0,0)[lb]{\smash{$v$}}}%
    \put(0.40964821,0.02424558){\color[rgb]{0,0,0}\makebox(0,0)[lb]{\smash{$u_2$}}}%
    \put(0.39078789,0.12281329){\color[rgb]{0,0,0}\makebox(0,0)[lb]{\smash{$v$}}}%
    \put(0.69445058,0.02424558){\color[rgb]{0,0,0}\makebox(0,0)[lb]{\smash{$u_1$}}}%
    \put(0.69529354,0.12281329){\color[rgb]{0,0,0}\makebox(0,0)[lb]{\smash{$v_1$}}}%
    \put(0.36128554,0.02424558){\color[rgb]{0,0,0}\makebox(0,0)[lb]{\smash{$u_1$}}}%
    \put(0.75176929,0.02424558){\color[rgb]{0,0,0}\makebox(0,0)[lb]{\smash{$u_2$}}}%
    \put(0.73649137,0.12281329){\color[rgb]{0,0,0}\makebox(0,0)[lb]{\smash{$v_2$}}}%
    \put(0.20007668,0.01349832){\color[rgb]{0,0,0}\makebox(0,0)[lb]{\smash{$,$}}}%
    \put(0.54040652,0.01349832){\color[rgb]{0,0,0}\makebox(0,0)[lb]{\smash{$,$}}}%
    \put(0.88073639,0.01349832){\color[rgb]{0,0,0}\makebox(0,0)[lb]{\smash{$,$}}}%
  \end{picture}%
\endgroup%
 \la{toreg}
\eeq
One important feature of the pentagon is that it does not preserve the three space-time symmetries of either the bottom or the top squares.\footnote{Note that thanks to the integrability of the flux-tube, the square Wilson loop has actually infinitely many (hidden or dynamical) symmetries. They are responsible for the separate conservation of all individual momenta and for the factorization of the S-matrix for instance.}  
As a result, the pentagon transition does not preserve the momentum and energy of the excitations. Even the number of excitations does not need to be conserved, see middle object in (\ref{toreg}).

In the pictures (\ref{toreg}) the insertions are assumed to be renormalized within some scheme, an example of which will be given shortly. There are accordingly no UV divergences in~(\ref{toreg}) coming from the insertions themselves. There are nonetheless well understood UV divergences accompanying the cusped Wilson loop~\cite{Polyakov:1980ca,KK} on which the insertions are living.
These ones are universal and can be removed by dividing each expression in (\ref{toreg}) by a pentagon Wilson loop of the same shape but without any insertions.
The finite and conformally invariant quantities so obtained are what we call \textit{pentagon transitions}. As an illustration, we have
\beq\la{thetransition}
\qquad\def\svgwidth{7.5cm} 
\begingroup%
  \makeatletter%
  \providecommand\color[2][]{%
    \errmessage{(Inkscape) Color is used for the text in Inkscape, but the package 'color.sty' is not loaded}%
    \renewcommand\color[2][]{}%
  }%
  \providecommand\transparent[1]{%
    \errmessage{(Inkscape) Transparency is used (non-zero) for the text in Inkscape, but the package 'transparent.sty' is not loaded}%
    \renewcommand\transparent[1]{}%
  }%
  \providecommand\rotatebox[2]{#2}%
  \ifx\svgwidth\undefined%
    \setlength{\unitlength}{551.97490234bp}%
    \ifx\svgscale\undefined%
      \relax%
    \else%
      \setlength{\unitlength}{\unitlength * \real{\svgscale}}%
    \fi%
  \else%
    \setlength{\unitlength}{\svgwidth}%
  \fi%
  \global\let\svgwidth\undefined%
  \global\let\svgscale\undefined%
  \makeatother%
  \begin{picture}(1,0.30726035)%
    \put(0,0){\includegraphics[width=\unitlength]{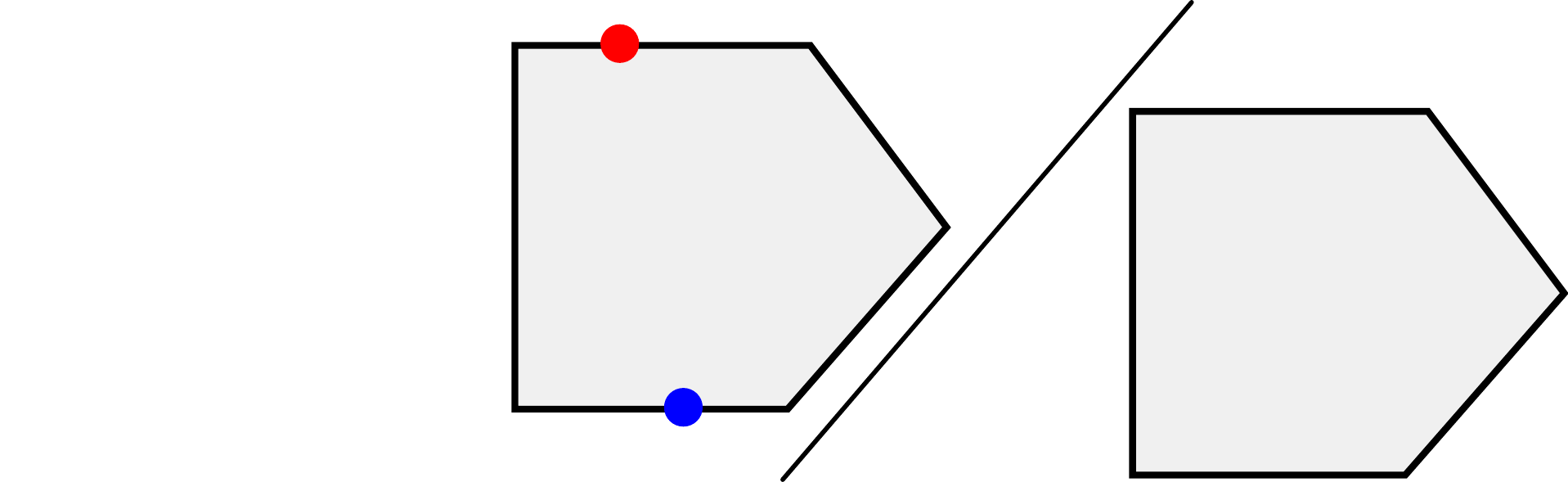}}%
    \put(-0.0018046,0.15310123){\color[rgb]{0,0,0}\makebox(0,0)[lb]{\smash{$\displaystyle P(u|v)\ =$}}}%
    \put(0.41818378,0.07449579){\color[rgb]{0,0,0}\makebox(0,0)[lb]{\smash{$u$}}}%
    \put(0.37896643,0.23110747){\color[rgb]{0,0,0}\makebox(0,0)[lb]{\smash{$v$}}}%
  \end{picture}%
\endgroup%
 
\eeq
and
\beq\la{thetransition2}
\qquad\def\svgwidth{9cm} 
\begingroup%
  \makeatletter%
  \providecommand\color[2][]{%
    \errmessage{(Inkscape) Color is used for the text in Inkscape, but the package 'color.sty' is not loaded}%
    \renewcommand\color[2][]{}%
  }%
  \providecommand\transparent[1]{%
    \errmessage{(Inkscape) Transparency is used (non-zero) for the text in Inkscape, but the package 'transparent.sty' is not loaded}%
    \renewcommand\transparent[1]{}%
  }%
  \providecommand\rotatebox[2]{#2}%
  \ifx\svgwidth\undefined%
    \setlength{\unitlength}{647.97490234bp}%
    \ifx\svgscale\undefined%
      \relax%
    \else%
      \setlength{\unitlength}{\unitlength * \real{\svgscale}}%
    \fi%
  \else%
    \setlength{\unitlength}{\svgwidth}%
  \fi%
  \global\let\svgwidth\undefined%
  \global\let\svgscale\undefined%
  \makeatother%
  \begin{picture}(1,0.26173853)%
    \put(0,0){\includegraphics[width=\unitlength]{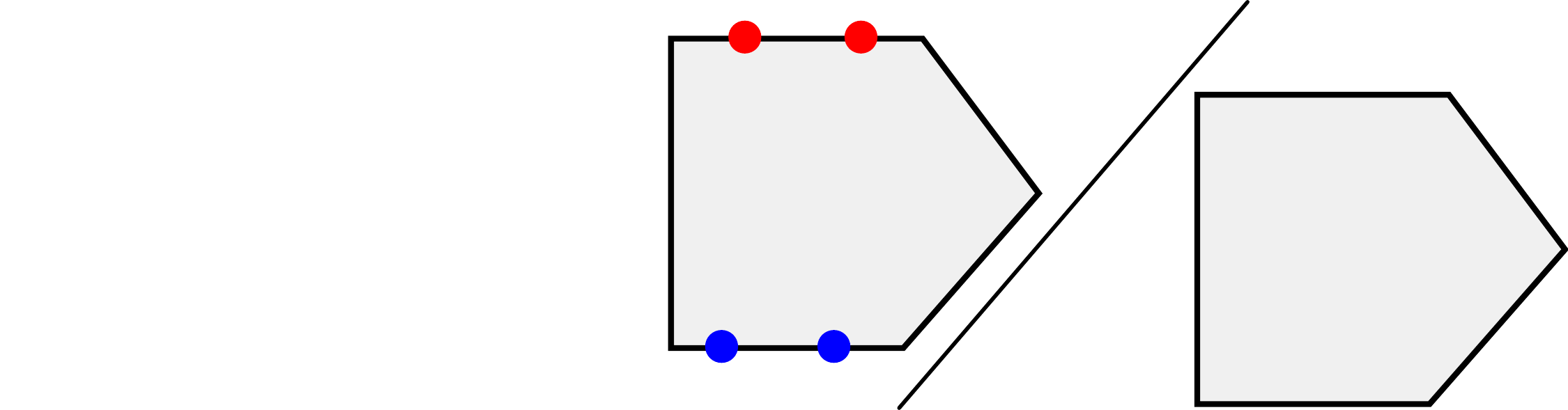}}%
    \put(-0.00153724,0.13041869){\color[rgb]{0,0,0}\makebox(0,0)[lb]{\smash{$\displaystyle P(u_1,u_2|v_1,v_2)\ =$}}}%
    \put(0.44512057,0.06345895){\color[rgb]{0,0,0}\makebox(0,0)[lb]{\smash{$u_1$}}}%
    \put(0.46109804,0.19933723){\color[rgb]{0,0,0}\makebox(0,0)[lb]{\smash{$v_1$}}}%
    \put(0.53517499,0.19933723){\color[rgb]{0,0,0}\makebox(0,0)[lb]{\smash{$v_2$}}}%
    \put(0.51672828,0.06345895){\color[rgb]{0,0,0}\makebox(0,0)[lb]{\smash{$u_2$}}}%
  \end{picture}%
\endgroup%
 
\eeq
for the one- and two-particle transitions, respectively. Note that for simplicity we depicted transitions for particles of the same kind at the top and bottom. We also used blue and red circles to indicate the two conjugate fields that respectively create and annihilate a given excitation. For instance, for a complex scalar excitation we would have $Z(u)$ at the bottom and $\bar{Z}(v)$ at the top, and the transition in~(\ref{thetransition}) is what we would denote as $P_{ZZ}(u|v)$.

An important aspect of our decomposition is that any two adjacent pentagons overlap on a middle square. Therefore, the product of two consecutive pentagon transitions accounts \textit{ twice} for the propagation of the eigenstate on this middle square. To remedy for this double counting, we must `divide' by the relevant \textit{square transition}, which is defined by an expression similar to (\ref{thetransition}) and (\ref{thetransition2}) but with a square in place of a pentagon. {This procedure can be thought of as the proper way of gluing two consecutive pentagons together through the matching on the square of their common excitations.}
In practice it means that whenever we integrate over the rapidities ${\bf u}$ of the intermediate states, which always happens on a middle square, we should include a measure $\mu({\bf u})$ to get rid of the effect mentioned before. If the spectrum were discrete the measure would be exactly the inverse of the square transition. Since we have a continuum of states the measure is in fact a density. For a single particle, it is defined as
\beq\la{singleparticlemeasure}
\qquad\def\svgwidth{8cm} 
\begingroup%
  \makeatletter%
  \providecommand\color[2][]{%
    \errmessage{(Inkscape) Color is used for the text in Inkscape, but the package 'color.sty' is not loaded}%
    \renewcommand\color[2][]{}%
  }%
  \providecommand\transparent[1]{%
    \errmessage{(Inkscape) Transparency is used (non-zero) for the text in Inkscape, but the package 'transparent.sty' is not loaded}%
    \renewcommand\transparent[1]{}%
  }%
  \providecommand\rotatebox[2]{#2}%
  \ifx\svgwidth\undefined%
    \setlength{\unitlength}{490.79995117bp}%
    \ifx\svgscale\undefined%
      \relax%
    \else%
      \setlength{\unitlength}{\unitlength * \real{\svgscale}}%
    \fi%
  \else%
    \setlength{\unitlength}{\svgwidth}%
  \fi%
  \global\let\svgwidth\undefined%
  \global\let\svgscale\undefined%
  \makeatother%
  \begin{picture}(1,0.33366802)%
    \put(0,0){\includegraphics[width=\unitlength]{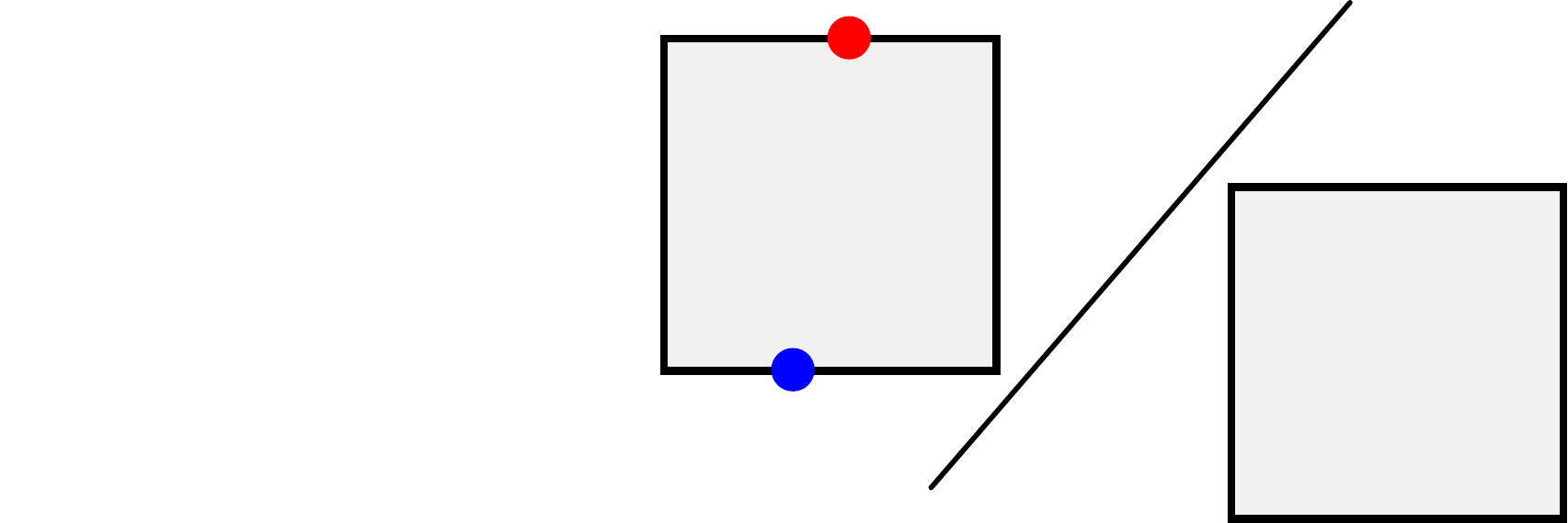}}%
    \put(-0.00202953,0.1798539){\color[rgb]{0,0,0}\makebox(0,0)[lb]{\smash{$\displaystyle{2\pi\over\mu(u)}\delta(u-v)\ =$}}}%
    \put(0.48925269,0.1279373){\color[rgb]{0,0,0}\makebox(0,0)[lb]{\smash{$u$}}}%
    \put(0.52338676,0.25842978){\color[rgb]{0,0,0}\makebox(0,0)[lb]{\smash{$v$}}}%
  \end{picture}%
\endgroup%

\eeq
where, to remove the cusp divergences, we divided by the expectation value of the bare square WL. This last step can actually be viewed as subtracting out the (infinite) energy of the flux tube vacuum. 
Similarly, for two particles, we have
 \beq\la{twoparticlesmeasure}
\def\svgwidth{18cm} 
\begingroup%
  \makeatletter%
  \providecommand\color[2][]{%
    \errmessage{(Inkscape) Color is used for the text in Inkscape, but the package 'color.sty' is not loaded}%
    \renewcommand\color[2][]{}%
  }%
  \providecommand\transparent[1]{%
    \errmessage{(Inkscape) Transparency is used (non-zero) for the text in Inkscape, but the package 'transparent.sty' is not loaded}%
    \renewcommand\transparent[1]{}%
  }%
  \providecommand\rotatebox[2]{#2}%
  \ifx\svgwidth\undefined%
    \setlength{\unitlength}{1203.81064453bp}%
    \ifx\svgscale\undefined%
      \relax%
    \else%
      \setlength{\unitlength}{\unitlength * \real{\svgscale}}%
    \fi%
  \else%
    \setlength{\unitlength}{\svgwidth}%
  \fi%
  \global\let\svgwidth\undefined%
  \global\let\svgscale\undefined%
  \makeatother%
  \begin{picture}(1,0.13603821)%
    \put(0,0){\includegraphics[width=\unitlength]{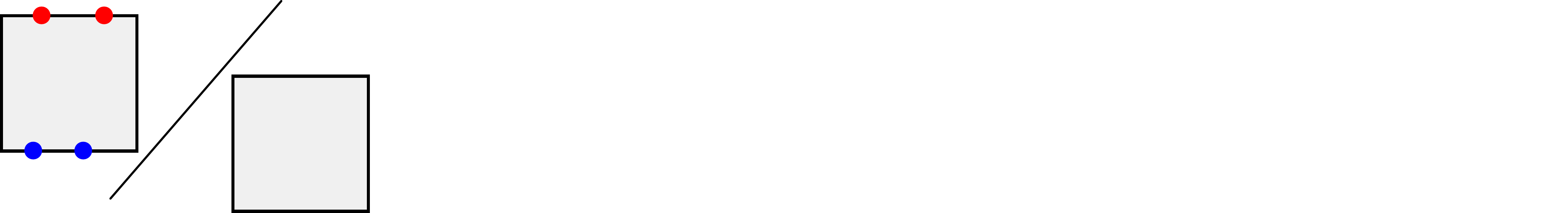}}%
    \put(0.24725123,0.06748278){\color[rgb]{0,0,0}\makebox(0,0)[lb]{\smash{$\displaystyle{={(2\pi)^2\over\mu(u_1)\mu(u_2)}\[\delta(u_1-v_1) \delta(u_2-v_2) +S(u_1,u_2)\delta(u_1-v_2) \delta(u_2-v_1)\]}$}}}%
    \put(0.01306287,0.05216071){\color[rgb]{0,0,0}\makebox(0,0)[lb]{\smash{$u_1$}}}%
    \put(0.01900484,0.1066923){\color[rgb]{0,0,0}\makebox(0,0)[lb]{\smash{$v_1$}}}%
    \put(0.04496158,0.05216071){\color[rgb]{0,0,0}\makebox(0,0)[lb]{\smash{$u_2$}}}%
    \put(0.05887822,0.1066923){\color[rgb]{0,0,0}\makebox(0,0)[lb]{\smash{$v_2$}}}%
  \end{picture}%
\endgroup%

\eeq
where $S(u_1,u_2)$ is the S-matrix between the two excitations. Notice that in our normalization the rapidities are ordered and states with different orderings are related by an S-matrix. The fact that the two-particle measure factorizes into the product of the single-particle measures, i.e., $\mu(u_1,u_2)=\mu(u_1)\mu(u_2)$, is an outcome of the coordinate Bethe ansatz normalization that we are using. We shall use this normalization for any number of particles and hence the $N$-particle integration measure reads
\beq\la{measuredefinition}
d {\bf u} ={\cal N}_{\bf a} \prod_{j=1}^N  \frac{du_j}{2\pi}\,\mu_{\bf a} (u_j) \, ,
\eeq
where ${\cal N}_{\bf a}$ is a symmetry factor (it is equal to $1/N!$ for identical particles for example).

Though they might look very different, the measure and the pentagon transitions are actually not independent. Instead,
 \beq\la{measureEq}
\underset{v=u}{\operatorname{{\rm residue}}}\, P_{aa}(u|v)= \frac{i}{\mu_a(u)}   \,.
 \eeq
This relation has a simple geometrical origin. In position space, the residue at $u=v$ of the pentagon transition controls the regime where $\sigma_1,\sigma_2\to-\infty$ with $\sigma_1-\sigma_2$ kept fixed. This limit corresponds to sending the bottom and top pentagon insertions to the edge opposite to the middle cusp, i.e., close to the left edge in (\ref{thetransition}). This on the other hand is conformally equivalent to flattening the rightmost cusp of the pentagon. In this way we end up with the square depicted in (\ref{singleparticlemeasure}). 

As mentioned above, beyond Born level, the insertions along null lines in (\ref{thetransition}--\ref{measuredefinition}) should be properly renormalized. There are of course many different schemes for doing this. Here we will briefly comment on the most natural one in our geometrical context. It is based on the observation that a possible way of regularizing an insertion on a null WL is to move it a bit away from the line -- the distance from the line playing the role of the regularization scale. Now there is a simple way of implementing this geometrical regularization by using the polygon Wilson loops with insertions given to us by the (component) expansion of the super Wilson loop. In a nutshell, we add a bump with an insertion at the top, by choosing carefully a component of the super loop, that we immediately flatten:
\beq
\centering
\def\svgwidth{8.5cm}
\begingroup%
  \makeatletter%
  \providecommand\color[2][]{%
    \errmessage{(Inkscape) Color is used for the text in Inkscape, but the package 'color.sty' is not loaded}%
    \renewcommand\color[2][]{}%
  }%
  \providecommand\transparent[1]{%
    \errmessage{(Inkscape) Transparency is used (non-zero) for the text in Inkscape, but the package 'transparent.sty' is not loaded}%
    \renewcommand\transparent[1]{}%
  }%
  \providecommand\rotatebox[2]{#2}%
  \ifx\svgwidth\undefined%
    \setlength{\unitlength}{393.55900879bp}%
    \ifx\svgscale\undefined%
      \relax%
    \else%
      \setlength{\unitlength}{\unitlength * \real{\svgscale}}%
    \fi%
  \else%
    \setlength{\unitlength}{\svgwidth}%
  \fi%
  \global\let\svgwidth\undefined%
  \global\let\svgscale\undefined%
  \makeatother%
  \begin{picture}(1,0.24646865)%
    \put(0,0){\includegraphics[width=\unitlength]{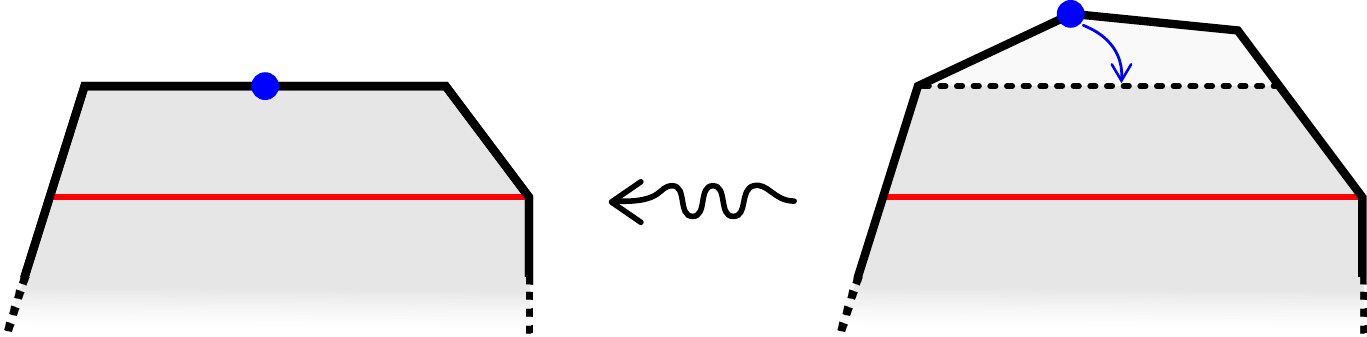}}%
  \end{picture}%
\endgroup%

\eeq
As it is, this is not yet the proper way of doing, because we introduced meantime an extra cusp in the problem, sitting in the rightmost picture above. We know how to handle this kind of divergence, however, by using squares and pentagons. Hence, we can insert a regularized excitation through the combination
\beq
\centering
\def\svgwidth{14cm}
\begingroup%
  \makeatletter%
  \providecommand\color[2][]{%
    \errmessage{(Inkscape) Color is used for the text in Inkscape, but the package 'color.sty' is not loaded}%
    \renewcommand\color[2][]{}%
  }%
  \providecommand\transparent[1]{%
    \errmessage{(Inkscape) Transparency is used (non-zero) for the text in Inkscape, but the package 'transparent.sty' is not loaded}%
    \renewcommand\transparent[1]{}%
  }%
  \providecommand\rotatebox[2]{#2}%
  \ifx\svgwidth\undefined%
    \setlength{\unitlength}{601.96000977bp}%
    \ifx\svgscale\undefined%
      \relax%
    \else%
      \setlength{\unitlength}{\unitlength * \real{\svgscale}}%
    \fi%
  \else%
    \setlength{\unitlength}{\svgwidth}%
  \fi%
  \global\let\svgwidth\undefined%
  \global\let\svgscale\undefined%
  \makeatother%
  \begin{picture}(1,0.19410394)%
    \put(0,0){\includegraphics[width=\unitlength]{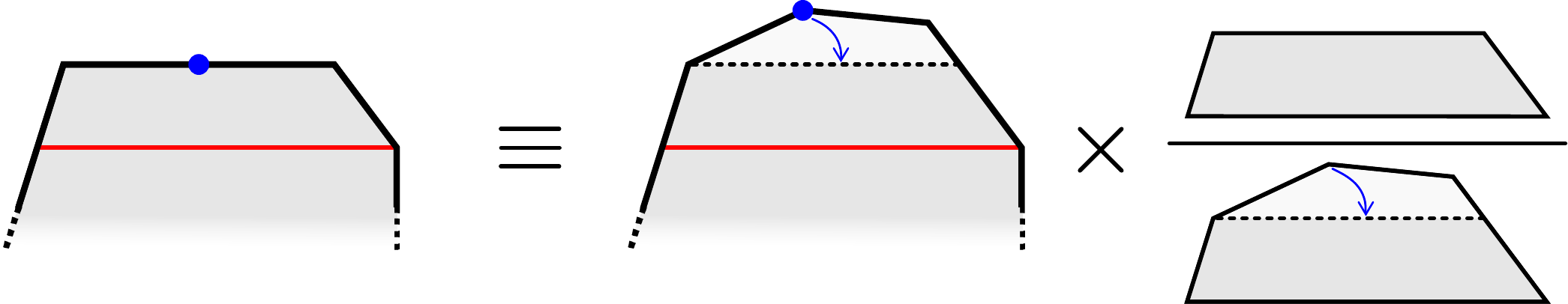}}%
    \put(0.10227009,0.17641327){\color[rgb]{0,0,0}\makebox(0,0)[lb]{\smash{$\text{reg}$}}}%
  \end{picture}%
\endgroup%
 \la{scalarFlat}
\eeq
When deforming the loop we also added one more edge and thus three new cross ratios appeared along the way. In other words, 
what was before the top square in the left hand side is now a middle square in the right hand side. This new middle square is equipped with its usual three parameters $\tau, \sigma$ and $\phi$. To flatten the bump and get a renormalized insertion, we take $\tau\to\infty$, then Fourier transform with respect to $\sigma$, and extract finally the leading piece that goes as $e^{-E(u)\tau+i m\phi}$. The multi-particle insertions can be treated similarly \cite{toappear}, see also the discussion section. 

Note that in (\ref{scalarFlat}) we added a blue circle at the top cusp. This one can represent a scalar insertion for example. If we want to insert a gluonic excitation instead, then we will have two options. We can insert it directly at a cusp, like for the scalar, using the super Wilson loop. Another option is to recall that a geometric deformation of the loop generates by itself a gauge field insertion. So we can simply flatten the loop without inserting any fields by hand. The latter option is the way gluons are created at the bottom and top of the bosonic Wilson loop, while the former occurs for some components of the super loop. They differ by a simple form factor which will be discussed in section \ref{formSec}.

\subsection{Combining the Pieces Together} \la{combiningSec}

We shall now combine all the pieces together and convert the somewhat formal sums over flux tube states in (\ref{decompositionIntro}) into precise integrals over Bethe rapidities and sums over particle quantum numbers. 

We start with the definition of our renormalized Wilson loop~$\mathcal{W}$. We saw above that each pentagon transition should be divided by a bare pentagon Wilson loop, in order to remove the UV divergences. Similarly each measure associated to a middle square should be multiplied by a bare square Wilson loop. Hence, the finite and conformal invariant object that admits the decomposition (\ref{decFormal}) is obtained by dividing the original Wilson loop or the component of the super loop by 
\beq
{\rm w} \equiv { \< W_{1^\text{st} \text{pent}}\>}\,{1 \over \<W_{1^\text{st} \text{middle sq}}\>}\,{\< W_{2^\text{nd} \text{pent}}\>}\,{1\over \<W_{2^\text{nd} \text{middle sq}}\>}\ \dots\ {\< W_{(n-4)^\text{th} \text{pent}}\>} \la{wdef} \,.
\eeq
For example, for the usual bosonic loop {we have} 
\beq
\cW=\mathcal{W}_\text{MHV}\equiv  \<W\>/{\rm w}\, , \la{calWeq}
\eeq
while for the component (\ref{TreeLevelScalar}) we get $\mathcal{W}=\mathcal{W}^{(7145)}_\text{NMHV}\equiv \<W^{(7145)}_\text{NMHV}\>/{\rm w}$.
We could as well define a superconformal invariant ratio $\mathbb{W}=\<W_\text{super}\>/{\rm w}$ containing all these cases at once\footnote{The $\eta$'s are the dual Grassmann variables \cite{etas,Dualsuperconforma}. They carry a lower edge index $i=1,\dots,n$ and an upper R-charge index $A=1,2,3,4$.}
\beqa\la{superW}
{\mathbb W}=\cW_\text{MHV}+ \eta_i^1 \eta_j^2 \eta_k^3 \eta_l^4\,\, \cW_\text{NMHV}^{(ijkl)} + \eta_i^1 \eta_j^2 \eta_k^3 \eta_l^4  \eta_{m}^1 \eta_{n}^2 \eta_{o}^3 \eta_{p}^4\,\, \cW_\text{N$^2$MHV}^{(ijkl)(mnop)}  + \dots   \la{superdefinition}
\eeqa
The ratio $\mathbb W$ is finite and conformal invariant because $W_\text{super}$ has the same UV divergences as the bosonic loop \cite{Dualsuperconforma}, and so is renormalized by the same ratio of pentagons and squares as in (\ref{calWeq}). In practice, we will always consider separately the different components of $\mathbb W$ since they are described by different OPE transitions. An interesting open question, which we leave for the future, is whether all components of $\mathbb W$ can be analyzed using the OPE.

We now re-write the decompositions (\ref{decFormal}) using the labels ${\bf a},{\bf b},\dots$ for the (discrete) isotopic degrees of freedom and ${\bf u},{\bf v},\dots$ for the (continuous) rapidities. To be as explicit as possible we will consider specific components of $\mathbb W$ and therefore fix the top and bottom {pentagon operators} in (\ref{decFormal}). We begin with the case where both of them stand for {the bosonic transitions}. Then $\mathcal{W}{=\cW_\text{MHV}}$ describes the bosonic Wilson loops dual to the MHV amplitudes. The simplest example is the hexagon for which we have a single middle square in (\ref{decompositionIntro}),(\ref{decFormal}). 
Given the discussion in the previous section we are led to 
\beq\la{Whex}
\def\svgwidth{21cm} 
\begingroup%
  \makeatletter%
  \providecommand\color[2][]{%
    \errmessage{(Inkscape) Color is used for the text in Inkscape, but the package 'color.sty' is not loaded}%
    \renewcommand\color[2][]{}%
  }%
  \providecommand\transparent[1]{%
    \errmessage{(Inkscape) Transparency is used (non-zero) for the text in Inkscape, but the package 'transparent.sty' is not loaded}%
    \renewcommand\transparent[1]{}%
  }%
  \providecommand\rotatebox[2]{#2}%
  \ifx\svgwidth\undefined%
    \setlength{\unitlength}{2419.71875bp}%
    \ifx\svgscale\undefined%
      \relax%
    \else%
      \setlength{\unitlength}{\unitlength * \real{\svgscale}}%
    \fi%
  \else%
    \setlength{\unitlength}{\svgwidth}%
  \fi%
  \global\let\svgwidth\undefined%
  \global\let\svgscale\undefined%
  \makeatother%
  \begin{picture}(1,0.16259258)%
    \put(0,0){\includegraphics[width=\unitlength]{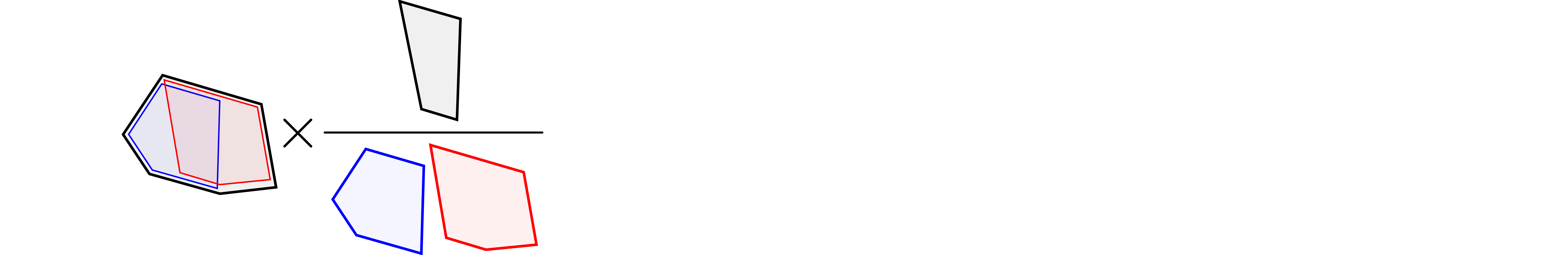}}%
    \put(-0.00109775,0.07415386){\color[rgb]{0,0,0}\makebox(0,0)[lb]{\smash{${\cal W}_\text{hex}\equiv$}}}%
    \put(0.35596854,0.07415386){\color[rgb]{0,0,0}\makebox(0,0)[lb]{\smash{$\displaystyle=\sum\limits_{{\bf a}} \int\!\! d{\bf u} \,P_{\,\bf a}(0|{\bf u})  \,e^{-E({\bf u})\tau
+ip({\bf u})\sigma+im\phi}P_{\,{\bf a}}(\bar{\bf u}|0)$}}}%
  \end{picture}%
\endgroup%

\eeq
which should be clear by itself right now. All the ingredients in this expression were discussed before, see (\ref{thetransition}--\ref{measuredefinition}), with the exception of the notation $\bar{\textbf{u}}$. It stands for $\bar {\bf u}=\{-u_N,\dots,-u_1\}$ and is an artifact of our conventions, on which we comment further in appendix \ref{FlippingAp}.\footnote{These minus signs could be absorbed into a redefinition of the pentagon transitions, which would just move the drawbacks elsewhere.}

As we are describing the bosonic Wilson loop, all states being summed over in (\ref{Whex}) have no overall R-charge.
The lightest states propagating in the middle square dominate the decomposition (\ref{Whex}) at large $\tau$. In the present case the lightest accessible state is the vacuum itself, which has zero energy, momentum and $U(1)$ charge. Furthermore the vacuum to vacuum transition is, by definition, equal to $1$ since in that case the numerator and denominator in the ratio (\ref{thetransition}) coincide. Hence $\mathcal{W}_\text{hex}=1+\dots$ where the dots stand for the contribution of the excited states. 

At small enough coupling, the lightest excited states are always single-particle and thus parametrized by a single rapidity~$u$. For the bosonic loop, we get the two conjugate gluonic excitations $\Blue{F}=F_{-z}$ and $\Red{\bar F}=F_{-\bar z}$.\footnote{\la{ft10} The subscript $\mu = -$ in $F_{\mu\nu}$ denotes the light-like direction identified with the $\sigma$ direction of the flux tube while $\nu = z = 1+i2$ and $\nu = \bar{z} = 1-i2$ stand for right- and left-handed polarizations with respect to the 1-2 plane transverse to the flux tube. } They carry no R-charge and opposite $U(1)$ charge $\pm 1$ with regard to rotations in the plane transverse to the flux tube. At weak coupling, their energy (twist) is equal to one. They contribute equally to the decomposition (\ref{Whex}) except that the contribution of $\Blue{F}$ is weighted by $\Blue{e^{+i \phi}}$ while the contribution of $\Red{\bar F}$ is dressed by $\Red{e^{-i \phi}}$. Hence 
\beqa
\mathcal{W}_\text{hex}&=&1+ {\color{blue} e^{+i \phi}}  \int \frac{du}{2\pi} P_{\color{blue} F}(0|u)\mu(u) P_{\color{blue} F}(-u|0) e^{-E(u) \tau+i p(u) \sigma}\nn \\
&&\,\,\,\,\,+\, {\color{red} e^{-i \phi}}  \int \frac{du}{2\pi} P_{\color{red} \bar F}(0|u)\mu(u) P_{\color{red} \bar  F}(-u|0) e^{-E(u) \tau+i p(u) \sigma} +\dots  \la{33}
\eeqa
where the dots stand for the contribution of heavier excitations (such as gluonic bound states) or multiparticle states. Here, $P_\Blue{F}(0|u)$ and $P_\Red{\bar{F}}(0|u)$ are form factors for creation of a gauge field from the vacuum. They are nothing else but a pentagon Wilson loop with an insertion of a Faraday tensor at the top edge. The two form factors are the same $P_\Blue{F}(0|u)=P_\Red{\bar{F}}(0|u)$ as a consequence of charge conjugation that swaps ${\color{blue} F}$ and ${\color{red} \bar F}$. The latter is a symmetry of the pentagon for the following reason.  A pentagon can always be embedded in an ${\mathbb R}^{1,2}\subset{\mathbb R}^{1,3}$ subspace and is therefore invariant under reflection of the extra direction which we denote as $x_2$. This translates into a $\mathbb{Z}_2$ symmetry of the pentagon which is nothing but charge conjugation, see footnote \ref{ft10}. 
Note finally that these form factors, $P_\Blue{F}(0|u)$ and $P_\Red{\bar{F}}(0|u)$, can be set to $1$ by an appropriate normalization of the single gauge field wave function (see end of this section and section \ref{beforeSec} for further discussion of this point). Under this convention the WL (\ref{33}) becomes
\beqa
\mathcal{W}_\text{hex}=1+  2\cos(\phi)f(\tau,\sigma)+\dots\qquad\text{with}\qquad f(\tau,\sigma)\equiv  \int \frac{du}{2\pi} \mu(u) e^{-E(u) \tau+i p(u) \sigma} \,.  \la{hexF}
\eeqa

Our next example is the MHV heptagon for which we have
\beqa\la{hepdecomposition}
&&
\def\svgwidth{11cm}
\begingroup%
  \makeatletter%
  \providecommand\color[2][]{%
    \errmessage{(Inkscape) Color is used for the text in Inkscape, but the package 'color.sty' is not loaded}%
    \renewcommand\color[2][]{}%
  }%
  \providecommand\transparent[1]{%
    \errmessage{(Inkscape) Transparency is used (non-zero) for the text in Inkscape, but the package 'transparent.sty' is not loaded}%
    \renewcommand\transparent[1]{}%
  }%
  \providecommand\rotatebox[2]{#2}%
  \ifx\svgwidth\undefined%
    \setlength{\unitlength}{1033.85878906bp}%
    \ifx\svgscale\undefined%
      \relax%
    \else%
      \setlength{\unitlength}{\unitlength * \real{\svgscale}}%
    \fi%
  \else%
    \setlength{\unitlength}{\svgwidth}%
  \fi%
  \global\let\svgwidth\undefined%
  \global\let\svgscale\undefined%
  \makeatother%
  \begin{picture}(1,0.38155191)%
    \put(0,0){\includegraphics[width=\unitlength]{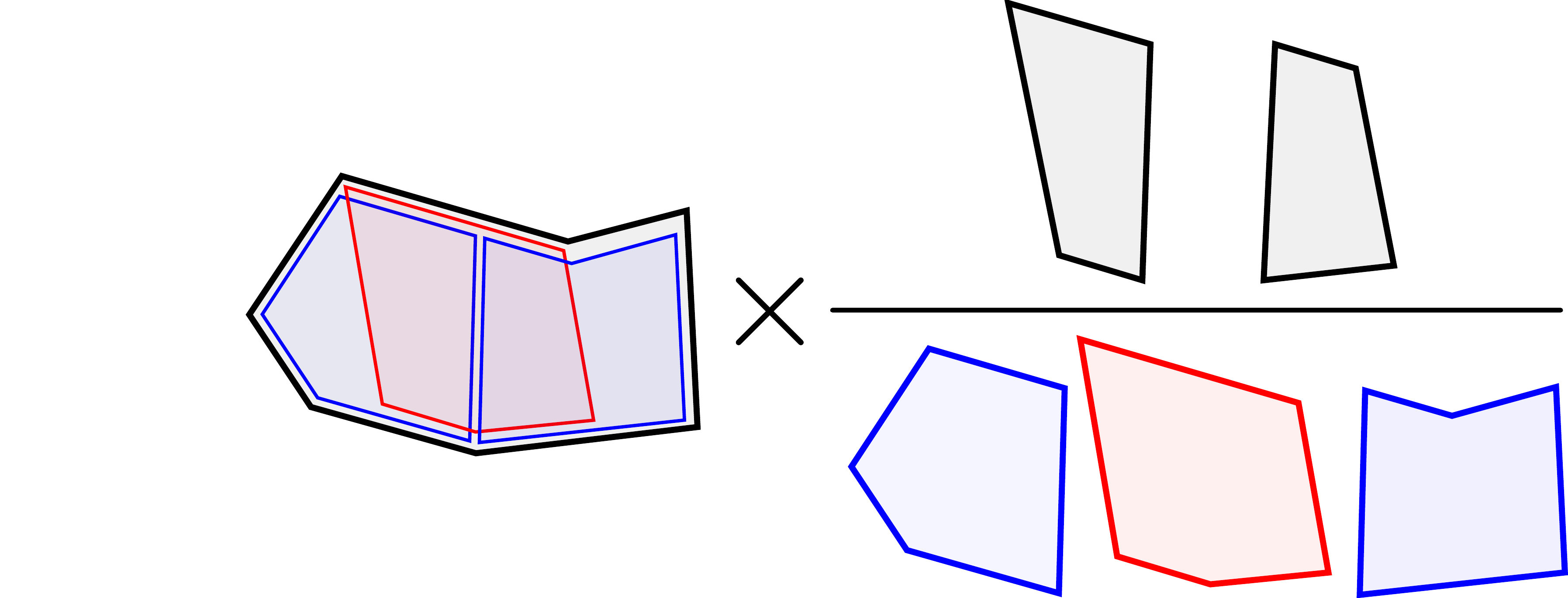}}%
    \put(-0.00256926,0.18075386){\color[rgb]{0,0,0}\makebox(0,0)[lb]{\smash{${\cal W}_\text{hep}\equiv$}}}%
  \end{picture}%
\endgroup%
  \la{Whept}\\
&&=\sum_{{\bf a},{\bf b}} \int \!\! d{\bf u}\,d{\bf v} \,P_{\,\bf a}(0|{\bf u})\,e^{-E({\bf u})\tau_1+ip({\bf u})\sigma_1+im_1\phi_1} P_{\,{\bf a}{\bf b}}(\bar{\bf u}|{\bf v})\,e^{-E({\bf v})\tau_2+ip({\bf v})\sigma_2+im_2\phi_2} P_{\,{\bf b}}(\bar{\bf v}|0) \nn
\eeqa
Again, the vacuum contribution gives $1$. Then, the leading processes at large $\tau_1,\,\tau_2$ are those involving a single gluonic excitation $\Blue{F}$ or $\Red{\bar F}$. Such gluonic excitation can be produced at the bottom and absorbed at the top.  It can also be produced latter or annihilated before. Furthermore, in the middle transition, it can change its nature. This is because the pentagon can absorb the $U(1)$ charge of the excitations and convert an $\Blue{F}$ into an $\Red{\bar{F}}$. We therefore have two possible gluonic transitions
\beq\label{PbarPold}
P(u|v) \equiv P_\Blue{FF}(u|v)=P_\Red{FF}(u|v)\qquad\text{and}\qquad \bar{P}(u|v) \equiv P_{\Blue{F}\Red{\bar{F}}}(u|v)=P_{\Red{\bar{F}\Blue{F}}}(u|v)\, , 
\eeq
associated respectively to the $U(1)$-preserving and $U(1)$-violating processes: $\Blue{F}\,\(\Red{\bar F}\)\rightarrow \Blue{F}\,\(\Red{\bar F}\)$ and $\Blue{F}\,\(\Red{\bar F}\)\rightarrow \Red{\bar F}\,\(\Blue{F}\)$. 
In sum, we have
\beqa
\mathcal{W}_\text{hep} &=&1 \qquad\qquad \qquad\qquad\qquad\qquad\qquad\,\,\,\, \, \text{vacuum}\to  \text{vacuum}  \to \text{vacuum} \to \text{vacuum} \nn \\ 
&+&2\cos(\phi_1) f(\tau_1,\sigma_1)  \qquad\qquad \qquad\qquad  \text{vacuum}\to  \,\,\, \Blue{F}  \,(\Red{ \bar F})\,\, \to \text{vacuum} \to \text{vacuum} \nn \\ 
&+&2 \cos(\phi_2) f(\tau_2,\sigma_2) \qquad\qquad \qquad\qquad  \text{vacuum}\to \text{vacuum} \to \,\,\, \Blue{F}  \,(\Red{ \bar F}) \,\, \to \text{vacuum} \nn \\ 
&+&2  \cos(\phi_1-\phi_2) \bar{h}(\tau_1,\tau_2,\sigma_1,\sigma_2)  \qquad\,\,\,\,\,\,\,   \text{vacuum}\to\,\,\,\Red{ \bar F} \,(\Blue{ \bar F}) \,\,\to  \,\,\,\Blue{F}  \,(\Red{ \bar F}) \,\, \to \text{vacuum} \nn \\ 
&+&2 \cos(\phi_1+\phi_2) h(\tau_1,\tau_2,\sigma_1,\sigma_2) \qquad\,\,\,\,\,\,\,   \text{vacuum}\to\,\,\,  \Blue{F} \,(\Red{ \bar F}) \,\,\to  \,\,\,\Blue{F}  \,(\Red{ \bar F})  \,\,\to \text{vacuum} \nn \\ 
&+&  \dots  \la{hepMHV0}
\eeqa
Here, $f$ is the same as before, see (\ref{hexF}), the helicity preserving transition contributes as 
\beq
h(\tau_1,\tau_2,\sigma_1,\sigma_2)= \int \frac{du}{2\pi}\int \frac{dv}{2\pi}  \,\mu(u)\,P(-u|v)\mu(v) \,e^{-\tau_1E (u)+i p(u)\sigma_1-\tau_2E(v)+i p(v)\sigma_2}\, ,  \la{hInt} 
\eeq
and the helicity violating transition $\bar h$ is given by the same expression with $P$ replaced by $\bar P$. 

An important comment about the heptagon concerns the integration contour in (\ref{hInt}). As explained before $P(u|v)$ has a pole at $u=v$ and this entails choosing an $i\epsilon$ prescription for properly integrating the pentagon transition in (\ref{hInt}). The residue at this pole controls at the end the collinear limit $\sigma_1,\sigma_2\to -\infty$ with $\sigma_1-\sigma_2$ held fixed, which was related to the measure in (\ref{measureEq}). This implies that we should shift slightly the integration contours in (\ref{hInt}) to the upper half plane such that ${\rm Im}(u)={\rm Im}(v)={0^+}$. 

We now move to a case where the {first and last transitions in (\ref{decFormal}) are charged}. This is relevant for the study of NMHV amplitudes. For illustration, we consider the NMHV heptagon component $\mathcal{W}^{(7145)}_\text{NMHV}$ introduced in (\ref{TreeLevelScalar}). For this component, a complex scalar $Z$ is inserted at the bottom cusp and the conjugate scalar $\bar Z$ is inserted at the top cusp.  
{These scalars carry one unit of R-charge implying that the first and last pentagon transitions, noted as $P_*$, are charged transitions carrying the corresponding R-charge. Since the middle} pentagon is not charged and thus invariant under $SU(4)$ rotations, the unit of R-charge ought to propagate all the way from the bottom to the top. As a result, the vacuum can no longer propagate at intermediate steps and the analogue of the first line in (\ref{hepMHV0}) is absent. The leading large $\tau$ contribution should carry one unit of R-charge and is just the scalar excitation $Z$ 
\beq
\text{vacuum}\rightarrow  Z(u) \to Z(v) \rightarrow  \text{vacuum}\, . \la{vacZZvac}
\eeq
That is, instead of (\ref{hepMHV0}), we simply have 
\beqa
\mathcal{W}_\text{hep}^{(7145)}  = \int \frac{du\,dv}{(2\pi)^2}  P_*(0|u)\mu(u)P(-u|v)\mu(v)P_*(-v|0)\,e^{-\tau_1E (u)+i p(u)\sigma_1-\tau_2E(v)+i p(v)\sigma_2} +\dots      \la{hepNMHV0} 
\eeqa
Note that there is no dependence on $\phi_1$ or $\phi_2$ in (\ref{hepNMHV0}) since scalars are neutral with respect to the $U(1)$ rotation. 
In this expression $P(u|v)=P_{ZZ}(u|v)$ stands for the scalar pentagon transition. {The first and last transitions $P_*(0|u)$ and $P_*(-v|0)$ represent the amplitude for creating a scalar excitation from the vacuum through a charged transition. These can again be removed by a judicious choice of normalization of the single particle wave function. A convenient one in this case is to set $P_*(0|u)=P_*(u|0)=1/g$, see also section \ref{beforeSec}. For a similar hexagon NMHV component, in this normalization, we simply have
\beq\la{hexNMHV0}
\def\svgwidth{9cm}
\begingroup%
  \makeatletter%
  \providecommand\color[2][]{%
    \errmessage{(Inkscape) Color is used for the text in Inkscape, but the package 'color.sty' is not loaded}%
    \renewcommand\color[2][]{}%
  }%
  \providecommand\transparent[1]{%
    \errmessage{(Inkscape) Transparency is used (non-zero) for the text in Inkscape, but the package 'transparent.sty' is not loaded}%
    \renewcommand\transparent[1]{}%
  }%
  \providecommand\rotatebox[2]{#2}%
  \ifx\svgwidth\undefined%
    \setlength{\unitlength}{733.57265625bp}%
    \ifx\svgscale\undefined%
      \relax%
    \else%
      \setlength{\unitlength}{\unitlength * \real{\svgscale}}%
    \fi%
  \else%
    \setlength{\unitlength}{\svgwidth}%
  \fi%
  \global\let\svgwidth\undefined%
  \global\let\svgscale\undefined%
  \makeatother%
  \begin{picture}(1,0.31626042)%
    \put(0,0){\includegraphics[width=\unitlength]{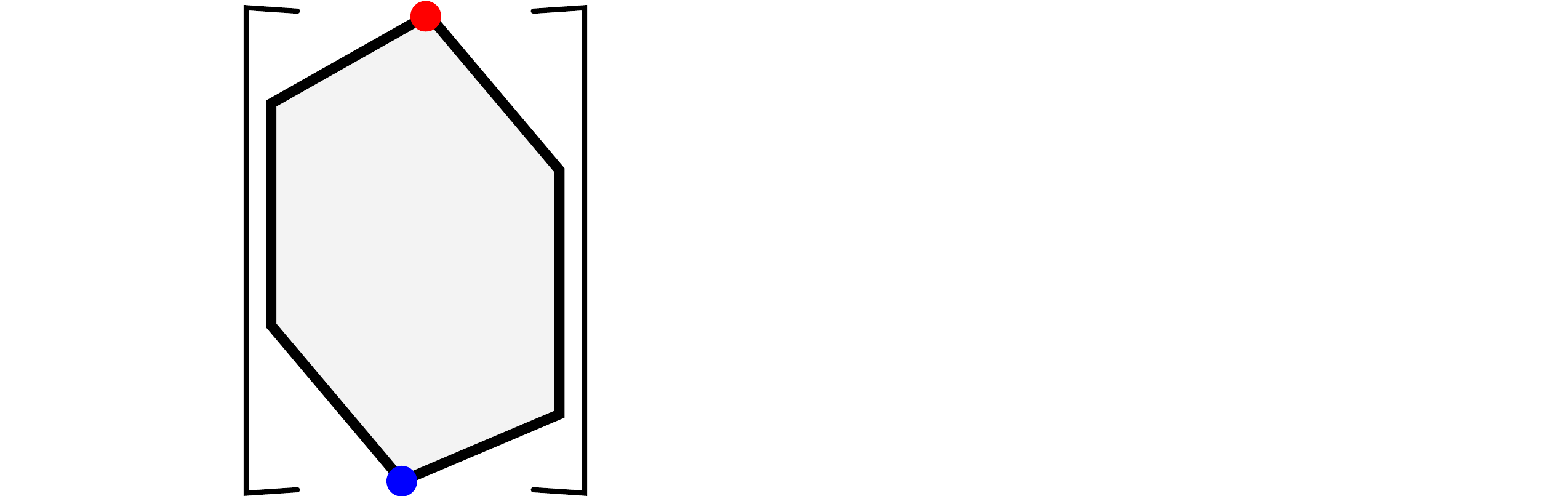}}%
    \put(0.29413351,0.05100252){\color[rgb]{0,0,0}\makebox(0,0)[lb]{\smash{$\,_1$}}}%
    \put(0.32427242,0.15500165){\color[rgb]{0,0,0}\makebox(0,0)[lb]{\smash{$\,_2$}}}%
    \put(0.29542228,0.23581474){\color[rgb]{0,0,0}\makebox(0,0)[lb]{\smash{$\,_3$}}}%
    \put(0.21012829,0.25740801){\color[rgb]{0,0,0}\makebox(0,0)[lb]{\smash{$\,_4$}}}%
    \put(0.17534614,0.17091917){\color[rgb]{0,0,0}\makebox(0,0)[lb]{\smash{$\,_5$}}}%
    \put(0.20835308,0.07619839){\color[rgb]{0,0,0}\makebox(0,0)[lb]{\smash{$\,_6$}}}%
    \put(-0.00141679,0.15143325){\color[rgb]{0,0,0}\makebox(0,0)[lb]{\smash{$\displaystyle \mathcal{W}_\text{hex}^{(6134)}$}}}%
    \put(0.40863121,0.15143325){\color[rgb]{0,0,0}\makebox(0,0)[lb]{\smash{$\displaystyle =\frac{1}{g^2} \int \frac{du}{2\pi}  \mu(u)  \,e^{-\tau E (u)+i p(u)\sigma} +\dots$}}}%
  \end{picture}%
\endgroup%

\eeq

We remind the reader that we are working in rapidity space $u$ because this variable is the most natural one from the integrability view point. We could of course use the momentum of the excitation $p(u)$, which is somewhat more `physical'. The two measures in both rapidity and momentum spaces would then be related as
\beq\la{tomomentum}
dp\,\hat\mu(p)=du\,\mu(u) \, .
\eeq
  
We conclude this subsection with a short comment on the normalization independence of `physical' quantities. The square measure and the pentagon transition enter `physical' amplitudes through the particular combinations that appear in the hexagon and heptagon expansion, see for example (\ref{33}) and (\ref{hepNMHV0}). In practice this implies that by analyzing the amplitudes we can only access  to
\beq
P(0|u)\mu(u)P(u|0) \qquad \text{and}\qquad  {P(u|v)\over P(0|u)P(v|0)} 
\eeq
for MHV Wilson loops (and similarly with a few stars for the supersymmetric case). Hence, only these ratios are physical for us. Our normalization, which corresponds to setting the single-particle creation amplitudes to a constant, is just equivalent to saying that we are only studying and interested in these (normalization independent) physical quantities.

\subsection{Flattening -- Extracting Pentagons and Square Transitions} \la{summarySection}

One of the goal of this paper is to confront predictions for pentagon transitions and measures with perturbative data at weak coupling. To set the ground work for this comparison we shall now go through some specifics of the weak coupling expansion. Our consideration will also illustrate the way one can extract the transitions and measures from the \textit{flattening} of the Wilson loops.

A distinctive feature of the weak coupling expansion is that the energies of the flux-tube excitations are dominated by their tree-level expressions which are independent of the momenta. For example, the gluonic and scalar excitations discussed above have both energy one in accord with their (bare) twist. More precisely, their energies take the form $E(u)=1+\gamma(u)$ where $\gamma(u)$ depends on the flavour but always starts at order $g^2$. Note on the other hand that their momenta read $p(u)=2u+O(g^2)$ in both cases. To isolate the single particle contributions from the full OPE expansion of the Wilson loops, it suffices then to take the large $\tau_i$ limit and look for the term(s) decaying like $e^{-\tau_i}$. 
Take for instance the hexagon (\ref{hexF}). We have
\beq
\mathcal{W}_\text{hex} =1+ 2\cos(\phi) e^{-\tau} \int \frac{du}{2\pi}  \mu(u) e^{-\gamma(u)\tau+i p(u)\sigma} + \dots \la{hexExp}
\eeq
at large $\tau$, and moreover both the measure $\mu(u)$ and the anomalous dimension $\gamma(u)$ are of order $g^2$. Hence, at the $l$-th loop order in perturbation theory, the function dressing $e^{-\tau}$ is a polynomial of degree $l-1$ in $\tau$. The terms multiplying positive powers of $\tau$ in this polynomial necessarily involve loop corrections to the energy \cite{OPEpaper} and thus the knowledge of the measure at the previous loop orders. The most interesting term at each loop order is then the one without any powers of $\tau$. It is this one which after Fourier transform yields directly the $l$-th loop measure in momentum space, or equivalently in rapidity space using (\ref{tomomentum}).
Of course, to generate new predictions for the Wilson loops or reciprocally to check the full consistency of the OPE, it is important to keep all powers of $\tau$. But to extract the measure we can just expand at large $\tau$ and drop all positive powers of $\tau$. We dub this procedure {\it flattening}.  The exact same discussion applies for NMHV hexagon components such as (\ref{hexNMHV0}) up to a shift of the overall normalization by~$g^2$. 

The strategy for extracting the pentagon transitions is similar and applies to both the bosonic heptagon (\ref{hepMHV0}) and the susy components like (\ref{hepNMHV0}). This time we look for the term that scales as $e^{-\tau_1-\tau_2}$ at large $\tau_{1}, \tau_{2}$ {and drop all} positive powers of $\tau_1,\tau_2$ in it. For the scalar NMHV component (\ref{hepNMHV0}) this is the end of the process. 
For the bosonic one in (\ref{hepMHV0}) we still need to separate the helicity preserving transition ($h$) from the helicity violating one ($\bar h$) as they come with different $\phi$'s dependence. 
For example, the contribution in the last line of (\ref{hepMHV0}) corresponds to the term in the heptagon bosonic Wilson loop $\mathcal{W}_\text{hept}$ that scales as 
\beq\la{leadingheptagon}
2 \cos(\phi_1+\phi_2) e^{-\tau_1-\tau_2} \Big[ \underbrace{\int \frac{du \,dv}{(2\pi)^2}  \mu(u)P(-u|v)\mu(v) e^{i p(u)\sigma_1+i p(v)\sigma_2 }}_{\displaystyle\equiv h(\sigma_1,\sigma_2)}+ \text{positive powers of }\tau_1,\tau_2  \Big] 
\eeq
and directly probes the helicity preserving pentagon transition $P(u|v)$. Similarly, the term that scales as $2 \cos(\phi_1-\phi_2) e^{-\tau_1-\tau_2} \bar h(\sigma_1,\sigma_2)$ probes the helicity violating one, with $\bar h(\sigma_1,\sigma_2)$ given by the same expression as $h(\sigma_1,\sigma_2)$ with $\bar P$ instead of $P$. 

In summary, by flattening hexagons and heptagons we have direct access to the measures and pentagon transitions respectively, 
\beq \la{flattening}
\def\svgwidth{12cm}
\begingroup%
  \makeatletter%
  \providecommand\color[2][]{%
    \errmessage{(Inkscape) Color is used for the text in Inkscape, but the package 'color.sty' is not loaded}%
    \renewcommand\color[2][]{}%
  }%
  \providecommand\transparent[1]{%
    \errmessage{(Inkscape) Transparency is used (non-zero) for the text in Inkscape, but the package 'transparent.sty' is not loaded}%
    \renewcommand\transparent[1]{}%
  }%
  \providecommand\rotatebox[2]{#2}%
  \ifx\svgwidth\undefined%
    \setlength{\unitlength}{983.70390625bp}%
    \ifx\svgscale\undefined%
      \relax%
    \else%
      \setlength{\unitlength}{\unitlength * \real{\svgscale}}%
    \fi%
  \else%
    \setlength{\unitlength}{\svgwidth}%
  \fi%
  \global\let\svgwidth\undefined%
  \global\let\svgscale\undefined%
  \makeatother%
  \begin{picture}(1,0.34952082)%
    \put(0,0){\includegraphics[width=\unitlength]{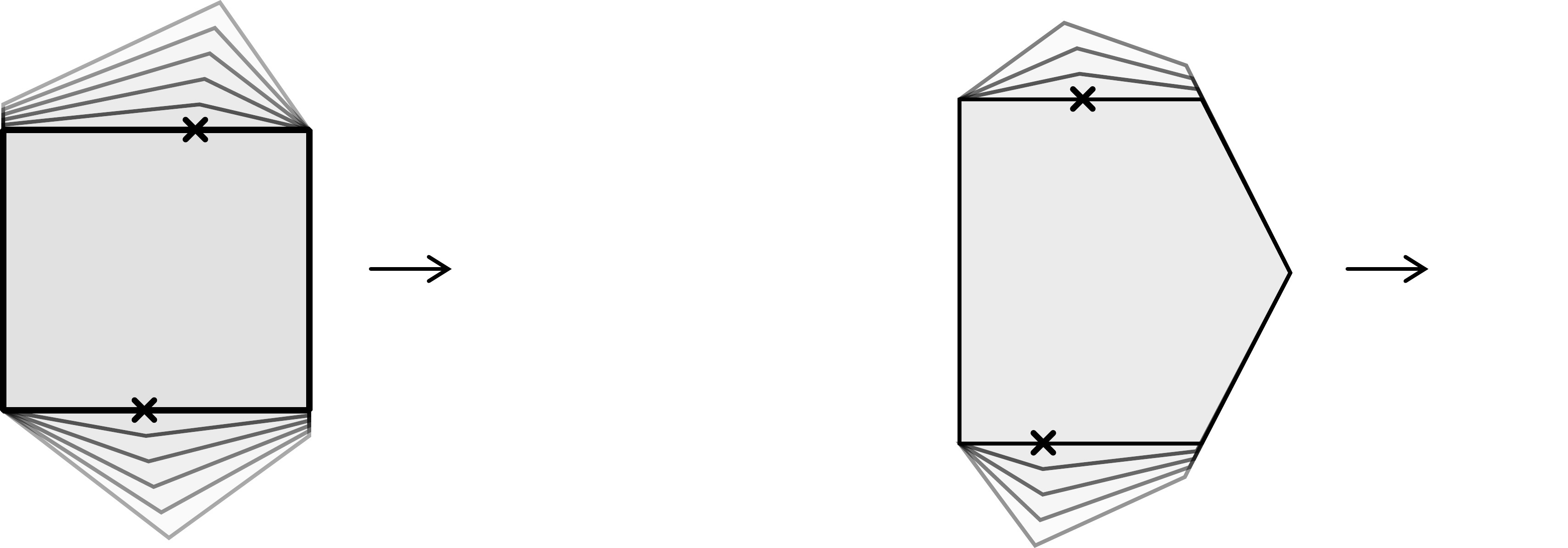}}%
    \put(0.31460406,0.16971178){\color[rgb]{0,0,0}\makebox(0,0)[lb]{\smash{$\mu(u)\,,$}}}%
    \put(0.93267632,0.16971178){\color[rgb]{0,0,0}\makebox(0,0)[lb]{\smash{$P(u|v)$}}}%
  \end{picture}%
\endgroup%

\eeq
Polygons with more edges are made out of the same building blocks and therefore do not involve new (independent) structures. Their analysis tests the way $\mu$ and $P$ are glued together into a big polygon, as illustrated in appendix \ref{gluingappendix} for the scalar NMHV component of an octagon WL at Born level. 

Our final comment concerns the reflection symmetry of an insertion obtained by the above flattening procedure. One would have noticed that the two edges on the top of the hexagon in (\ref{flattening}), or equivalently on the bottom, are not on an equal footing since only one of them ends on a cusp of the square.
This implies that the hexagon is not invariant under $\sigma \rightarrow -\sigma$. Instead, under a reflection interchanging the left and right sides of the hexagon, we have that $\sigma\to-\sigma+\log(1+e^{-2\tau})$, while $\tau$ and $\phi$ stay the same. What is important here is that the symmetry breaking term $\log(1+e^{-2\tau})$ is suppressed at large $\tau$ and cannot affect the leading twist-one states. In other words, the insertions of scalar and gauge fields considered in this paper are invariant under the $\sigma\to-\sigma$ transformation or equivalently under the flipping of the signs of the momenta. For the measure we can derive from it that $\mu(u)=\mu(-u)$ while for the transition we get
\beq
P(-u|-v)=P(v|u) \, ,
\eeq 
which is the first axiom for the pentagon transitions in \cite{short}.

\subsection{Revisiting Scalars at Born Level} \la{beforeSec}
To get familiarized with the square and pentagon transitions it helps computing them directly from their definitions (\ref{thetransition}) and (\ref{singleparticlemeasure}). This is what we shall do here at the leading order in perturbation theory and for the simplest possible insertions, i.e. for scalars. 

We first prepare two GKP eigenstates $|u\>$ and $\<v|$ for a single scalar with rapidity $u$ or $v$. At Born level, these states are simply obtained by inserting a scalar field at the bottom or top edge of the loop, see (\ref{scalarinsertion}). For example, a single particle state with momentum $p=2u$ is given by\footnote{Note that the insertion $|\sigma\>$ would look different in different conformal frames. For example, in the $(0,\infty)$ frame we have $x(\sigma)=e^{2\sigma}$ and therefore $|\sigma\> \propto \star\sqrt{x}\,Z(x)\star\nn$. 
In the $(0,1)$ frame we have $x(\sigma)=e^{2\sigma}/(1+e^{2\sigma})$ instead and then the same state takes the form $|\sigma\> \propto \star\sqrt{x(1-x)}\,Z\(x\)\star$.
}
\beq\la{momentumstate}
|u\>\equiv{\cN(u)\over\sqrt2}\int\limits_{-\infty}^\infty\!\! d\sigma\, e^{2iu\sigma}|\sigma\>, \qquad \text{where} \qquad 
|\sigma\>=\star\,\sqrt{|\partial_\sigma x(\sigma)|} \,{Z}(x(\sigma))\,\star\, ,
\eeq
with $x=x^-$ the coordinate along the null line (which is the only direction relevant here) and $\cN(u)$ a normalization factor. Note that the factor $\sqrt2$ could be absorbed in $\cN$ but is introduced here to make the following formulae look nicer.

We can now use these states to compute the square and pentagon transitions by a direct Feynman diagram algebra. We start with a square and insert the state $|\sigma_1\>$ for the complex scalar $Z$ at the bottom edge and the state $\<\sigma_2|$ for the conjugate $\bar Z$ field at the top edge. At tree level, we just draw a free propagator between the two fields, which are dressed by their conformal factors, and get
\beq\la{squaretransition}
\def\svgwidth{9cm}
\begingroup%
  \makeatletter%
  \providecommand\color[2][]{%
    \errmessage{(Inkscape) Color is used for the text in Inkscape, but the package 'color.sty' is not loaded}%
    \renewcommand\color[2][]{}%
  }%
  \providecommand\transparent[1]{%
    \errmessage{(Inkscape) Transparency is used (non-zero) for the text in Inkscape, but the package 'transparent.sty' is not loaded}%
    \renewcommand\transparent[1]{}%
  }%
  \providecommand\rotatebox[2]{#2}%
  \ifx\svgwidth\undefined%
    \setlength{\unitlength}{979.56923828bp}%
    \ifx\svgscale\undefined%
      \relax%
    \else%
      \setlength{\unitlength}{\unitlength * \real{\svgscale}}%
    \fi%
  \else%
    \setlength{\unitlength}{\svgwidth}%
  \fi%
  \global\let\svgwidth\undefined%
  \global\let\svgscale\undefined%
  \makeatother%
  \begin{picture}(1,0.27928117)%
    \put(0,0){\includegraphics[width=\unitlength]{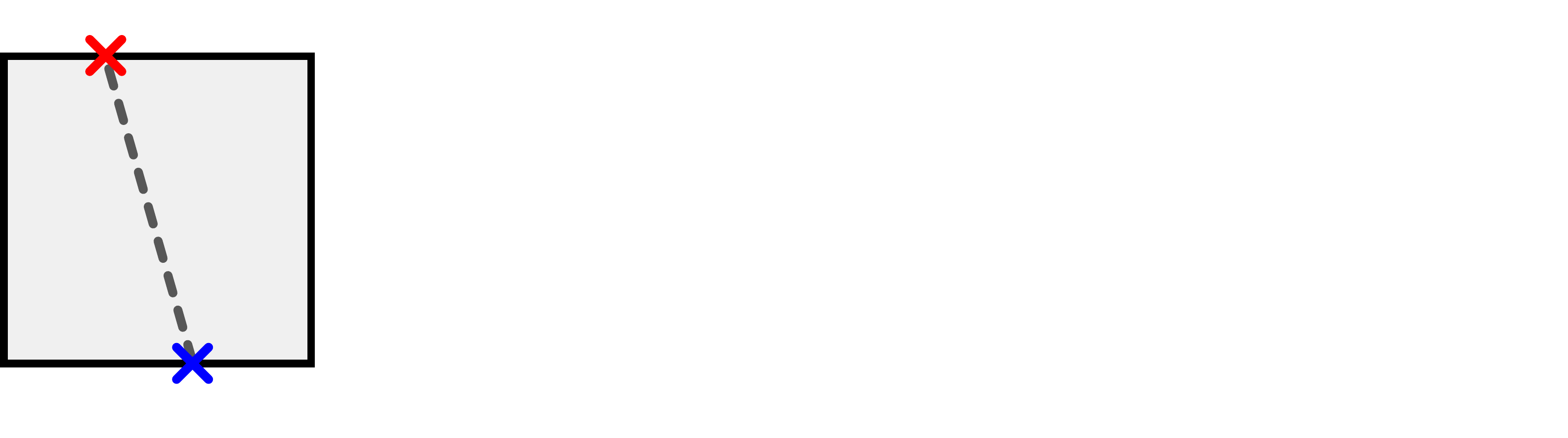}}%
    \put(0.10746312,0.00347624){\color[rgb]{0,0,0}\makebox(0,0)[lb]{\smash{$\sigma_1$}}}%
    \put(0.05192851,0.26808239){\color[rgb]{0,0,0}\makebox(0,0)[lb]{\smash{$\sigma_2$}}}%
    \put(0.21853235,0.1406794){\color[rgb]{0,0,0}\makebox(0,0)[lb]{\smash{$\displaystyle =\sqrt{|\d_{\sigma_1}x|}\times{g^2\over x-y}\times\sqrt{|\d_{\sigma_2}y|}={2g^2\over 
e^{\sigma_2-\sigma_1}+e^{\sigma_1-\sigma_2}}$}}}%
  \end{picture}%
\endgroup%

\eeq
where we have suppressed the trivial $x^+$ dependence.\footnote{We work in the Euclidian kinematics where $x$ and $y$ are space-like separated for all values of $\sigma_{1,2}$. For example, in the $(0,\infty)$ frame, this choice corresponds to $x(\sigma)=e^{2\sigma}$ and $y(\sigma)=-e^{2\sigma}$. Please see figure 3 in \cite{AMApril} or the discussion around (75) in \cite{Wang} for simple examples illustrating some differences between Euclidean and Lorentzian kinematics.}  Using~(\ref{momentumstate}) we can convert the position space relation (\ref{squaretransition}) into its momentum space counterpart and extract the measure by comparing with (\ref{singleparticlemeasure}). This leads to
\beq\la{squareinmomentum}
\def\svgwidth{12.5cm}
\begingroup%
  \makeatletter%
  \providecommand\color[2][]{%
    \errmessage{(Inkscape) Color is used for the text in Inkscape, but the package 'color.sty' is not loaded}%
    \renewcommand\color[2][]{}%
  }%
  \providecommand\transparent[1]{%
    \errmessage{(Inkscape) Transparency is used (non-zero) for the text in Inkscape, but the package 'transparent.sty' is not loaded}%
    \renewcommand\transparent[1]{}%
  }%
  \providecommand\rotatebox[2]{#2}%
  \ifx\svgwidth\undefined%
    \setlength{\unitlength}{729.41933594bp}%
    \ifx\svgscale\undefined%
      \relax%
    \else%
      \setlength{\unitlength}{\unitlength * \real{\svgscale}}%
    \fi%
  \else%
    \setlength{\unitlength}{\svgwidth}%
  \fi%
  \global\let\svgwidth\undefined%
  \global\let\svgscale\undefined%
  \makeatother%
  \begin{picture}(1,0.15683708)%
    \put(0,0){\includegraphics[width=\unitlength]{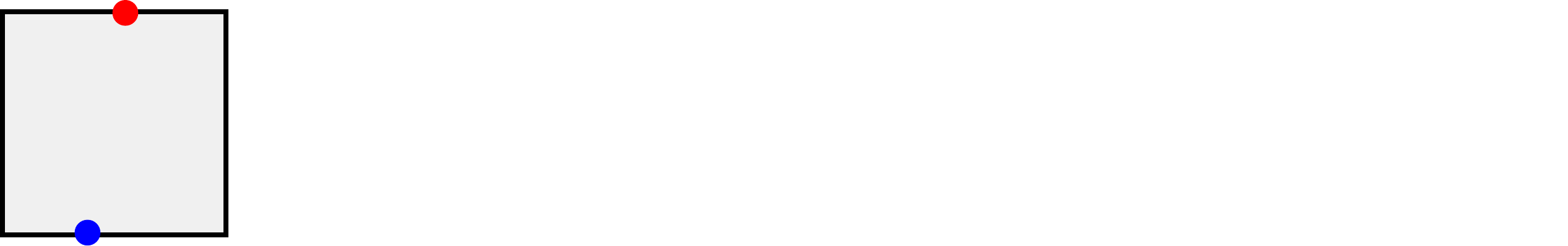}}%
    \put(0.16159489,0.07198731){\color[rgb]{0,0,0}\makebox(0,0)[lb]{\smash{$\displaystyle =\cN(u)^2\,{g^2\pi^2\over2\cosh(\pi u)}\,\delta(u-v)\qquad\Rightarrow\qquad\mu(u)=
{4\cosh(\pi u)\over\pi g^2\cN(u)^2}$}}}%
    \put(0.04568733,0.02530423){\color[rgb]{0,0,0}\makebox(0,0)[lb]{\smash{$u$}}}%
    \put(0.07304197,0.11749497){\color[rgb]{0,0,0}\makebox(0,0)[lb]{\smash{$v$}}}%
  \end{picture}%
\endgroup%

\eeq
where we used that at tree level the vacuum square transition in the denominator of (\ref{singleparticlemeasure}) is equal to one.  

The computation of the pentagon transition at Born level is no more difficult. We now insert the states $|\sigma_2\>$ and $\<\sigma_1|$ on the bottom and top edges of a pentagon and find
\beq\la{pentagontransition}
\def\svgwidth{11cm}
\begingroup%
  \makeatletter%
  \providecommand\color[2][]{%
    \errmessage{(Inkscape) Color is used for the text in Inkscape, but the package 'color.sty' is not loaded}%
    \renewcommand\color[2][]{}%
  }%
  \providecommand\transparent[1]{%
    \errmessage{(Inkscape) Transparency is used (non-zero) for the text in Inkscape, but the package 'transparent.sty' is not loaded}%
    \renewcommand\transparent[1]{}%
  }%
  \providecommand\rotatebox[2]{#2}%
  \ifx\svgwidth\undefined%
    \setlength{\unitlength}{1172.77392578bp}%
    \ifx\svgscale\undefined%
      \relax%
    \else%
      \setlength{\unitlength}{\unitlength * \real{\svgscale}}%
    \fi%
  \else%
    \setlength{\unitlength}{\svgwidth}%
  \fi%
  \global\let\svgwidth\undefined%
  \global\let\svgscale\undefined%
  \makeatother%
  \begin{picture}(1,0.22781479)%
    \put(0,0){\includegraphics[width=\unitlength]{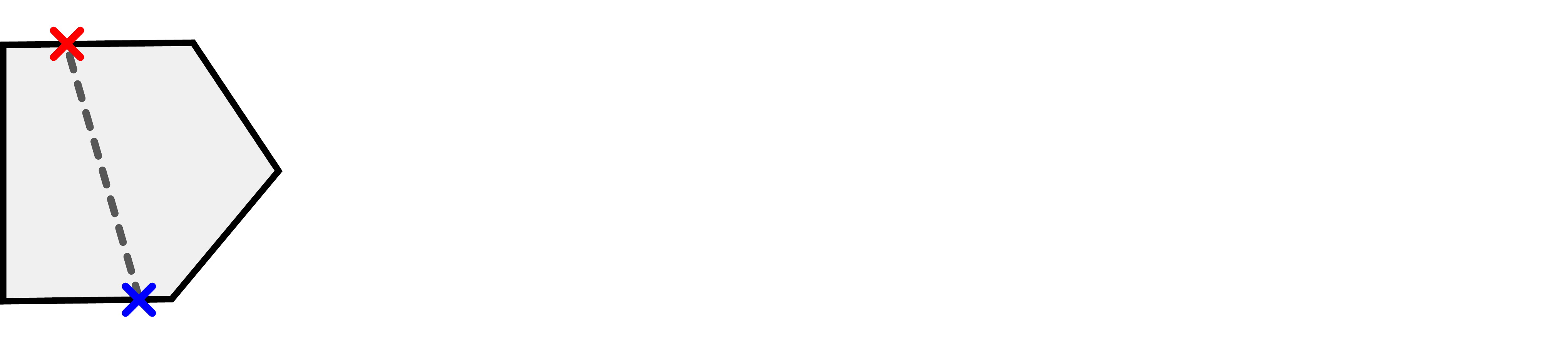}}%
    \put(0.08157375,0.00290356){\color[rgb]{0,0,0}\makebox(0,0)[lb]{\smash{$\sigma_1$}}}%
    \put(0.03382372,0.21846091){\color[rgb]{0,0,0}\makebox(0,0)[lb]{\smash{$\sigma_2$}}}%
    \put(0.19617383,0.11341079){\color[rgb]{0,0,0}\makebox(0,0)[lb]{\smash{$\displaystyle =\sqrt{|\d_{\sigma_1}x|}\times{g^2\over x-y}\times\sqrt{|\d_{\sigma_2}y|}={2g^2\over e^{\sigma_2-\sigma_1}
+e^{\sigma_1-\sigma_2}+e^{\sigma_2+\sigma_1}}$}}}%
  \end{picture}%
\endgroup%

\eeq
The reader may have noticed that at this order in perturbation theory the excitations do not probe all of the geometry of the pentagon. Still they know about it since they are in different conformal frames
\footnote{For example, we can take $y(\sigma)=-e^{2\sigma}$ and $x(\sigma)=e^{2\sigma}/(1+e^{2\sigma})$ in (\ref{pentagontransition}).},
in contrast with the square transition where they are both in the same frame. This is the reason for the difference between (\ref{squaretransition}) and (\ref{pentagontransition}). We note also that we can always shift the origin of $\sigma_1$ or $\sigma_2$ by a constant. We use the natural choice for the origin that is induced from the pentagon geometry, see appendix \ref{geomeryappendix} for details. We now use (\ref{momentumstate}) to convert the position space transition (\ref{pentagontransition}) into its momentum space counterpart (\ref{thetransition}) and conclude that 
\beq\la{scalarpentagontransition}
P(u|v)={g^2\over4}\,\cN(u)\,\cN(v)\,\Gamma\(\tfrac{1}{2}-iu\)\Gamma\(iu-iv\)\Gamma\(\tfrac{1}{2}+iv\) \, .
\eeq

Let us add a comment here on the square limit of the tree level pentagon transition. 
If we send both excitations in the pentagon transition (\ref{pentagontransition}) to the left, $\sigma_1,\sigma_2\to-\infty$, while holding $\sigma=\sigma_1-\sigma_2$ fixed, we arrive at the position space square transition (\ref{squaretransition}). This is what we explained before.
The position space behavior at large $\sigma_1+\sigma_2$ translates, in momentum space, to the limit $u\to v$. More precisely, in prefect agreement with (\ref{measureEq}) we observe that (\ref{scalarpentagontransition}) and (\ref{squareinmomentum}) are related through
\beq
\underset{v=u}{\operatorname{{\rm Res}}}\, P(u|v)= \frac{i}{\mu(u)} \, .
\eeq

The remaining single scalar transition entering our decomposition is the amplitude $P_*(0|u)$ for creating an excitation from the vacuum through the charged pentagon, see (\ref{fP}) for instance. {At tree level, }the charged {transition} is nothing else than a pentagon with a scalar insertion at the right bottom cusp. Accordingly, a direct way to $P_*(0|u)$ is to start from the pentagon transition (\ref{pentagontransition}), send the bottom excitation to the latter cusp using $\sigma_1\to\infty$ and extract the term scaling like $e^{-\sigma_1}$. After multiplying the result by $1/(2g)$, to match with the normalization~\cite{superloopsimon} of the scalar insertion in the super loop, and Fourier transforming, we arrive at 
\beq\la{Pstarposition}
\def\svgwidth{10.5cm}
\begingroup%
  \makeatletter%
  \providecommand\color[2][]{%
    \errmessage{(Inkscape) Color is used for the text in Inkscape, but the package 'color.sty' is not loaded}%
    \renewcommand\color[2][]{}%
  }%
  \providecommand\transparent[1]{%
    \errmessage{(Inkscape) Transparency is used (non-zero) for the text in Inkscape, but the package 'transparent.sty' is not loaded}%
    \renewcommand\transparent[1]{}%
  }%
  \providecommand\rotatebox[2]{#2}%
  \ifx\svgwidth\undefined%
    \setlength{\unitlength}{1117.51513672bp}%
    \ifx\svgscale\undefined%
      \relax%
    \else%
      \setlength{\unitlength}{\unitlength * \real{\svgscale}}%
    \fi%
  \else%
    \setlength{\unitlength}{\svgwidth}%
  \fi%
  \global\let\svgwidth\undefined%
  \global\let\svgscale\undefined%
  \makeatother%
  \begin{picture}(1,0.21434474)%
    \put(0,0){\includegraphics[width=\unitlength]{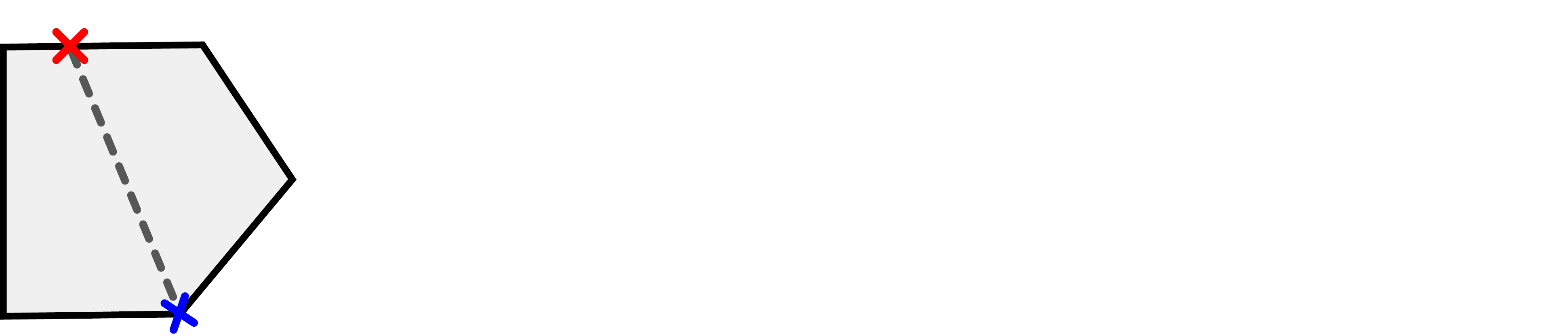}}%
    \put(0.03549623,0.20452834){\color[rgb]{0,0,0}\makebox(0,0)[lb]{\smash{$\sigma$}}}%
    \put(0.20587422,0.0942837){\color[rgb]{0,0,0}\makebox(0,0)[lb]{\smash{$\displaystyle ={g\over e^\sigma+e^{-\sigma}}\qquad\Rightarrow\qquad P_*(0|u)=\cN(u){g\pi\over2\cosh(\pi u)}$}}}%
  \end{picture}%
\endgroup%

\eeq

We can test now how these various ingredients combine together when constructing physical quantities (i.e., those entering the computation of scattering amplitudes), by looking at the integrand for the heptagon super Wilson loop, for instance. To leading order at large~$\tau_i$, it describes the process `$\text{vacuum }\rightarrow Z(u) \to Z(v) \rightarrow \text{ vacuum}$'
and it is given by 
\beq
P_*(0|u)\,\mu(u)\,P(-u|v)\,\mu(v)\, P_*(-v|0)  = \Gamma\(\tfrac{1}{2}+iu\)\Gamma\(-iu-iv\)\Gamma\(\tfrac{1}{2}+iv\)\, .  \la{integrand2}
\eeq
We verify here that the unphysical normalization $\mathcal{N}(u)$ drops out completely at the end, as it should be.\footnote{An even simpler example would be the hexagon integrand $P_*(0|u)\,\mu(u)\,P_*(-u|0)  =\pi g^2\textrm{sech}{(\pi u)}$ where again the normalization neatly drops out. } We further observe that the integrand (\ref{integrand2}) matches perfectly with the one (\ref{integrand}) computed in the introduction (by decomposing a tree level scattering amplitude) once we identity $p_1=2u$ and $p_2=2v$. Finally, we already pointed out that we can always fix our normalization, i.e. $\mathcal{N}(u)$, such that 
\beq\la{normalization}
P_*(0|u)=1/g\,.
\eeq
We can impose this normalization at any loop order as we will always do it in this paper. Then
\beq\label{mu-P-BL}
\mu(u)=\frac{\pi g^2}{\cosh(\pi u)}\,,\qquad P(u|v)={\Gamma\(iu-iv\)\over g^2\Gamma\(\tfrac{1}{2}+iu\)\Gamma\(\tfrac{1}{2}-iv\)}\, ,
\eeq
to leading order at weak coupling. Note that in this normalization the hexagon integrand is now governed by the measure $\mu(u)$ and nothing else, see (\ref{hexNMHV0}).

\section{Scalars} \la{scalarSec}

In this section we introduce and motivate our conjecture for the transition of a single scalar excitation over the pentagon Wilson loop at finite coupling. As for the gluonic excitations discussed in~\cite{short} and later on in this paper, our construction relies on two fundamental axioms. They encode the properties of the transition under permutation and crossing of the excitations respectively.
Their main ingredients are the S-matrix for scattering on top of the GKP flux tube and the mirror rotation that we shall first review. At the end, part of the evidence for our expressions comes from the matching with higher-loop data that shall be performed here for both the hexagon and heptagon Wilson loops.

\subsection{The Scalar S-matrix and the Mirror Rotation}

One feature of the scalars is that they are charged under the $SU(4)$ symmetry of the gauge theory. The scalars $\Phi_i$ come in 6 different flavours $i=1, \ldots , 6$ and transform in the vector representation of $SU(4)$. When inserted with definite rapidities along the flux tube direction, they scatter among themselves and exchange their flavour indices. The most elementary process involves the scattering of two scalars only 
\beq\label{2to2process}
\Phi_{i}(u)\Phi_{j}(v) \rightarrow \Phi_{l}(v)\Phi_{k}(u)
\eeq
and is fully characterized by the S-matrix $S_{ij}^{kl}(u, v)$. The latter matrix was extracted in~\cite{toappearAdam} from the all-loop Bethe ansatz equations for the super-spin-chain of the $\mathcal{N}=4$ gauge theory~\cite{BS, BES}. It has the expected form for a factorized S-matrix having the $O(6)$ symmetry, being proportional to the rational R-matrix $R_{ij}^{kl}(u-v)$ in the vector representation~\cite{Zamolodchikov} (see also~\cite{toappearAdam} for the case at hand)
\beq
S_{ij}^{kl}(u, v) = S(u, v)R_{ij}^{kl}(u-v)\, , \qquad R_{ij}^{kl}(u-v) = \frac{u-v}{u-v-i}\delta_{i}^{k}\delta_{j}^{l} + \ldots\, .
\eeq
The object of interest here is the scalar factor $S(u, v)$ that multiplies this R-matrix. This phase encodes the dynamical information of the GKP background that distinguishes it from other integrable models sharing the same $O(6)$ symmetry. It can be read directly from the scattering of two scalars in the symmetric channel or, equivalently, from the scattering of two scalars $Z$, with $Z \propto \Phi_1+i\Phi_2$. The scattering phases in the other two channels appearing in~(\ref{2to2process}), i.e., the adjoint and singlet channels, differ from it by simple rational prefactors. For instance, the scattering phase in the singlet channel reads
\beq\label{singlet}
S_{\textrm{singlet}}(u, v) \equiv \frac{1}{6}S_{ij}^{kl}(u, v)\delta^{ij}\delta_{kl} = \frac{(u-v+i)(u-v+2i)}{(u-v-i)(u-v-2i)} \, S(u, v)\, .
\eeq

At generic values of the coupling constant, the scattering phase $S(u, v)$ is a complicated function of the two rapidities and in particular it is not just a function of their difference. It is only known implicitly through the solution to a linear integral equation (see~\cite{toappearAdam} and Appendix~\ref{ansatz-App}). This equation is however easily solved iteratively at weak coupling hence providing explicit expression for $S(u, v)$ order by order in perturbation theory. To the leading order the relevant expression was already given in~(\ref{scalarSmatrix}) with $s=1/2$ for a scalar and $p_{1}=2u, p_{2}=2v$. The computation of the higher-loop corrections is explained in Appendix~\ref{ansatz-App} and implemented in the notebook attached to this paper.

In what follows we shall also need some non-perturbative information about $S(u, v)$. This one is provided by~\cite{toappearAdam}
\beq\label{main-properties}
\def\svgwidth{15.3cm}
\begingroup%
  \makeatletter%
  \providecommand\color[2][]{%
    \errmessage{(Inkscape) Color is used for the text in Inkscape, but the package 'color.sty' is not loaded}%
    \renewcommand\color[2][]{}%
  }%
  \providecommand\transparent[1]{%
    \errmessage{(Inkscape) Transparency is used (non-zero) for the text in Inkscape, but the package 'transparent.sty' is not loaded}%
    \renewcommand\transparent[1]{}%
  }%
  \providecommand\rotatebox[2]{#2}%
  \ifx\svgwidth\undefined%
    \setlength{\unitlength}{1723.13398438bp}%
    \ifx\svgscale\undefined%
      \relax%
    \else%
      \setlength{\unitlength}{\unitlength * \real{\svgscale}}%
    \fi%
  \else%
    \setlength{\unitlength}{\svgwidth}%
  \fi%
  \global\let\svgwidth\undefined%
  \global\let\svgscale\undefined%
  \makeatother%
  \begin{picture}(1,0.39277271)%
    \put(0,0){\includegraphics[width=\unitlength]{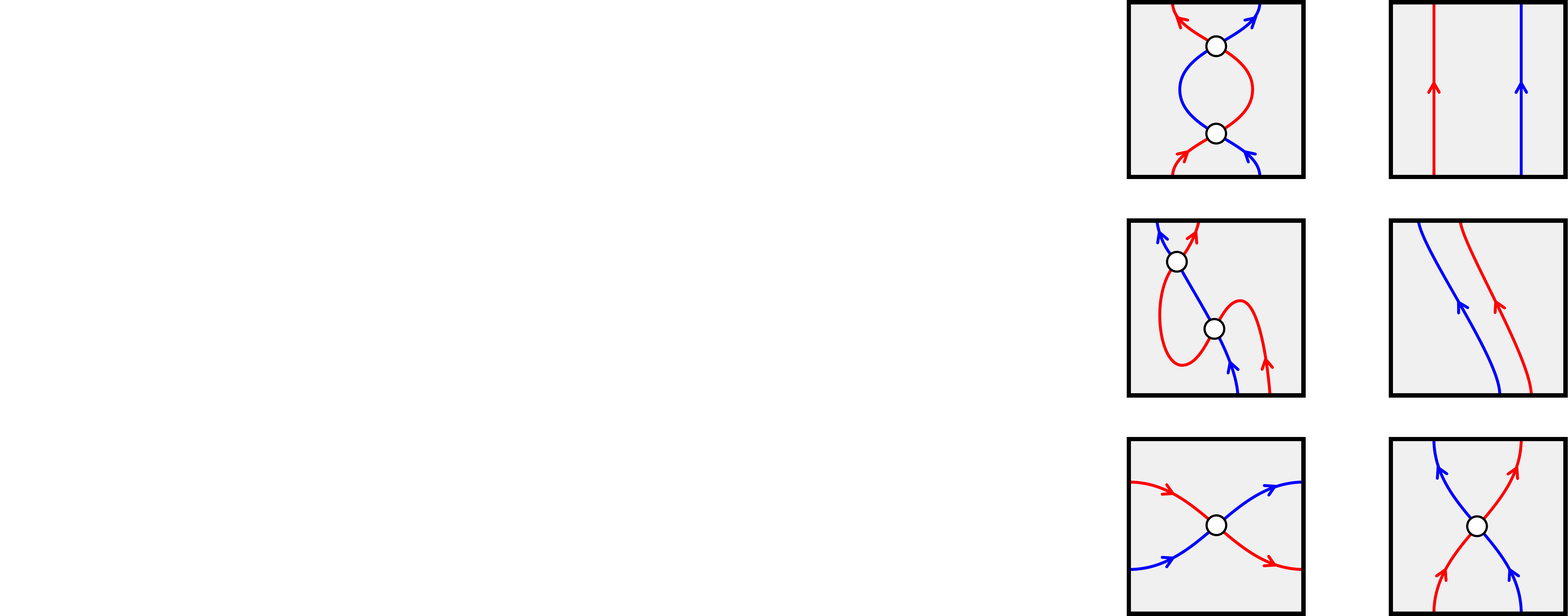}}%
    \put(-0.00069554,0.32835823){\color[rgb]{0,0,0}\makebox(0,0)[lb]{\smash{$\displaystyle \textrm{A. Unitarity :} \quad \qquad\qquad  S(u, v)S(v, u) = 1$}}}%
    \put(0.84799064,0.32835823){\color[rgb]{0,0,0}\makebox(0,0)[lb]{\smash{$\displaystyle =$}}}%
    \put(-0.00069554,0.18907712){\color[rgb]{0,0,0}\makebox(0,0)[lb]{\smash{$\displaystyle \textrm{B. Crossing :} \, \quad \qquad\qquad  S(u^{2\gamma}, v)S(u, v) = \frac{u-v}{u-v+2i}$}}}%
    \put(0.84799064,0.18907712){\color[rgb]{0,0,0}\makebox(0,0)[lb]{\smash{$\displaystyle \propto$}}}%
    \put(-0.00069554,0.04979602){\color[rgb]{0,0,0}\makebox(0,0)[lb]{\smash{$\displaystyle \textrm{C. Mirror symmetry :} \, \, \quad S(u^{\gamma}, v^{\gamma}) = S(u, v)$}}}%
    \put(0.84799064,0.04979602){\color[rgb]{0,0,0}\makebox(0,0)[lb]{\smash{$\displaystyle =$}}}%
  \end{picture}%
\endgroup%

\eeq
The first equation above is easily grasped since it expresses the unitarity of the theory. In an integrable theory all scattering processes are elastic hence resulting in simple unitarity relation like equation A in~(\ref{main-properties}). Of particular importance for the remaining two equations is the mirror transformation $\gamma$ that maps the rapidity plane back to itself, $\gamma : u \rightarrow u^{\gamma}$. This map takes an excitation from the real kinematics, that is where its energy and momentum are real, and brings it to the so-called mirror kinematics, where its energy and momentum are both purely imaginary. Physically it is a transformation that swaps the role of the (flux-tube) space and time directions $(\sigma, \tau)\rightarrow (-\tau, \sigma)$. The mirror excitation can then be thought as propagating with energy $-ip(u^{\gamma})$ along the time $\sigma$ and with momentum $-iE(u^{\gamma})$ with respect to the space $-\tau$. Pictorially, if we think of a real particle as evolving from bottom to top then we can view the mirror one as evolving from left to right,
\beq\label{mirror}
\def\svgwidth{15.3cm}
\begingroup%
  \makeatletter%
  \providecommand\color[2][]{%
    \errmessage{(Inkscape) Color is used for the text in Inkscape, but the package 'color.sty' is not loaded}%
    \renewcommand\color[2][]{}%
  }%
  \providecommand\transparent[1]{%
    \errmessage{(Inkscape) Transparency is used (non-zero) for the text in Inkscape, but the package 'transparent.sty' is not loaded}%
    \renewcommand\transparent[1]{}%
  }%
  \providecommand\rotatebox[2]{#2}%
  \ifx\svgwidth\undefined%
    \setlength{\unitlength}{1726.35839844bp}%
    \ifx\svgscale\undefined%
      \relax%
    \else%
      \setlength{\unitlength}{\unitlength * \real{\svgscale}}%
    \fi%
  \else%
    \setlength{\unitlength}{\svgwidth}%
  \fi%
  \global\let\svgwidth\undefined%
  \global\let\svgscale\undefined%
  \makeatother%
  \begin{picture}(1,0.1175085)%
    \put(0,0){\includegraphics[width=\unitlength]{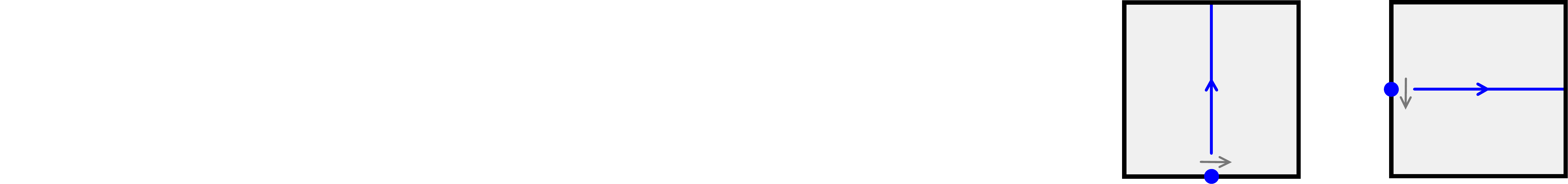}}%
    \put(0.2942403,0.05301839){\color[rgb]{0,0,0}\makebox(0,0)[lb]{\smash{$\gamma{\rm :}\  u\to u^\gamma$}}}%
    \put(0.8447633,0.05301839){\color[rgb]{0,0,0}\makebox(0,0)[lb]{\smash{$\displaystyle =$}}}%
    \put(0.90129851,0.0687741){\color[rgb]{0,0,0}\makebox(0,0)[lb]{\smash{$u$}}}%
    \put(0.77803324,0.02428739){\color[rgb]{0,0,0}\makebox(0,0)[lb]{\smash{$u^\gamma$}}}%
  \end{picture}%
\endgroup%

\eeq
One important point here is that the GKP background is expected to be mirror invariant. This symmetry was formulated in~\cite{AldayMaldacena,Bootstrapping} and is common to relativistic models. It means that the theory should look the same before and after the $\gamma$ rotation. At the level of the dispersion relation it translates into
\beq
p(u^{\gamma}) = iE(u)\, , \qquad E(u^{\gamma}) = ip(u)\, ,
\eeq
and was checked for a scalar in~\cite{MoreDispPaper}. For this excitation the transformation $\gamma$ is especially simple~\cite{MoreDispPaper} as it amounts to shifting the rapidity $u$ by the imaginary unit : $u^{\gamma} = u+i$. This is quite transparent at strong coupling where the theory becomes relativistic and the rapidity $u$ becomes the hyperbolic rapidity $\theta = \pi u/2$, such that $\theta^{\gamma} = \theta +i\pi/2$. For more generic values of the coupling it is primordial to specify the path along which the shift $u^{\gamma} = u+i$ is performed. The reason behind this subtle step is that observables like the energy, the momentum, or the S-matrix are multivalued functions of their rapidities. Notably, they all have square-root branch points at $u = \pm 2g + i/2$. To implement correctly the mirror rotation $\gamma$ it is necessary to pass in-between these two branch points.
\footnote{At weak coupling these two branch points collide leaving some singularities behind them at $u=i/2$. For this reason the mirror rotation is non-perturbative in nature and requires a non-perturbative control on the observables under study.} Equation C in~(\ref{main-properties}) reflects the mirror invariance at the level of the scalar S-matrix. It implies indeed that the scattering of two scalars propagating in the mirror channel is identical to the scattering of two scalars in the real kinematics.  Finally we note that when combining two mirror rotations we get a transformation that changes the sign of both the energy and the momentum. This is the more familiar crossing transformation that maps a particle into an anti-particle. For a scalar excitation particle and anti-particle are the same and the equation B in~(\ref{main-properties}) simply results from this identification (see~\cite{toappearAdam,Zamolodchikov} for more details). 

We stress that the equations A-B-C  in~(\ref{main-properties}) are valid regardless of the value of the coupling constant. They are actually much more general and would be the same in any factorizable O(6) model with mirror symmetry, like the (relativistic) 2d non-linear O(6) sigma model for instance.

Equipped with these equations we can now turn to the consideration of the scalar pentagon transition.

\subsection{The Pentagon Transition at Finite Coupling}

We are interested in the transition from a state $\Phi_{i}(u)$ at the bottom to a state $\Phi_{j}(v)$ at the top of the pentagon. The matrix structure for this transition is very simple. Since the R-symmetry is preserved by the pentagon
\footnote{Note that of all the continuous symmetries of the underlying theory this is the only one preserved by the pentagon.}
the scalar transition ought to be proportional to the Kronecker delta
\beq
\left<\Phi_{j}(v)\right|\widehat{\mathcal{P}}\left|\Phi_{i}(u)\right> = P(u|v)\delta_{ij}\, .
\eeq
In other words there is only one possible channel for the transition of a single scalar through the pentagon and it is characterized by a single dynamical function $P(u|v)$. This is the latter object our fundamental axioms apply to. They read
\beqa\label{f-axioms}
&&\!\!\!\!\!\!\!\!\!\!\!\!\!\!\!\!\!\!\!\!\def\svgwidth{15.3cm}
\begingroup%
  \makeatletter%
  \providecommand\color[2][]{%
    \errmessage{(Inkscape) Color is used for the text in Inkscape, but the package 'color.sty' is not loaded}%
    \renewcommand\color[2][]{}%
  }%
  \providecommand\transparent[1]{%
    \errmessage{(Inkscape) Transparency is used (non-zero) for the text in Inkscape, but the package 'transparent.sty' is not loaded}%
    \renewcommand\transparent[1]{}%
  }%
  \providecommand\rotatebox[2]{#2}%
  \ifx\svgwidth\undefined%
    \setlength{\unitlength}{1536.1296875bp}%
    \ifx\svgscale\undefined%
      \relax%
    \else%
      \setlength{\unitlength}{\unitlength * \real{\svgscale}}%
    \fi%
  \else%
    \setlength{\unitlength}{\svgwidth}%
  \fi%
  \global\let\svgwidth\undefined%
  \global\let\svgscale\undefined%
  \makeatother%
  \begin{picture}(1,0.06209411)%
    \put(0,0){\includegraphics[width=\unitlength]{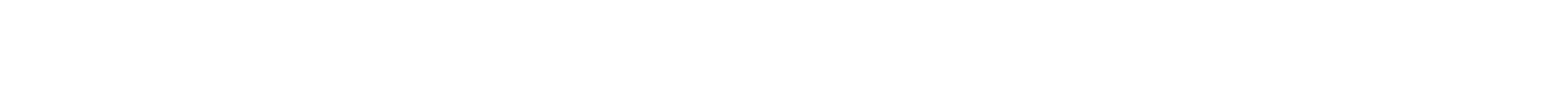}}%
    \put(0.11409918,0.02776364){\color[rgb]{0,0,0}\makebox(0,0)[lb]{\smash{$\displaystyle \textrm{I}.\,  \qquad  P(u|v) = S(u,v)P(v|u)$}}}%
    \put(0.81195693,0.02776364){\color[rgb]{0,0,0}\makebox(0,0)[lb]{\smash{$\text{see (\ref{funrel})}$}}}%
  \end{picture}%
\endgroup%
\\
&&\!\!\!\!\!\!\!\!\!\!\!\!\!\!\!\!\!\!\!\!\def\svgwidth{15.3cm}
\begingroup%
  \makeatletter%
  \providecommand\color[2][]{%
    \errmessage{(Inkscape) Color is used for the text in Inkscape, but the package 'color.sty' is not loaded}%
    \renewcommand\color[2][]{}%
  }%
  \providecommand\transparent[1]{%
    \errmessage{(Inkscape) Transparency is used (non-zero) for the text in Inkscape, but the package 'transparent.sty' is not loaded}%
    \renewcommand\transparent[1]{}%
  }%
  \providecommand\rotatebox[2]{#2}%
  \ifx\svgwidth\undefined%
    \setlength{\unitlength}{1718.65878906bp}%
    \ifx\svgscale\undefined%
      \relax%
    \else%
      \setlength{\unitlength}{\unitlength * \real{\svgscale}}%
    \fi%
  \else%
    \setlength{\unitlength}{\svgwidth}%
  \fi%
  \global\let\svgwidth\undefined%
  \global\let\svgscale\undefined%
  \makeatother%
  \begin{picture}(1,0.12224396)%
    \put(0,0){\includegraphics[width=\unitlength]{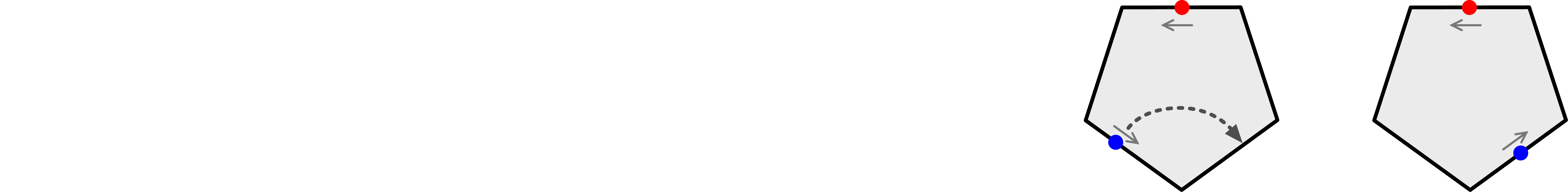}}%
    \put(0.6753225,0.00546282){\color[rgb]{0,0,0}\makebox(0,0)[lb]{\smash{$u^{-\gamma}$}}}%
    \put(0.73390761,0.06476332){\color[rgb]{0.30588235,0.30588235,0.30588235}\makebox(0,0)[lb]{\smash{$-\gamma$}}}%
    \put(0.97650242,0.011457){\color[rgb]{0,0,0}\makebox(0,0)[lb]{\smash{$u$}}}%
    \put(0.83469543,0.05804308){\color[rgb]{0,0,0}\makebox(0,0)[lb]{\smash{$\displaystyle =$}}}%
    \put(0.7468583,0.12679028){\color[rgb]{0,0,0}\makebox(0,0)[lb]{\smash{$v$}}}%
    \put(0.93089437,0.12679028){\color[rgb]{0,0,0}\makebox(0,0)[lb]{\smash{$v$}}}%
    \put(0.10198135,0.05804308){\color[rgb]{0,0,0}\makebox(0,0)[lb]{\smash{$\displaystyle \textrm{II}.\, \qquad P(u^{-\gamma}|v) = P(v|u)$}}}%
  \end{picture}%
\endgroup%
\nn
\eeqa
where $S(u, v)$ is the scattering phase introduced before and $-\gamma$ stands for the inverse of the mirror rotation. These axioms are quite similar to the ones proposed in~\cite{short} for gluons. Axiom I is especially important since it relates the pentagon transition $P(u, v)$ to the S-matrix $S(u, v)$.
Axiom II states that the pentagon transition goes back to itself if the inverse mirror transformation is performed on the bottom rapidity $u$. Behind it is the idea that the very same transformation that rotates an excitation from one edge to another on the square would apply to the pentagon as well. Under this assumption the excitation $u^{-\gamma}$ on the bottom edge of the pentagon can be viewed as an excitation $u$ now living on the right neighbouring edge (see picture in~(\ref{f-axioms})). Under a cyclic rotation of the pentagon this is nothing else that describing the pentagon transition $P(v|u)$ as written in axiom II. Note that there is a slight difference here with the case of gauge fields where this transformation also changes the (relative) helicity of the gluons~\cite{short}. 

It was pointed out in~\cite{short} that Axiom I is reminiscent of the Watson's equation for form factors in integrable theory~\cite{Watson}. This analogy is sometimes deceptive and shall be made more precise shortly. 
To motivate our axioms we can perform the following consistency checks:
\begin{description}
\item\textit{Unitarity:} It is nice to observe that Axiom I directly implies the unitarity of the S-matrix,
\beq
S(u,v)S(v,u) = \frac{P(u|v)}{P(v|u)}\times \frac{P(v|u)}{P(u|v)} = 1\, ,
\eeq
in agreement with property A in~(\ref{main-properties}).
\item\textit{Mirror symmetry:}  Combining twice Axiom II we find
\beq
P(u^{-\gamma}|v^{-\gamma}) = P(v^{-\gamma}|u) = P(u|v)\, .
\eeq
It means that the pentagon transition is mirror invariant. This relation has a simple geometric understanding since it follows from the cyclicity property of the pentagon (i.e., the invariance under an overall $\gamma$-rotation of the pentagon). We notice that this property of the pentagon transition would be in conflict with Axiom I if the S-matrix $S(u, v)$ was not itself invariant under a mirror rotation, see property C in~(\ref{main-properties}).
\item\textit{Watson's equation:} We can consider the process where a pair of scalars is produced at the top of the pentagon. It is described by the form factor
\beq
\left<\Phi_{i}(u)\Phi_{j}(v)\right|\widehat{\mathcal{P}}\left|0\right> = P(0|u,v)\delta_{ij}\, ,
\eeq
and should satisfy the Watson's equation
\beq\label{formfactor}
\def\svgwidth{15.3cm}
\begingroup%
  \makeatletter%
  \providecommand\color[2][]{%
    \errmessage{(Inkscape) Color is used for the text in Inkscape, but the package 'color.sty' is not loaded}%
    \renewcommand\color[2][]{}%
  }%
  \providecommand\transparent[1]{%
    \errmessage{(Inkscape) Transparency is used (non-zero) for the text in Inkscape, but the package 'transparent.sty' is not loaded}%
    \renewcommand\transparent[1]{}%
  }%
  \providecommand\rotatebox[2]{#2}%
  \ifx\svgwidth\undefined%
    \setlength{\unitlength}{1714.8328125bp}%
    \ifx\svgscale\undefined%
      \relax%
    \else%
      \setlength{\unitlength}{\unitlength * \real{\svgscale}}%
    \fi%
  \else%
    \setlength{\unitlength}{\svgwidth}%
  \fi%
  \global\let\svgwidth\undefined%
  \global\let\svgscale\undefined%
  \makeatother%
  \begin{picture}(1,0.15378129)%
    \put(0,0){\includegraphics[width=\unitlength]{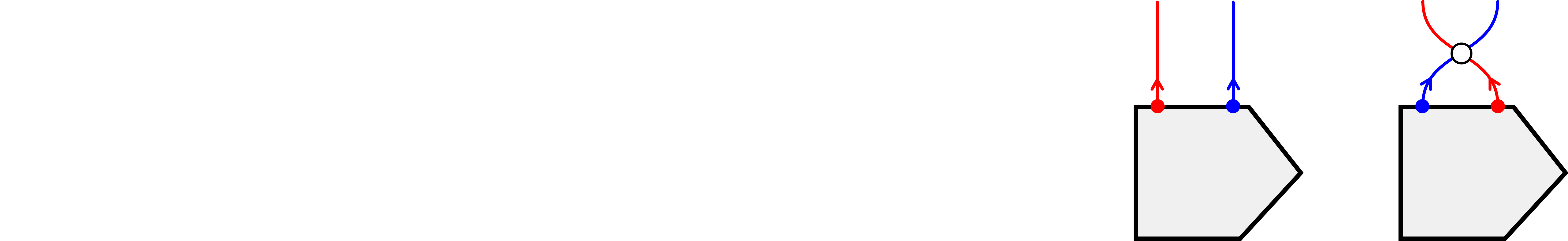}}%
    \put(0.84215594,0.09859463){\color[rgb]{0,0,0}\makebox(0,0)[lb]{\smash{$\displaystyle =$}}}%
    \put(-0.00042503,0.09859463){\color[rgb]{0,0,0}\makebox(0,0)[lb]{\smash{$\displaystyle \left<\Phi_{i}(u)\Phi_{j}(v)\right|\widehat{\mathcal{P}}\left|0\right> = S_{ij}^{kl}(v, u)\left<\Phi_{l}(v)\Phi_{k}(u)\right|\widehat{\mathcal{P}}\left|0\right>$}}}%
    \put(0.73059794,0.06529516){\color[rgb]{0,0,0}\makebox(0,0)[lb]{\smash{$u$}}}%
    \put(0.77911579,0.06529516){\color[rgb]{0,0,0}\makebox(0,0)[lb]{\smash{$v$}}}%
    \put(-0.00042503,0.02395178){\color[rgb]{0,0,0}\makebox(0,0)[lb]{\smash{$\displaystyle \Rightarrow\qquad\ \,P(0|u,v) = S_{\textrm{singlet}}(v, u)P(0|v,u)$}}}%
    \put(0.89947738,0.06529516){\color[rgb]{0,0,0}\makebox(0,0)[lb]{\smash{$v$}}}%
    \put(0.94799526,0.06529516){\color[rgb]{0,0,0}\makebox(0,0)[lb]{\smash{$u$}}}%
  \end{picture}%
\endgroup%

\eeq
with $S_{\textrm{singlet}}(u, v)$ defined previously in~(\ref{singlet}). The Watson's relation is easy to understand: reordering two adjacent scalars within a state (here at the top of the pentagon) is equivalent to acting with the S-matrix (see picture in~(\ref{formfactor})).  The point we would like to stress is that the Watson's equation is not independent of our axioms~(\ref{f-axioms}). In fact it is a consequence of them. To see it, we observe that we can access to $P(0|u,v)$ by performing on the bottom excitation either two mirror rotations or three inverse mirror rotations,
\beq
\def\svgwidth{15.3cm}
\begingroup%
  \makeatletter%
  \providecommand\color[2][]{%
    \errmessage{(Inkscape) Color is used for the text in Inkscape, but the package 'color.sty' is not loaded}%
    \renewcommand\color[2][]{}%
  }%
  \providecommand\transparent[1]{%
    \errmessage{(Inkscape) Transparency is used (non-zero) for the text in Inkscape, but the package 'transparent.sty' is not loaded}%
    \renewcommand\transparent[1]{}%
  }%
  \providecommand\rotatebox[2]{#2}%
  \ifx\svgwidth\undefined%
    \setlength{\unitlength}{1716.55488281bp}%
    \ifx\svgscale\undefined%
      \relax%
    \else%
      \setlength{\unitlength}{\unitlength * \real{\svgscale}}%
    \fi%
  \else%
    \setlength{\unitlength}{\svgwidth}%
  \fi%
  \global\let\svgwidth\undefined%
  \global\let\svgscale\undefined%
  \makeatother%
  \begin{picture}(1,0.12142322)%
    \put(0,0){\includegraphics[width=\unitlength]{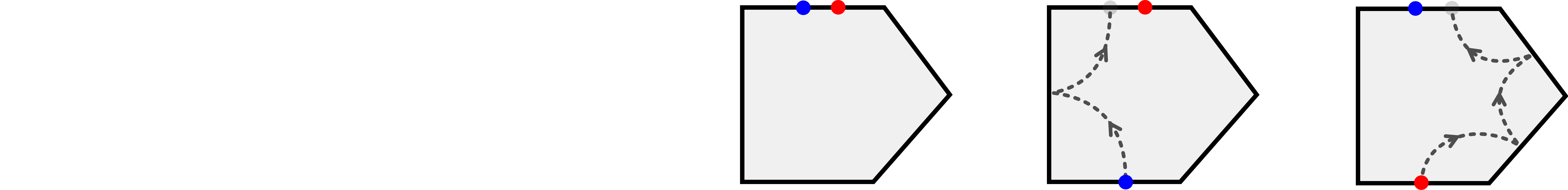}}%
    \put(0.62692822,0.05556034){\color[rgb]{0,0,0}\makebox(0,0)[lb]{\smash{$\displaystyle =$}}}%
    \put(-0.00037477,0.05556034){\color[rgb]{0,0,0}\makebox(0,0)[lb]{\smash{$\displaystyle P(0|u, v) = P(u^{2\gamma}|v)= P(v^{-3\gamma}|u)$}}}%
    \put(0.72375287,0.01315721){\color[rgb]{0,0,0}\makebox(0,0)[lb]{\smash{$u^{2\gamma}$}}}%
    \put(0.72188867,0.09052147){\color[rgb]{0,0,0}\makebox(0,0)[lb]{\smash{$v$}}}%
    \put(0.52801197,0.09052147){\color[rgb]{0,0,0}\makebox(0,0)[lb]{\smash{$v$}}}%
    \put(0.50377738,0.09052147){\color[rgb]{0,0,0}\makebox(0,0)[lb]{\smash{$u$}}}%
    \put(0.87081291,0.02670307){\color[rgb]{0,0,0}\makebox(0,0)[lb]{\smash{$v^{-3\gamma}$}}}%
    \put(0.82266912,0.05556034){\color[rgb]{0,0,0}\makebox(0,0)[lb]{\smash{$\displaystyle =$}}}%
    \put(0.89619128,0.09052147){\color[rgb]{0,0,0}\makebox(0,0)[lb]{\smash{$u$}}}%
  \end{picture}%
\endgroup%

\eeq
It allows us to write
\beq
\frac{P(0|u, v)}{P(0|v, u)} = \frac{P(u^{2\gamma}|v)}{P(u^{-3\gamma}|v)} = \frac{S(u^{2\gamma}, v)S(u^{\gamma}, v)}{S(v, u^{-2\gamma})S(v, u^{-\gamma})S(v, u)}\, ,
\eeq
where in the last equality we made use of our two axioms~(\ref{f-axioms}) to reexpress the ratio on the left-hand side in terms of S-matrices only. After using the unitarity property of $S(u, v)$ and comparing with~(\ref{formfactor}), we conclude that our axioms will be consistent with the Watson's equation if and and only if the pentagon identity
\beq\label{Watson-scalar}
S(u^{2\gamma}, v)S(u^{\gamma}, v)S(u, v)S(u^{-\gamma}, v)S(u^{-2\gamma}, v) = S_{\textrm{singlet}}(v, u)
\eeq
is observed. A simple algebra based on the property B of the scalar S-matrix reveals that this identity is indeed correct! 

It is interesting to notice that the consistency between the fundamental relation and the Watson's equations strongly relies on the fact that we are dealing with an $O(6)$ invariant S-matrix. Curiously, the same algebra would not work out correctly if we were using an S-matrix with  $O(N)$ symmetry with $N\neq 6$. 
\end{description}
This series of checks illustrate the nice interplay between the axioms for the pentagon transition and the general properties of the flux tube S-matrix.

The fundamental axioms~(\ref{f-axioms}) allow us to make an educated guess for what the pentagon transition should be. Looking at Axiom I, for instance, we realize that the pentagon transition $P(u|v)$ is essentially the square-root of the S-matrix $S(u, v)$. More precisely and with help of property A, we get
\beq
\frac{P(u|v)^2}{P(v|u)^2} = S(u, v)^2 = \frac{S(u, v)}{S(v, u)}\, ,
\eeq
that is solved by
\beq\label{PsS}
P(u|v)^2 = z(u, v)S(u, v)\, ,
\eeq
with $z(u, v) = z(v, u)$ a symmetric function.
Plugging~(\ref{PsS}) into Axiom II and using symmetry of $z(u, v) $, we arrive at
\beq
\frac{z(u, v)}{z(u^{-\gamma}, v)} =  \frac{S(u^{-\gamma}, v)}{S(v, u)} = \frac{S(v, u^{\gamma})}{S(v, u)}\frac{(u-v-i)}{(u-v+i)}\times \frac{u-v}{u-v}\, ,
\eeq
where in the last step we used the crossing property B of the S-matrix. Recalling that $u^{\gamma} = u+i$ for a scalar, we see that the above relation is equivalently written as
\beq
z(u, v) = \frac{w(u, v)S(v, u^{\gamma})}{(u-v)(u-v+i)} = \frac{w(u, v)}{(u-v)(u-v+i)S(u^{\gamma}, v)}\, ,
\eeq
where $w(u, v)$ is, by construction, invariant under mirror rotation of its rapidity, $w(u^{\gamma}, v) = w(u, v)$. It is also symmetric under exchange of the two rapidities $w(u, v) = w(v, u)$.
\footnote{This follows both form the property of $z(u, v)$ and from unitarity and crossing, which can be combined into
\beq
(u-v)(u-v+i)S(u^{\gamma}, v) = (v-u)(v-u+i)S(v^{\gamma}, u)\, . 
\eeq}
Clearly the simplest possible solution for $w$ is that it is a constant. Our conjecture is that it is exactly equal to $1/g^2$ (within the normalization assumed in this paper).

Combining everything together our proposal for the transition of a single scalar over the pentagon is
\beq\label{conj-scalar}
P(u|v)^2 = \frac{S(u, v)}{g^2(u-v)(u-v+i)S(u^{\gamma}, v)}\, . 
\eeq
We see that it is expressed directly in terms of the scalar S-matrix which can be constructed exactly using integrability. The only ambiguity that is left over is the choice of the branch when taking the square-root of~(\ref{conj-scalar}). This one is easily fixed by comparison with data, at tree level already.

To complete our construction we also need to get the expression for the measure. According to our previous discussion, see~(\ref{measureEq}), the measure $\mu(u)^2$ is readable from the double pole of $P(u|v)^2$ at $u=v$. Looking at~(\ref{conj-scalar}) and using that $S(u, u) = -1$, it should then be true that
\beq\label{mu2}
\mu(u)^2 = g^2\lim_{v\rightarrow u}\,  \frac{i}{u-v}S(u^{\gamma}, v)\,.
\eeq
At first sight it is not obvious that this limit exists at all. Remarkably enough, it appears that $S(u^{\gamma}, v)$ has a simple zero for $u\sim v$ and this regardless of the value of the coupling constant. The expression~(\ref{mu2}) thus determines the measure unambiguously, up to an overall sign which is easily fixed from data. 

Equations~(\ref{conj-scalar}) and~(\ref{mu2}) conclude our discussion of the all-loop scalar pentagon transition.

\subsection{Weak Coupling Expansion}

We shall now confront our OPE predictions with explicit data extracted from perturbative scattering amplitudes at weak coupling. It is quite straightforward to expand all our building blocks (i.e., the transition, measure, energy and momentum) in perturbation theory. This is explained in appendix \ref{ansatz-App} and implemented in a \verb"mathematica" notebook attached to this submission. For instance, one easily verifies that the scalar pentagon transition takes the form
\beq
P(u|v) = \frac{\Gamma(iu-iv)}{g^2\Gamma(\frac{1}{2}+iu)\Gamma(\frac{1}{2}-iv)}\bigg[1+ g^2\alpha(u, v) + O(g^4)\bigg]\,, \la{Ppert}
\eeq
at weak coupling. Here the function $\alpha(u, v)$ is given by
\beq
\begin{aligned}
\alpha(u, v)\, \, \, \,  = &\, \, \, \, \, \, \,  \psi^{(1)}{(\ft{1}{2}+iu)} + \psi^{(1)}{(\ft{1}{2}-iv)}-\frac{\pi^2}{3}\\
&+H_{iu-\ft{1}{2}}H_{iv-\ft{1}{2}}+H_{-iu-\ft{1}{2}}H_{-iv-\ft{1}{2}}+H_{-iu-\ft{1}{2}}H_{iv-\ft{1}{2}}-H_{iu-\ft{1}{2}}H_{-iv-\ft{1}{2}}\, ,  \\
\end{aligned} \la{alpha}
\eeq
where $H_z = \psi(z+1)-\psi(1)$, $\psi(z) = \partial_z \log{\Gamma(z)},$ and $\psi^{(1)}(z) = \partial_z \psi(z)$. 

We immediately notice that the leading order expression in (\ref{Ppert}) agrees perfectly with the pentagon transition computed at Born level from Feynman diagrams in (\ref{mu-P-BL}). This was already matched successfully against a particular component of the super Wilson loop in the introduction. 

Higher loops are similar to (\ref{alpha}) but contain higher derivatives of the polygamma function. In fact, one neat feature of our finite coupling conjecture is the presence of a sort of exponentiation: The \textit{logarithm} of the square bracket in (\ref{Ppert}) is much nicer than the bracket itself. Namely, it is at most quadratic in $\psi^{(n)}$ -- derivatives of the polygamma function -- with arguments $1/2\pm iu$ or $1/2\pm iv$ only! This follows from the expansion of the functions $f_1,f_2,f_3,f_4$ that appear in the construction of the finite-coupling expression and that only generate quadratic monomials of this sort in perturbation theory (see appendix \ref{ansatz-App} and the discussion surrounding (\ref{gammaInt}) for more details). 

The measure is related to the pole of the pentagon transition at $u=v$, see (\ref{measureEq}) and (\ref{mu2}). This yields
\beq\label{measure-NLO}
\mu(u) = \frac{\pi g^2}{\cosh{(\pi u)}}\bigg[1+ g^2\bigg(\frac{4\pi^2}{3}-\frac{2\pi^2}{\cosh^2{(\pi u)}}-2H_{iu-\ft{1}{2}}H_{-iu-\ft{1}{2}}\bigg) + O(g^4)\bigg]\, .
\eeq
It follows from what was said before that the logarithm of this square bracket is also a nicer quantity at any loop order. Namely, it is at most quadratic in $\psi^{(n)}(1/2\pm iu)$. 

Finally, we recall that for a scalar the momentum and energy are given by
\beqa
p(u)& = &2u -2\pi g^2\textrm{tanh}(\pi u) + O(g^4)\, ,\\
E(u)=1+\gamma(u)& =& 1 + 2g^2(\psi(\ft{1}{2}+iu)+\psi(\ft{1}{2}-iu)-2\psi(1))+O(g^4)\, .
\eeqa
It is worth mentioning that the way the energy and the momentum are computed is not disconnected from the way we compute the S-matrices or the pentagon transitions. On the contrary, when computing the S-matrices we solve a set of integral equations for an infinite series of auxiliary variables, see appendix \ref{ansatz-App}. These auxiliary variables are the higher conserved charges of the flux tube, which guarantees its integrability. The energy and momentum are just two special representatives of these charges, see (\ref{EnergyMomenta}). 

We now have all the ingredients to predict the leading OPE behavior for the scalar component of the NMHV hexagon and the NMHV heptagon to any loop order. In the following subsections we will check them against  perturbation theory to provide support for our conjectures. 

\subsection{The NMHV Hexagon and the Scalar Measure} \la{hexScalarSec}
We start by matching our predictions with perturbative data for the NMHV hexagon, see~(\ref{hexNMHV0}). 
Unfortunately, the perturbative data is conventionaly given in position space ($\sigma$) rather than in rapidity space ($u$). Hence, in order to compare it with our predictions, we have to perform the Fourier transform from $u$ to $\sigma$ in (\ref{hexNMHV0}) or reciprocally inverse-Fourier transform the perturbative data. 
At the first few loops it is rather straightforward to do either and perform the relevant integrals explicitly, since we deal with relatively simple functions. At higher loops this becomes quite tedious and one has to rely on alternative strategies. We will present one of them in the following.

We begin with the Born level where we can easily Fourier transform our expression~(\ref{measure-NLO}) for the measure,
\beqa
\nn \mathcal{W}^{(6134)}_\text{tree} &=& e^{-\tau} \frac{1}{g^2}\int\limits\frac{du}{2\pi}\mu(u)e^{ip(u)\sigma-\gamma(u) \tau} +\dots \\
\nn &=&e^{-\tau}  \int\limits\frac{du}{2\pi} \frac{\pi}{\cosh(\pi u)} e^{2iu \sigma} +\dots \\
\la{hexagonTree} &=& e^{-\tau} \frac{1}{2\cosh(\sigma)}  +\dots 
\eeqa
It agrees perfectly with the direct computation of the Born level amplitude:
\beq\la{hexagonTree2}
\def\svgwidth{12cm}
\begingroup%
  \makeatletter%
  \providecommand\color[2][]{%
    \errmessage{(Inkscape) Color is used for the text in Inkscape, but the package 'color.sty' is not loaded}%
    \renewcommand\color[2][]{}%
  }%
  \providecommand\transparent[1]{%
    \errmessage{(Inkscape) Transparency is used (non-zero) for the text in Inkscape, but the package 'transparent.sty' is not loaded}%
    \renewcommand\transparent[1]{}%
  }%
  \providecommand\rotatebox[2]{#2}%
  \ifx\svgwidth\undefined%
    \setlength{\unitlength}{894.39716797bp}%
    \ifx\svgscale\undefined%
      \relax%
    \else%
      \setlength{\unitlength}{\unitlength * \real{\svgscale}}%
    \fi%
  \else%
    \setlength{\unitlength}{\svgwidth}%
  \fi%
  \global\let\svgwidth\undefined%
  \global\let\svgscale\undefined%
  \makeatother%
  \begin{picture}(1,0.25581476)%
    \put(0,0){\includegraphics[width=\unitlength]{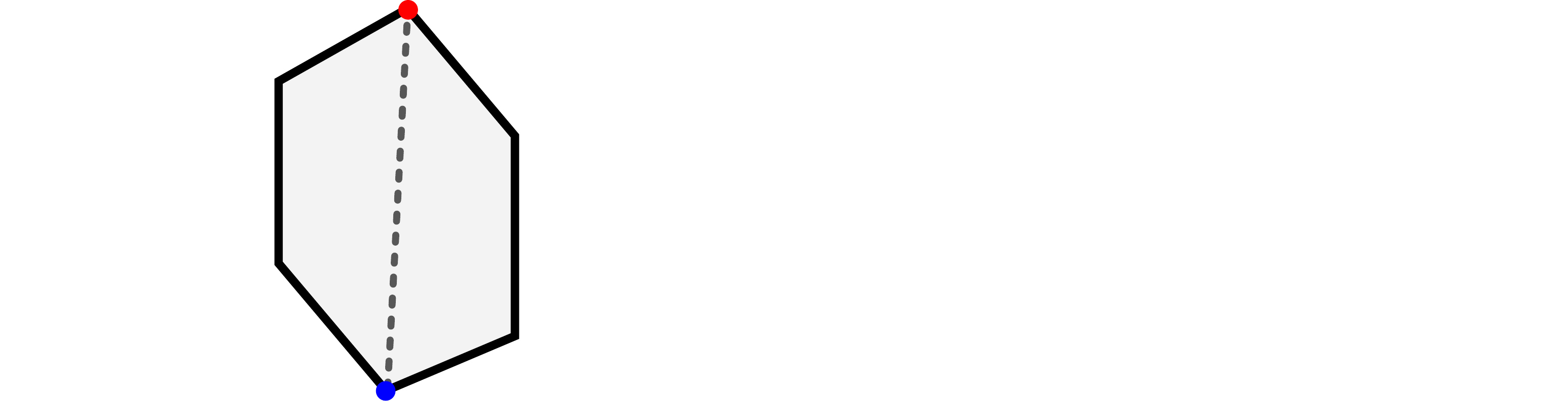}}%
    \put(0.27702277,0.0404899){\color[rgb]{0,0,0}\makebox(0,0)[lb]{\smash{$\,_1$}}}%
    \put(0.3017423,0.11147729){\color[rgb]{0,0,0}\makebox(0,0)[lb]{\smash{$\,_2$}}}%
    \put(0.2780798,0.19207043){\color[rgb]{0,0,0}\makebox(0,0)[lb]{\smash{$\,_3$}}}%
    \put(0.20812281,0.20978094){\color[rgb]{0,0,0}\makebox(0,0)[lb]{\smash{$\,_4$}}}%
    \put(0.17959495,0.13884394){\color[rgb]{0,0,0}\makebox(0,0)[lb]{\smash{$\,_5$}}}%
    \put(0.2066668,0.06115522){\color[rgb]{0,0,0}\makebox(0,0)[lb]{\smash{$\,_6$}}}%
    \put(-0.001162,0.10855054){\color[rgb]{0,0,0}\makebox(0,0)[lb]{\smash{$\displaystyle \mathcal{W}^{(6134)}_\text{tree} = $}}}%
    \put(0.35662094,0.10855054){\color[rgb]{0,0,0}\makebox(0,0)[lb]{\smash{$\displaystyle = \frac{1}{\<6,1,3,4\>} =e^{-\tau} \frac{1}{2\cosh(\sigma)}  +\dots$}}}%
  \end{picture}%
\endgroup%

\eeq
where we used the hexagon twistors written down in appendix \ref{geomeryappendix}, see (\ref{hexagontwistors}). In section~\ref{Intropart2} and~\ref{beforeSec} we performed a similar check for the heptagon. In that case we inverse Fourier transformed the (leading collinear limit of the) heptagon amplitude and matched it against the bootstrap predictions. Of course, the heptagon check includes the hexagon one, since the two are related by a collinear limit. It is nice to verify it explicitly nevertheless. The match of (\ref{hexagonTree}) with (\ref{hexagonTree2}) is also useful to align the sign ambiguity discussed below (\ref{mu2}) with our convention for the normalization of the twistors (\ref{hexagontwistors}).

From the bootstrap viewpoint higher loops are not conceptually more involved. The only difficulty, as alluded to before, is technical and related to the Fourier transformation to position space.
The simplest way of guessing these Fourier integrals is to {assume} -- based on empirical evidence -- that at any loop order {the hexagon contribution} (\ref{hexNMHV0}) can be parametrized in terms of so called Harmonic Polylogs (HPLs). Precisely we write 
\beq
\mathcal{W}^{(6134)}= \frac{e^{-\tau}}{2\cosh(\sigma)} \sum_{l=0}^\infty g^{2l} \sum_{n=0}^{l} \tau^n F^{(l)}_n(\sigma) \,+\,\cO(e^{-2\tau})\, , \la{expanded}
\eeq
where $F_{n}^{(l)}$ are linear combinations of HPLs in $x=e^{2\sigma}$ of maximum degree $2l-n$ and with slots $0$ or $1$ only,\footnote{There is an even more convenient ansatz: one applies \texttt{HPLLogExtract@ansatz/.HPL[\{0\},-x]->HPL[\{0\},x]} to the \texttt{ansatz}  (\ref{ansatz}). This new ansatz is a bit more convenient: It is more manifestly real for $x\ll 1$, provided the coefficients $\{a, a_i, \ldots\}$ are all real.} 
\beq
F^{(l)}_n(\sigma) =a+\sum_{i=0}^1 \,a_{i} H_{i}(-x)+\sum_{i,j=0}^1 a_{i,j} H_{i,j}(-x)\,+\, \dots\,+\!\!\!\!\!\!\sum_{i_1,\dots,i_{2l-n}=0}^1\!\!\!\!\!\!\!\! a_{i_1,\dots,i_{2l-n}} H_{i_1,\dots,i_{2l-n}}(-x)\, . \la{ansatz}
\eeq
Here $a$, $a_{0}$, $a_{1}$, $a_{0,0}$, $a_{0,1}$, etc., are just constants, which depend implicitly on $l$ and $n$. For example, at tree level $l=n=0$ and all that we have is a constant $a$ which in our normalization is just $1$. Note that the harmonic polylogs are generalizations of the usual polylogs. In fact, up to degree three, they can all be re-expressed as familiar polylogs ${\rm Li}_{1,2,3}$. More generally, they can all be conveniently manipulated using the \verb"mathematica" package \verb"HPL" \cite{HPLpackage}, where, for illustration, $H_{1,0}(x)$ is entered as \texttt{HPL[\{1,0\},x]}. For further details on this interesting class of functions please see~\cite{HPLs,HPLpackage}. 

Equipped with the ansatz (\ref{ansatz}) the next step is to fix the constants $a,a_0,\dots$. One way of doing is by comparing the Taylor expansion in $e^{\sigma}$ of the ansatz (\ref{ansatz}) with the one of the integral in (\ref{hexNMHV0}) at the given loop order. What is nice about it is that both expansions are very simple to perform. To expand (\ref{ansatz}) we can use the \verb"HPL" package, for instance, while to expand (\ref{hexNMHV0}) we just need to extract residues of the integrand in the lower half of the $u$ plane. In the latter case, each residue is in correspondence with a power of $e^{\sigma}$. 

Once all the constants have been fixed using the first few terms in the expansion, one can test a few more residues as self-consistency checks. Typically, it is relatively easy to perform the analysis for the first few hundred powers of $e^{\sigma}$. Then, as an extra verification, one can compare (\ref{ansatz}) and~(\ref{hexNMHV0}) numerically with high precision.

This algorithm works perfectly up to the maximal order we have checked, which is three loops. The functions $F_{n}^{(l)}(\sigma)$  that we obtained are summarized in Appendix \ref{FnlAppendix} and given in a companying notebook. The simplest ones, at one loop, are given by
\beq
F_{0}^{(1)} =  -2\log(1+x)\log(1+1/x) \,, \qquad F^{(1)}_1= 4 \log(\sqrt{x}+1/\sqrt{x}) \,. \la{1loopPrediction}
\eeq  
At two loops the most transcendental contribution is 
\beqa
F_{0}^{(2)} &=& 4 H_1 \bar{H}_0 H_{0,1}+\frac{1}{2} \bar{H}_0^2 H_{0,1}-4 \bar{H}_0 H_{0,1,1}+2 \zeta (3) \bar{H}_0+4
   H_1^3 \bar{H}_0+H_1^2 \bar{H}_0^2+\frac{4}{3} \pi ^2 H_1 \bar{H}_0\nn \\
   &+&\frac{1}{12} \pi ^2
   \bar{H}_0^2+\frac{1}{6} \pi ^2 H_{0,1}+4 H_{0,0,0,1}+2 H_{0,1,0,1}-4\zeta (3) H_1 +2 H_1^4+\pi ^2
   H_1^2+\frac{\pi ^4}{36} \nn
\eeqa
where $H$ is short for $H(-x)$ and $\bar H$ for $H(+x)$. For the other functions see appendix \ref{FnlAppendix}.

The next and most important step is to compare all the functions $F_{n}^{(l)}(\sigma)$ with Scattering Amplitudes. 
Up to two loops there are the 6 functions:
\beqa
&&F_{0}^{(0)}\,, \nn \\
&&F_{0}^{(1)}, F_{1}^{(1)}\,,\nn \\ 
&&F_{0}^{(2)}, F_{1}^{(2)},F_{2}^{(2)} \,,
\eeqa
to match against perturbative results for the ratio function $\mathcal{R}^{(6134)}$, which is the same as our $\mathcal{W}^{(6134)}$ to leading order in the OPE (i.e., they differ only for higher powers of $e^{-\tau}$). 

Note that the mere fact that the perturbative results organize as in (\ref{expanded}) is in itself a prediction coming from the OPE approach which is far from obvious from the amplitude side. Checking that this structure comes right is the zero-th order step before matching the precise form of the functions $F_n^{(l)}$. 

The first function $F_0^{(0)}=1$, at Born level, was already matched before. We can extract $F_{n}^{(1)}$ by expanding the one-loop Hexagon ratio function to leading order at large~$\tau$. The six gluons NMHV amplitude at one loop was computed in \cite{Bern:1994cg} and written as a ratio function in terms of conformal cross-ratios in \cite{Dualsuperconforma}. A very convenient source for \textit{any} one loop N$^k$MHV n-point Amplitude is the package \cite{Bourjaily:2013mma} by Bourjaily, Caron-Huot and Trnka. For example, to extract the Hexagon one-loop ratio function using this package we simply define the \verb"Zs" to be given by the hexagon twistors (\ref{hexagontwistors}), run the command \verb"evaluate@superComponent[{1},{},{2},{3},{},{4}]@ratioIntegral[6,1]" and finally  \verb"Series" expand the result. In this way we read off both $F_{0}^{(1)}$ and $F_{1}^{(1)}$ and find that they agree precisely with our predictions (\ref{1loopPrediction}). At two loops we can use the results of Dixon, Drummond and Henn in \cite{Dixon:2011nj}. In their notations, the component we are considering is given by 
\beq
\mathcal{R}^{(6134)}={g^2\over\<6,1,3,4\>}V_3(u_1,u_2,u_3)\,\qquad\text{with}\quad u_i=\small
 {\< i,i+1,i+2,i+3\>  \< i,i-1,i-2,i-3\>\over \<i,i+1,i-2,i-3\> \< i,i-1,i+2,i+3\> } \,,
\eeq
see (\ref{theus}). Here also, after OPE expanding their result, we find a perfect match with the bootstrap predictions.
\footnote{We thank Lance Dixon, Matt von Hippel and Jeffrey Pennington for sharing with us a notebook with the near collinear expansion of the NMHV functions in \cite{Dixon:2011nj} up to $\cO(e^{-2\tau})$.} 

At three loops the ratio function is not known. Formulae (\ref{HexPredictions}) in appendix \ref{FnlAppendix} are new predictions from the OPE approach. It would be very interesting to use them to constrain the three-loop result or to check them against an independent computation.

\subsection{The NMHV Heptagon and the Scalar Transitions} \la{hepScalarSec}

We will now check the bootstrap predictions for the pentagon transition. As explained in section \ref{summarySection}, the pentagon transition governs the term that decays as $e^{-\tau_1-\tau_2}$ in the heptagon amplitude.
It can read from the component $\mathcal{R}^{(7145)}$ for which a scalar is inserted at the bottom cusp and absorbed at a top cusp, see (\ref{TreeLevelScalar}). At tree level we already observed a perfect match between our expression and the one coming from the Amplitude side, see sections \ref{Intropart2} and \ref{beforeSec}. We now move to loops. 

At any loop order, the OPE decomposition states that 
\beq
\mathcal{R}^{(7145)} \simeq e^{-\tau_1-\tau_2} \[\,\int\limits_{\mathbb{R}+i0}\int\limits_{\mathbb{R}+i0}\frac{du dv}{g^2(2\pi)^2}\mu(u)\mu(v)P(-u|v)e^{ip(u)\sigma_1+ip(v)\sigma_2} + \text{higher powers of $\tau_1$, $\tau_2$}\]\, .
\la{whatWeWant}
\eeq
At $l$ loops order we have at most $l$ total powers of the $\tau$'s. The terms with positive powers of $\tau_i$ are obtained by dressing previous loop orders by the anomalous energy of the excitations. Hence they probe the structure of the OPE but they do not probe the $l$-th loop correction to the transition (which is what we are most interested in, in this paper). At each loop order, the quantum correction to the transition appears in the term with no powers of $\tau_1,\tau_2$. Furthermore, the more powers of $\tau_i$ we consider the simpler (less transcendental) the functions get (see for example (\ref{HexPredictions}) for the hexagon).
Hence the most non-trivial checks are, by far, those involving no powers of $\tau_i$. We are now going to check those terms up to two loops.

We start by writing our predictions~(\ref{Ppert}) in positions space,
\beq
\int\limits_{\mathbb{R}+i0}\int\limits_{\mathbb{R}+i0}\frac{du dv}{(2\pi)^2}{1\over g^2}\mu(u)\mu(v)P(-u|v)e^{ip(u)\sigma_1+ip(v)\sigma_2}  = \frac{1+ g^2\alpha(\sigma_1, \sigma_2)+g^4\beta(\sigma_1, \sigma_2)+\dots }{e^{\sigma_1-\sigma_2}+e^{\sigma_2-\sigma_1}+e^{\sigma_1+\sigma_2}} \,. \la{todo}
\eeq
The challenge is to compute the functions $\alpha,\beta$ from~(\ref{Ppert}) and match them against perturbative data. The one loop correction is easy enough to be done without any fancy techniques. It reads
\beq
\begin{aligned}
&\alpha = \log{(1+e^{2\sigma_1})}\log{(1+e^{2\sigma_2})}-\log{\left[\frac{e^{2\sigma_1}(1+e^{2\sigma_2})}{e^{2\sigma_1}+e^{2\sigma_2}+e^{2\sigma_1+2\sigma_2}}\right]}\log{\left[\frac{e^{2\sigma_2}(1+e^{2\sigma_1})}{e^{2\sigma_1}+e^{2\sigma_2}+e^{2\sigma_1+2\sigma_2}}\right]} \\
&+\left[\text{Li}_2\left(\frac{e^{2\sigma_1}}{e^{2\sigma_1}+e^{2\sigma_2}+e^{2\sigma_1+2\sigma_2}}\right)+ \text{Li}_2\left(\frac{e^{2\sigma_1}}{1+e^{2\sigma_1}}\right) + \sigma_1\leftrightarrow \sigma_2\right]-\frac{\pi^2}{6}\, .
\end{aligned}
\eeq
We can compare it with data as before: we use the \verb"Loop_Amplitudes.m" package \cite{Bourjaily:2013mma} to extract the corresponding heptagon NMHV one-loop component and expand it at large $\tau_i$ using the heptagon twistors (\ref{Heptagon2}). We find a perfect match. 

The computation of the two-loop prediction $\beta$ is considerably more difficult. After some work\footnote{We looked at the two loop symbol of \cite{Qbar} to get some inspiration about what kind of ansatz to use. In particular, it was very useful to use a projector \cite{VerguPaper} that removes all terms that can be written as classical polylogs therefore isolating the hardest parts of the result. We thank Cristian Vergu for his explanations about how this projector works.} one gets
\beqa
\beta&=& -3 \,\text{Li}_{2,2}\left(-x,-\frac{y}{y(1+x)}\right)+2\,
   \text{Li}_{3,1}\left(-x,-\frac{y}{x y+x}\right)+(x\leftrightarrow y) \la{beta}\\
   &+&\verb"huge expression with classical polylogs up to weight 4, "\pi\verb"'s and "\zeta(3) \,,\nn
\eeqa
where $x=e^{2\sigma_1}$ and $y=e^{2\sigma_2}$. The full expression can be found in a notebook attached to this submission. In the first line we have the functions 
 \beq
Li_{a,b}(x,y) \equiv \sum_{n=1}^\infty \sum_{m=1}^n \frac{x^n y^m}{n^a m^b} \,,
\eeq
which can not be written in terms of classical polylogs. We have checked the validity of (\ref{beta}) by matching the expansion of the first few hundred terms in small $x$ and large $y$ with the expansion computed by picking residues in the Fourier integral in the left hand side of (\ref{todo}). As a further check we also computed the left and right hand side of (\ref{todo}) numerically.

It is worth emphasizing that the expressions obtained from the bootstrap are way simpler when written in momentum space at finite coupling, see e.g.~(\ref{conj-scalar}). It is only when we expand the integrand in small $g$ and insist on going to position space that we get enormous expressions as in (\ref{beta}). Of course this is a necessary evil to compare with perturbation theory. 

We can now confront our predictions with the two-loop result obtained from the $Q$-bar approach of \cite{Qbar}.\footnote{We thank Simon Caron-Huot and Song He for sharing with us a notebook with their heptagon result  \cite{Qbar}.}  
To do so we discard all $\pi$'s and $\zeta(3)$'s in (\ref{beta}), since unfortunately the symbol is insensitive to those, and then compute the symbol of whatever remains.\footnote{We thank Song He for teaching us about the functions $Li_{2,2}$ and $Li_{3,1}$ and their symbols.} Then we compare the result with the $\tau_i\to\infty$ limit of the symbol \cite{Qbar} itself 
\footnote{Taking a collinear limit inside a symbol is not totally straightforward. For each slot of the symbol we write $\dots \otimes f \otimes\dots=a( \dots \otimes (e^{-\tau_1}) \otimes \dots)+b  (\dots \otimes (e^{-\tau_2}) \otimes \dots)+(\dots \otimes f_{reg} \otimes\dots)$ where, for large $\tau_i$, $f=e^{-a \tau_1-b \tau_2} (f_{reg}+\text{positive powers of $e^{-\tau_i}$})$. For this NMHV example, all terms with $e^{-\tau_i}$ as slots are nothing but the shuffle product of (the symbol of powers of) $\log(e^{\tau_i})= \tau_i$ with a $\tau_i$ independent symbol. Hence they only contribute to the higher powers of $\tau_i$ in (\ref{whatWeWant}) which are not what we are after here. Hence, in practice, we can drop them all and keep $f_{reg}$ only. In this way we obtain a symbol with four slots which depends on $\sigma_1$ and $\sigma_2$ only and which can be matched with the symbol of $\beta$. For a more complete discussion on comparisons of our predictions with symbols please see sections \ref{matchHex} and \ref{matchHep}.}
and find a perfect match! An interesting question is whether the leading collinear behaviour (\ref{beta}) can be used to construct the full NMHV heptagon two-loop ratio function from its  symbol. 

\section{Gluons} \la{gluonSec}

In this section we shall motivate our ansatz for the gluonic transitions and measure presented in \cite{short}. We will then extract the same transitions from known MHV amplitudes and match them against our expressions. Finally, we will comment on a different way of inserting a gluonic excitation, which is related to the MHV procedure by a form factor.

\subsection{Bootstrapping the Pentagon Transitions} \la{IntG}

The gauge field excitations come in two types: a particle ${F}$ and its anti-particle ${\bar{F}}$. They are associated to the twist-one components of the Faraday tensor $F_{-z}$ and $F_{-\bar{z}}$, respectively,
and they carry opposite $U(1)$ charge $\pm 1$ with regard to rotations in the plane transverse to the flux tube.

A $U(1)$ symmetric S-matrix has three independent components. One of them simply accounts for the scattering phase between two ${F}$s. The remaining two components of the S-matrix describe the backward and forward scattering of an ${F}$ with an ${\bar{F}}$. All other processes can be obtained from them by charge conjugation, that is by invoking the symmetry under the exchange ${F}\leftrightarrow{\bar{F}}$. An important observation regarding the scattering of gauge field excitations in the flux-tube theory is that it is reflectionless. It means that there is no backward scattering among gluonic excitations carrying opposite $U(1)$ charge: when an ${F}$ and an ${\bar{F}}$ scatter they do it without exchanging their individual momenta. The gluonic S-matrix is then fully specified by the two phases associated to the two processes
\beq
\begin{aligned}
S(u, v) :\qquad \Blue{F}(u)\Blue{F}(v)\rightarrow \Blue{F}(v)\Blue{F}(u)\, , \\
\bar{S}(u, v) :\qquad\Red{\bar{F}}(u)\Blue{F}(v)\rightarrow\Blue{F}(v)\Red{\bar{F}}(u)\, . \\
\end{aligned}
\eeq
These two scattering phases are actually not independent. They are related by
\beq\label{SbarS-rel}
S(u, v) = \frac{u-v+i}{u-v-i}\, \bar{S}(u, v) \, .
\eeq
Accordingly, the gauge-field S-matrix is known once $S(u,v)$ is known. The leading order expression for $S(u,v)$ at weak coupling is given by~(\ref{scalarSmatrix}) with the conformal spin $s=3/2$ appropriate for a gauge field excitation. The higher-loop corrections and more generally the finite-coupling prediction from integrability are detailed in Appendix~\ref{ansatz-App}.

The origin or explanation of the relation~(\ref{SbarS-rel}) in the flux-tube theory, and more generally of the absence of backward scattering, are not clear to us. They stand as predictions that we extracted from the spin-chain description, where ${F}$ and ${\bar{F}}$ are viewed as excitations of the large spin background. In this picture, the entire $\bar{S}(u, v)$ scattering originates from the back reaction of the large spin background, while the simple rational factor in~(\ref{SbarS-rel}) accounts for an extra bare or spin-chain scattering between two ${F}$s with same $U(1)$ charge. One interesting question is whether the dual string description can shed light on these features of the gluon S-matrix.

The gluon S-matrix has further remarkable properties, which are similar to~(\ref{main-properties}) and valid at any coupling. We list them here,
\beq\label{ABC-gf}
\begin{aligned}
&\textrm{A. Unitarity:} \qquad \qquad \qquad S(u, v)S(v, u) \, \, \, \, \, = 1\, ,\\
&\textrm{B. Crossing:} \qquad \qquad \qquad \,\, S(u^{2\gamma}, v)\bar{S}(u, v) = 1\, , \\
&\textrm{C. Mirror symmetry:}\qquad \,\,\, S(u^{\gamma}, v^{\gamma})S(v, u)  = 1\, ,
\end{aligned}
\eeq
together with the three equations obtained upon exchanging $S\leftrightarrow \bar{S}$. These equations have the same physical contents as the ones for scalars. The only difference is technical and resides in the mirror path $\gamma$ which is different from the one previously encountered. For a gauge field excitation the path $\gamma$ describes a loop in rapidity space~\cite{MoreDispPaper} and at the end of day maps a rapidity $u$ to the same rapidity $u^{\gamma} = u$. The subtlety is that the final rapidity is not lying on the same sheet as the original one. To properly rotate a gauge field into the mirror kinematics we should first cross the cut stretching between $u=\pm 2g +i/2$ in the upper-half plane and then go through the one connecting $u=\pm 2g -i/2$ in the lower-half of this new plane. This has to be done in this order as the order matters. For the inverse mirror rotation $-\gamma$ one would have to go counterclokwise and first cross the cut in the lower-half plane and so on. This is all we need to know about the mirror rotation of a gauge field. We stress that when applied to functions with no cuts in the $u$ plane (like the rational factor in~(\ref{SbarS-rel}) for instance) the mirror rotation boils down to the trivial transformation $u^{\gamma} = u$.

Equipped with the previous understanding of the scattering among gauge field excitations we now turn to the discussion of their pentagon transitions. Recall that we have two gluonic pentagon transitions
\beq\label{PbarP}
P(u|v) \equiv P_\Blue{FF}(u|v)\, , \qquad \bar{P}(u|v) \equiv P_{\Blue{F}\Red{\bar{F}}}(u|v)\, , 
\eeq
associated respectively to the $U(1)$-preserving and $U(1)$-violating processes, see discussion around (\ref{PbarPold}). 
Our conjectures for these transitions were already presented in~\cite{short}. Here we would like to elaborate on them.

The first step is to relate the pentagon transitions~(\ref{PbarP}) to the gluon S-matrix. This is the content of the fundamental axiom proposed in~\cite{short},
\beq\label{fund-axiom-gf}
P(u|v) = S(u,v)P(v|u)\, , \qquad \bar{P}(u|v) = \bar{S}(u,v)\bar{P}(v|u)\, .
\eeq
On their own they are not enough to fix completely the form of the transitions~(\ref{PbarP}). The extra information is provided by the transformation law under a mirror rotation. As already explained in the scalar section we expect that the transformation $\gamma$ introduced before will map an excitation on one edge of the pentagon to an excitation on the neighbouring right edge, see (\ref{mirror}) and (\ref{f-axioms}). For a gauge field excitation ${F}(u^{-\gamma})$ there are two possible excitations on the neighbouring edge with the same dispersion relation: ${F}(u)$ and ${\bar F}(u)$, and they carry opposite $U(1)$ charge $m=\pm1$. As explained in details in appendix \ref{FmirrorApp}, a mirror rotated gauge field is a gauge field with opposite $U(1)$ charge. That is, $\gamma:\ \Blue{F}\to\Red{\bar F}$ and similarly $\gamma:\ \Red{\bar F}\to\Blue{F}$.  We conclude therefore that
\beq\label{crossing-gf}
P(u^{-\gamma}|v) = \bar{P}(v|u)\, , \qquad \bar{P}(u^{-\gamma}|v) = P(v|u)\, .
\eeq

We can test the consistency of our axioms as we did before for the scalars. Starting with the helicity violating transition, we can bring the bottom excitation to the top using two mirror rotations and get $\bar P(u^{2\gamma}|v)$. Alternatively, starting with the helicity conserving transition, we can bring the bottom excitation to the top using three (inverse) mirror rotations and end up with $P(u^{-3\gamma}|v)$. Either way we find ourselves with two ${F}$ excitations on the top but in with opposite orderings. Using the two axioms for the transitions we find that 
\beq
{\bar P(u^{2\gamma}|v)\over P(u^{-3\gamma}|v)} = S(v,u)\, , \qquad\text{and similarly}\qquad 
{P(u^{2\gamma}|v)\over\bar P(u^{-3\gamma}|v)} = \bar S(v,u)\, ,
\eeq
in perfect agreement with the Watson's equation, see discussion below (\ref{formfactor}) and \cite{short}.

We will now explain how Axioms (\ref{fund-axiom-gf}) and (\ref{crossing-gf}) lead us to the proposal made in \cite{short} for the gluonic pentagon transitions.

By going along with an argumentation similar to the one presented before scalar, one would easily convince oneself that
\beq
P(u|v)\bar{P}(u|v) = \frac{S(u, v)}{S(u^{\gamma}, v)}
\eeq
is consistent with our axioms for the product of the two gluonic pentagon transitions. Of course this solution is not unique but it constrained as much as the scalar transition was. The choice of the normalization is at the end of the day motivated by the matching with data at weak coupling. What remains to be done is to fix the ratio
\beq\label{ruv1}
r(u,v) = P(u|v)/\bar{P}(u|v)\, .
\eeq
This is where the analysis slightly differs from scalar. This ratio satisfies the simple relation $(u-v-i)r(u, v) = (u-v+i)r(v,u)$ which is easily solved by
\beq\label{ruv2}
r(u, v) = \frac{f(u, v)}{g^2(u-v)(u-v-i)}\ ,
\eeq
where the coupling $g^2$ was introduced for later convenience and where $f(u, v)$ is a symmetric function. The latter function is not arbitrary and has to fulfill the equation
\beq\label{mirrorf}
f(u^{-\gamma}, v)f(u, v) = g^4(u-v)^2(u-v+i)(u-v-i)\, .
\eeq
The main difference between this relation and the one we had to solve before is that it involves simpler functions of the rapidities. In fact, if we were tracing back the origin of the right-hand side in~(\ref{mirrorf}), we would find that is associated to the spin-chain factor that distinguishes between $S$ and $\bar{S}$ in~(\ref{SbarS-rel}). In light of this remark it seems natural to look for a spin-chain-like solution to the `crossing' equation~(\ref{mirrorf}). This equation implies for instance that the function $f$ has a nontrivial monodromy under the mirror path $\gamma$. In the spin-chain context, the simplest way to place cuts of the right type and at the right place is through the use of the Zhukowki variable
\beq
x(u) = \frac{u+\sqrt{u^2-(2g)^2}}{2}\, .
\eeq
It was found as the natural deformation of the spectral parameter of the SYM spin chain, $x(u) = u + O(g^2)$. In these terms, a simple solution to~(\ref{mirrorf}) with the minimal amount of branch-point singularities is
\beq\label{f-sol}
f(u, v) = (x^{+}-g^2/y^{-})(x^{-}-g^2/y^{+})(y^{+}-g^2/x^{+})(y^{-}-g^2/x^{-})\, ,
\eeq
where $x^{\pm} = x(u\pm \ft{i}{2})$ and $y^{\pm} = x(v\pm \ft{i}{2})$. It is easily seen to solve~(\ref{mirrorf}) taking into account that under the mirror path $x^{\pm}(u^{-\gamma}) = g^2/x^{\pm}$ and using simple identities among Zhukowki variables like
\beq
(x^{+}-y^{-})(1-g^2/x^{+}y^{-}) = u-v+i\, , 
\eeq
for instance. Combining everything together we obtain the proposal~\cite{short}
\beq
P(u, v)^2 = \frac{f(u, v)}{g^2(u-v)(u-v-i)}\frac{S(u, v)}{S(u^{\gamma}, v)}\, , \la{finalP}
\eeq
with $f(u, v)$ the symmetric function~(\ref{f-sol}). The transitions $\bar{P}(u|v)$ is then obtained by using~(\ref{ruv1}) and (\ref{ruv2}).  The measure $\mu(u)$ is extracted from the residue of the transition as for the scalar, see (\ref{measureEq}). 

We close this section by presenting explicit expressions for $\mu(u)$, $\mu(u)P(-u|v)\mu(v)$, and $\mu(u)\bar P(-u|v)\mu(v)$ at leading order in perturbation theory. These combinations are nothing but the hexagon and heptagon leading OPE integrands,  see (\ref{hexF}-\ref{hInt}). We find
\beq\la{gluonpert}
\begin{array}{ll}
\vspace{4mm}
\mu(u)&\displaystyle=-g^2 \frac{\Gamma \left(\frac{1}{2}-iu\right) \Gamma \left(\frac{1}{2}+iu\right)}{u^2+\frac{1}{4}}+\dots\\
\vspace{4mm}
\mu(u)P(-u|v)\mu(v)&\displaystyle=-g^2\frac{ \Gamma \left(\frac{3}{2}+iu\right) \Gamma \left(\frac{3}{2}+iv\right) \Gamma (-iu-iv)}{\left(u^2+{1\over4}\right) \left(v^2+{1\over4}\right)}+\dots\\
\mu(u)\bar P(-u|v)\mu(v)&\displaystyle=\,\,\,g^4\frac{\Gamma \left(\frac{3}{2}+iu\right) \Gamma \left(\frac{3}{2}+iv\right) \Gamma (2-iu-iv)}{\left(u^2+\frac{1}{4}\right)^2
   \left(v^2+\frac{1}{4}\right)^2}+\dots
\end{array}
\eeq
See appendix \ref{ansatz-App} for the full finite-coupling formulae. In the next section we shall match these expressions -- and their higher loops generalizations  -- against available perturbative data.

\subsection{Matching with Data: $\mathcal{W}$ versus Remainder Function}
 \la{PredictionsSec}

To the first orders at weak coupling, the expansion of the gluonic transitions, measure and dispersion relation were presented in \cite{short}, see also (\ref{gluonpert}). It is straightforward to generate further terms in these expansions for the higher orders in perturbation theory. The procedure is explained in appendix \ref{ansatz-App} and automatized in the attached \verb"mathematica" notebook. With this in hand, we can proceed with the Fourier transform and obtain arbitrarily many predictions for the leading collinear behavior of Wilson loops in perturbation theory. 

Prior to compare bootstrap predictions and known amplitudes, we recall that the OPE computes the ratio ${\cW}$ defined in~(\ref{wdef}) and (\ref{calWeq}). On the other hand, the relevant data for bosonic Wilson loops (dual to MHV amplitudes) is usually expressed in terms of the so-called remainder function~\cite{Remainder}. It is quite simple to establish the dictionary between these two finite and conformally invariant quantities. The remainder function $\log \cR$ is defined as the logarithm of the ratio between the amplitude and the BDS ansatz \cite{BDS,Remainder}
\beq\la{remainder}
W=\cR\times W^\text{BDS}\, ,
\eeq
where $W^\text{BDS}$ is {given by} the Wilson loop expectation value computed in a $U(1)$ theory with coupling $4g^2\rightarrow \Gamma_\text{cusp}(g)$.
\footnote{Strictly speaking, this equality between $W^\text{BDS}$ and the v.e.v. of the Wilson loop in an abelian theory holds only up to scheme dependent quantities, which drop out when considering our ratios.}
At one loop $W^\text{BDS}$ is equal to $W$ and therefore the remainder functions starts at two loops only, $\cR=1+\cO(g^4)$. To convert the remainder function into our ${\cW}$, we plug  (\ref{remainder}) into the finite ratio of polygons (\ref{calWeq}) and obtain that
\beq\la{RtoW}
\cW=\cR\times\cW^\text{BDS}\, ,
\eeq
where $\cW^\text{BDS}$ is given by the same ratio of polygons as in (\ref{calWeq}), but computed in the $U(1)$ theory. For any given number of edges, the quantity $\mathcal{\cW^\text{BDS}}$ can be  explicitly  written in terms of the conformal cross-ratios and the cusp anomalous dimension. For example, for an hexagon we find (\ref{hexBDS}).
(Please note that in arriving at (\ref{RtoW}) we used that  the square and the pentagon Wilson loops are correctly captured by the BDS ansatz.)

The bootstrap predictions for the measure and the pentagon transitions can now be matched against the flattening of the data for $\mathcal{W}_\text{hex}$ and $\mathcal{W}_\text{hep}$, as explained in section \ref{summarySection} and worked out in greater detail in what follows.

\subsection{The MHV Hexagon and the Gluonic Measure} \la{matchHex}

We begin the discussion with the hexagon. In perturbation theory, its leading behaviour at large $\tau$ is given in (\ref{hexExp}), 
\beq
\mathcal{W}_\text{hex}=1+2 \cos(\phi)e^{-\tau}\Big[ f(\sigma) + \text{positive powers of $\tau$} \Big] + \dots\, , \la{repOfhexExp}  
\eeq
where 
\beq
f(\sigma)= \int \frac{du}{2\pi}  \,\mu(u)\, e^{i p(u)\sigma}=g^2 f_1(\sigma)+ g^4 f_2(\sigma) +g^6 f_3(\sigma)+g^8 f_4(\sigma)+\dots  \la{fInt}
\eeq
Clearly, to test the square measure it suffices to analyze the contribution $f(\sigma)$ which comes without powers of $\tau$, see also section \ref{summarySection}. 
We computed the latter function up to $\mathcal{O}(g^8)$ by Fourier transforming the expression for $\mu(u)$. We used here the same strategy as presented in sections \ref{hexScalarSec} and \ref{hepScalarSec} to perform the relevant integrals.\footnote{The only difference is that in sections \ref{hexScalarSec} and \ref{hepScalarSec} our ansatz involved summing over products of transcendental functions with constant coefficients, see for example (\ref{ansatz}). Here, it turns out that one must allow for rational pre-factors as well. Namely, at any given loop order, $f(\sigma)$ in (\ref{fInt}) can be written using an ansatz similar to (\ref{ansatz}) provided we replace $a_{\dots} \to a_{\dots}^+ \,e^{+\sigma}+a_{\dots}^- \,e^{-\sigma}$. \la{footnote32}}
At leading order we found
\beq
f_1(\sigma)=\(e^\sigma-e^{-\sigma}\)\sigma-\(e^\sigma+e^{-\sigma}\)\log(e^\sigma+e^{-\sigma})\, . \la{f1h1}
\eeq
The higher loops corrections $f_{2,3,4}(\sigma)$ are presented in a companying \verb"mathematica" notebook. 

To match these predictions with data we consider the bosonic hexagonal Wilson loop. 
Since the remainder function starts at two loops only, $\mathcal{W}_\text{1-loop}=\mathcal{W}_\text{BDS}$, see (\ref{RtoW}). We can easily compute the ratio of amplitudes $\mathcal{W}_\text{BDS}$ using the BDS ansatz \cite{BDS}. Alternatively, we can compute it by noticing that Wilson loops in the abelian theory are given by the exponential of the free propagator integrated along the loop. In particular, the ratio ${\cal W}^\text{BDS}_\text{hex}$ is the exponential of the one loop correlator between the bottom and top squares of the hexagon. This correlator is manifestly finite and can be easily written in terms of the conformal cross ratios \cite{Straps}
\beq\la{hexBDS}
\def\svgwidth{20cm}
\begingroup%
  \makeatletter%
  \providecommand\color[2][]{%
    \errmessage{(Inkscape) Color is used for the text in Inkscape, but the package 'color.sty' is not loaded}%
    \renewcommand\color[2][]{}%
  }%
  \providecommand\transparent[1]{%
    \errmessage{(Inkscape) Transparency is used (non-zero) for the text in Inkscape, but the package 'transparent.sty' is not loaded}%
    \renewcommand\transparent[1]{}%
  }%
  \providecommand\rotatebox[2]{#2}%
  \ifx\svgwidth\undefined%
    \setlength{\unitlength}{2391.54863281bp}%
    \ifx\svgscale\undefined%
      \relax%
    \else%
      \setlength{\unitlength}{\unitlength * \real{\svgscale}}%
    \fi%
  \else%
    \setlength{\unitlength}{\svgwidth}%
  \fi%
  \global\let\svgwidth\undefined%
  \global\let\svgscale\undefined%
  \makeatother%
  \begin{picture}(1,0.09878754)%
    \put(0,0){\includegraphics[width=\unitlength]{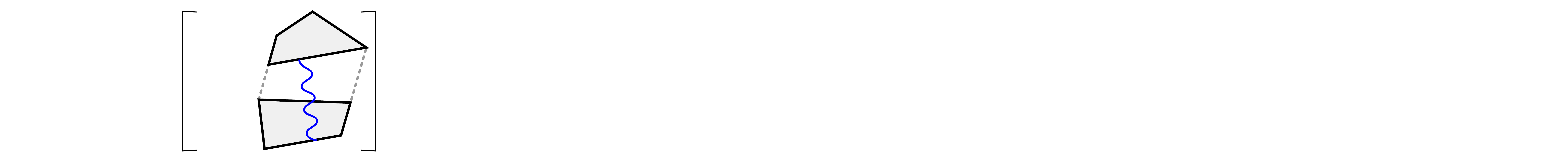}}%
    \put(-0.00057042,0.04711418){\color[rgb]{0,0,0}\makebox(0,0)[lb]{\smash{$\displaystyle{\cal W}_\text{hex}^\text{BDS}=\exp\, \Gamma_\text{cusp}$}}}%
    \put(0.24696793,0.04711418){\color[rgb]{0,0,0}\makebox(0,0)[lb]{\smash{$\displaystyle=\exp\[{\Gamma_\text{cusp}\over4}\{\text{Li}_2\left(u_2\right)-\text{Li}_2\left(1-u_1\right)-\text{Li}_2\left(1-u_3\right)+\log ^2\left(1-u_2\right)\right.$}}}%
    \put(0.28777831,0.00162064){\color[rgb]{0,0,0}\makebox(0,0)[lb]{\smash{$\displaystyle \left.-\log \left(u_1\right) \log\left(u_3\right)-\log\(u_1/ u_3\)\log\(1-u_2\)+\frac{\pi ^2}{6}\}\]$}}}%
  \end{picture}%
\endgroup%

\eeq
where $\Gamma_\text{cusp}=4g^2+\cO(g^4)$ is the cusp anomalous dimension. Expanding this result at large~$\tau$ we find 
\beq
\mathcal{W}_\text{1-loop}=1+ 2 \cos(\phi) e^{-\tau} g^2 \[\(e^\sigma-e^{-\sigma}\)\sigma-\(e^\sigma+e^{-\sigma}\)\log(e^\sigma+e^{-\sigma}) \]+\dots
\eeq
in perfect agreement with the bootstrap prediction (\ref{f1h1}).

We repeated this hexagon flattening exercise at two and three loops. At two loops, we used the result of \cite{twoloopshexagon,VerguPaper} for the two loops remainder function. We converted it into $\mathcal{W}_\text{2-loop}$ using (\ref{RtoW}) and expanded the result at large $\tau$. We found a precise match with our prediction for $f_2(\sigma)$ here also. 

At three loops, the data extraction is a bit more involved because the three loops MHV hexagon is only known at the level of the symbol \cite{Lance,Qbar}. 
At a given loop order, MHV amplitudes are expected to have uniform transcendentality. Accordingly, they are completely determined by their symbols, up to contributions multiplied by transcendental numbers like powers of $\pi$ or odd $\zeta$ functions. Nevertheless, even if we discard the latter contributions, converting the symbol of the three loops hexagon into a function of $\tau$, $\sigma$ and $\phi$ remains difficult. This eventually renders the comparison with the bootstrap predictions considerably more challenging than at the lower-loop orders. 

Fortunately, we are only interested in the large $\tau$ expansion for which there is a big simplification. 
The symbol of the three loops remainder function is a linear combination of monomials with six slots:
\beqa
S[\mathcal{R}_\text{3-loop}] =\sum_i C_i \bigotimes_{n=1}^6 a_i^{(n)}
\eeqa
where the slots $a_i^{(n)}$ depend on $\tau,\sigma$ and $\phi$. Using the method developed in \cite{DixonCorrespondence}, the symbol of a pure function with uniform transcendentally can be expanded at large $\tau$.\footnote{We thank Jeffrey Pennington for explaining to us a simple recursive algorithm for performing this kind of expansions.} The outcome is a linear combination of symbols of different lengths.\footnote{Note that there is no ambiguity when adding up symbols of different length if (and only if) we set all $\pi$'s and $\zeta$'s to zero. } Using this technique we find that
\beqa
S[\mathcal{R}_\text{3-loop}]&=& e^{-\tau} \sum_{k=0}^2 \tau^k  \sum_{l=1}^{6-k}  \sum_i C_i^{(l,k)} \bigotimes_{n=1}^l a_i^{(n,l,k)} +\ldots\, ,\la{Rexpanded}
\eeqa
where now the slots $a_i^{(n,l,k)}$ no longer depend on $\tau$. In principle, they could depend on both $\phi$ and $\sigma$. However, from the OPE approach it is clear that they should only depend on $\sigma$  as the dependence on the angle should factorize as in (\ref{repOfhexExp}). In other words, all the dependence on $\phi$ should appear as  $C_i^{(l,k)}=\cos(\phi) c_i^{(l,k)}(\sigma)$ and, indeed, this turns out to be the case. We observe furthermore that the slots of the symbol obtained after expanding in $\tau$ are simple functions of $\sigma$: they are either $x$ or $1+x$ where $x=e^{2\sigma}$. 
Symbols with such slots are in one-to-one correspondence  {with} the Harmonic Polylogs (HPL) with indices $0$ or $1$ that we encountered before.\footnote{We read the HPL from the symbol by reading its slots in reverse order and multiplying the result by $(-1)$ if the number of $1+x$ slots is odd. For example,
$x\otimes(1+x)\otimes(1+x)\otimes x\otimes x=S[ H_{0,0,1,1,0}(-x)] $ while $(1+x)\otimes(1+x)\otimes(1+x)\otimes x\otimes x=S[ - H_{0,0,1,1,1}(-x)]$. }
This is reassuring since we have found before that such functions form a good basis for writing {our conjectured} measure in position space. 
To summarize, from the analysis of (\ref{Rexpanded}) we conclude that
\beqa
\mathcal{R}_\text{3-loop}= 2 \cos(\phi) e^{-\tau} \times\, \text{linear combination of HPL's ($H_{0,1,\dots}(-x),\dots$)} +\dots 
\eeqa
where the dots stand for both the contributions proportional to $\pi$'s or $\zeta$'s, that the symbol can not probe, and the higher-twist corrections. We can now transform this into a prediction for $\mathcal{W}$ by using (\ref{RtoW}) and compare the outcome with $f_3(\sigma)$ obtained by using the bootstrap. We observe a perfect match. Nicely, we were informed by Lance Dixon et al. that this match can be upgraded to functional level such that even the $\pi$'s and $\zeta$'s contributions agree precisely~\cite{DixonCorrespondence,DixonTalk,Dixon}.

Finally, let us add a comment about the four loops function $f_4(\sigma)$ in~(\ref{fInt}). It turns out to be a powerful constraint on the four loops hexagon and, when combined with the techniques of \cite{Lance} (see also \cite{Dixon:2012yy}), it allowed the authors of \cite{Lance-Drummond} to successfully determine the symbol of the four loops remainder function.
 
 \subsection{The MHV Heptagon and the Gluonic Transition} \la{matchHep}
 The Fourier transform of the two gluonic pentagon transitions 
yield the functions $h(\sigma_1,\sigma_2)$ and $\bar h(\sigma_1,\sigma_2)$ which govern the leading collinear behaviour of the heptagon Wilson loop at any loop order, see discussion around (\ref{leadingheptagon}). At weak coupling, we have that
\beqa
h(\sigma_1,\sigma_2)&=&g^2 h_1(\sigma_1,\sigma_2)\,+ g^4 h_2(\sigma_1,\sigma_2)\,+\dots \nn \\ 
\bar h(\sigma_1,\sigma_2)&=&0\qquad\qquad \ \,\,+ g^4\, \bar{h}_2(\sigma_1,\sigma_2)\,+\dots \, ,\la{Job}
\eeqa
where the contribution $\bar h$ starts one loop later than $h$ since it comes from a non-helicity preserving transition, see (\ref{gluonpert}).  We computed the functions $h_1$, $h_2$ and $\bar h_2$ by Fourier transforming the conjecture~(\ref{finalP}) and found that 
\beq\la{h1prediction}
h_1(\sigma_1,\sigma_2)=\frac{e^{\sigma _1+\sigma _2}}{2} \log \frac{\left(e^{2 \sigma
   _1}+1\right) \left(e^{2 \sigma _2}+1\right)}{e^{2\sigma_1}+e^{2 \sigma _2}+e^{2 \sigma _1+2 \sigma _2}}+e^{\sigma _2-\sigma _1} \log \frac{e^{2 \sigma _2} \left(e^{2 \sigma_1}+1\right)}{e^{2 \sigma _1}+e^{2 \sigma _2}+e^{2 \sigma _1+2 \sigma_2}}+(\sigma_1\leftrightarrow\sigma_2)\, .
\eeq
The expressions for $h_2,\bar h_2$ are more bulky and given in the companying notebook. Here we just quote few terms of $\bar h_2$ for illustration,
 \beq
\bar h_2(\sigma_1,\sigma_2)=e^{\sigma_1+\sigma_2} \text{Li}_2\left(\frac{-1}{e^{2 \sigma_1}+
 e^{2\sigma_1-2\sigma_2}}\right) + 2 e^{\sigma_1-\sigma_2} \log(1+e^{2\sigma_2})+ \dots + \frac{\pi^2}{3} e^{-\sigma_1-\sigma_2} \,. \la{hb2prediction}
 \eeq
This should be contrasted with its remarkably simple expression in momentum space~(\ref{gluonpert}). This remark concludes the bootstrap side of the story and we now move to the comparison with available data.

As previously explained, to match the above predictions with the data we should first compute the product $\cW_\text{hep}=\cR_7\times\cW^\text{BDS}_\text{hep}$. The ratio ${\cal W}^\text{BDS}_\text{hep}$ is the exponential of the sum of three one-loop correlators
\beq\la{hepBDS}
\def\svgwidth{11cm}
\begingroup%
  \makeatletter%
  \providecommand\color[2][]{%
    \errmessage{(Inkscape) Color is used for the text in Inkscape, but the package 'color.sty' is not loaded}%
    \renewcommand\color[2][]{}%
  }%
  \providecommand\transparent[1]{%
    \errmessage{(Inkscape) Transparency is used (non-zero) for the text in Inkscape, but the package 'transparent.sty' is not loaded}%
    \renewcommand\transparent[1]{}%
  }%
  \providecommand\rotatebox[2]{#2}%
  \ifx\svgwidth\undefined%
    \setlength{\unitlength}{1121.27314453bp}%
    \ifx\svgscale\undefined%
      \relax%
    \else%
      \setlength{\unitlength}{\unitlength * \real{\svgscale}}%
    \fi%
  \else%
    \setlength{\unitlength}{\svgwidth}%
  \fi%
  \global\let\svgwidth\undefined%
  \global\let\svgscale\undefined%
  \makeatother%
  \begin{picture}(1,0.27015636)%
    \put(0,0){\includegraphics[width=\unitlength]{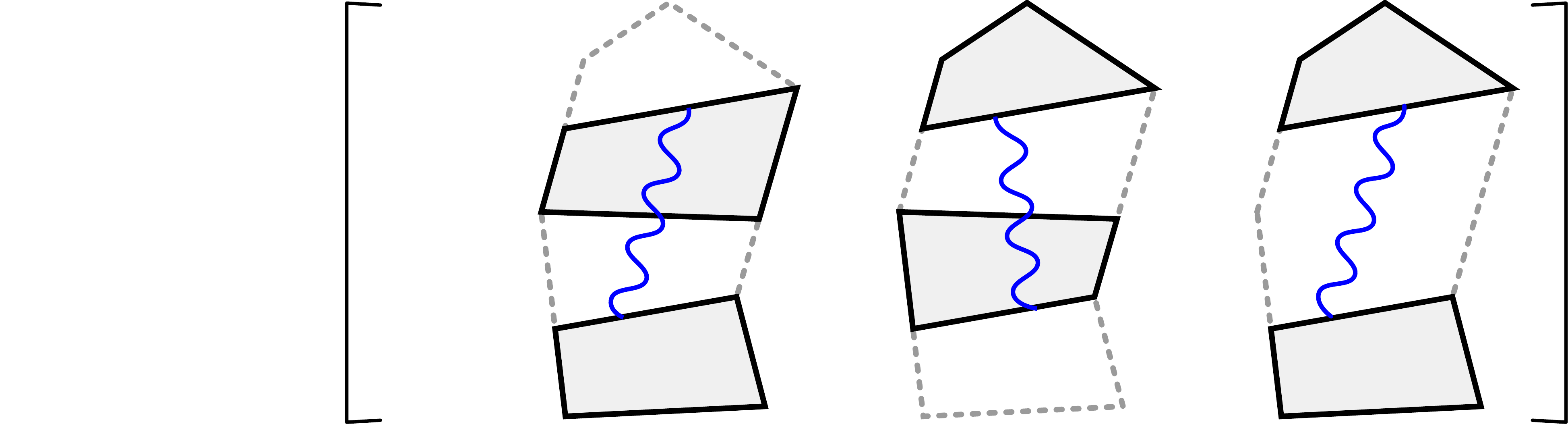}}%
    \put(-0.00130345,0.12118282){\color[rgb]{0,0,0}\makebox(0,0)[lb]{\smash{${\cal W}_\text{hep}^\text{BDS}=\exp\ \ \Gamma_\text{cusp}(\qquad\quad\qquad +\qquad\quad\qquad\,+\qquad\qquad\quad\ )$}}}%
  \end{picture}%
\endgroup%

\eeq
with the gluon propagator stretching between two non-neighbouring squares in the decomposition of the heptagon. The first two terms are nothing but the hexagon $\mathcal{W}_\text{hex}^\text{BDS}$ computed in~(\ref{hexBDS}) with arguments $\tau_1,\sigma_1,\phi_1$ and $\tau_2,\sigma_2,\phi_2$ respectively. The last term is a correlator between the bottom and the top squares of the heptagon and as such it depends on all the six cross-ratios of the heptagon. This object is actually not new and coincides precisely with the ratio $\widetilde r_\text{hep}$ introduced in~\cite{heptagonPaper}, see figure 5 of that paper. In sum, we get
\beq\la{hepBDS1}
\!\!\!\mathcal{W}_\text{hep}=\mathcal{R}_7(\tau_1,\sigma_1,\phi_1,\tau_2,\sigma_2,\phi_2) \mathcal{W}^\text{BDS}_\text{hex}(\tau_1,\sigma_1,\phi_1) \mathcal{W}^\text{BDS}_\text{hex}(\tau_2,\sigma_2,\phi_2) \tilde r_\text{hep}(\tau_1,\sigma_1,\phi_1,\tau_2,\sigma_2,\phi_2)\, .
\eeq
Since at one loop the remainder function is zero, ${\cal W}_\text{1-loop}$ is given by the last three terms in this expression only. In the large $\tau_{1,2}$ limit, we find
\beq\la{hepBDS2}
\mathcal{W}_\text{hep}=1+g^2 \Big[\underbrace{ 2\cos(\phi_1) e^{-\tau_1}f(\sigma_1)}_\text{from $\mathcal{W}^\text{BDS}_\text{hex}(\tau_1,\sigma_1,\phi_1)$}+\underbrace{ 2\cos(\phi_2) e^{-\tau_2}f(\sigma_2)}_\text{from $\mathcal{W}^\text{BDS}_\text{hex}(\tau_2,\sigma_2,\phi_2)$}+\underbrace{ 2\cos(\phi_1+\phi_2) e^{-\tau_1-\tau_2} f(\sigma_1,\sigma_2)}_\text{from $\widetilde r_\text{hep}(\tau_1,\tau_2,\sigma_1,\sigma_2,\phi_1,\phi_2)$} +\dots\Big]
\eeq
and observe that the three contributions are clearly distinct. Each of them correspond to a different sequence of transitions in (\ref{hepMHV0}).\footnote{At this loop order we see only three contributions in (\ref{hepBDS2}) while generically there are four processes in (\ref{hepMHV0}); this is expected since the helicity violating transition only starts at two loops as mentioned above.} For instance, from the viewpoint of~(\ref{hepMHV0}), the first term in (\ref{hepBDS2}) describes a transition where a particle is produced at the bottom square and annihilated at the second middle square. This is in line with the one-loop contribution to $\mathcal{W}^\text{BDS}_\text{hex}(\tau_1,\sigma_1,\phi_1)$ which involves a gluon exchange between the bottom and second middle square, see the first term in (\ref{hepBDS}). Of all the contributions in~(\ref{hepBDS2}) the most interesting one is certainly the last one, which describes the propagation of a gluon all the way from the bottom to the top. It is this contribution that probes the pentagon transition happening in the middle of the heptagon. By directly expanding $\tilde r_\text{hep}(\tau_1,\tau_2,\sigma_1,\sigma_2,\phi_1,\phi_2)$ in \cite{heptagonPaper} we find 
\beq
f(\sigma_1,\sigma_2)=\frac{e^{\sigma _1+\sigma _2}}{2} \log \frac{\left(e^{2 \sigma
   _1}+1\right) \left(e^{2 \sigma _2}+1\right)}{e^{2\sigma_1}+e^{2 \sigma _2}+e^{2 \sigma _1+2 \sigma _2}}+e^{\sigma _2-\sigma _1} \log \frac{e^{2 \sigma _2} \left(e^{2 \sigma_1}+1\right)}{e^{2 \sigma _1}+e^{2 \sigma _2}+e^{2 \sigma _1+2 \sigma_2}}+(\sigma_1\leftrightarrow\sigma_2) \nn
\eeq
which agrees beautifully with the bootstrap prediction (\ref{h1prediction}).

At two loops we can proceed similarly after noting that the remainder function in (\ref{hepBDS1}) needs now to be taken into account. The latter is not known yet as a function but fortunately its symbol has already been found~\cite{simonHep}. The situation is then analogous to the problem we handled before for the three-loop hexagon. Here again we can use the technology developed in~\cite{Lance-Drummond,DixonCorrespondence} to expand the symbol, this time at large $\tau_{1, 2}$. We then extract from it the term going like $e^{-\tau_1-\tau_2}$ and coming with \textit{no} powers of $\tau_1$ nor $\tau_2$. 
Eventually, we compare this contribution with the predictions $h_2$ and $\bar h_2$ obtained from the bootstrap and find a perfect match.\footnote{We recall that all the terms proportional to $\pi$'s or $\zeta$'s must be dropped when performing these checks, since the symbol is blind to them. In this repect, the last term in (\ref{hb2prediction}) stands as a new prediction or constraint from the bootstrap for the full two loops heptagon remainder \textit{function}.}

{Let us mention finally that in the course of our study of the leading OPE behaviour of the hexagon and heptagon Wilson loop we encountered numerous symbols of various complexity. Still, a pattern seemed to emerge as their slots were always more or less the same. It would be interesting to see if this can be a reflection of the recent motivic classification in \cite{Golden:2013xva} at the level of the collinear limit. }

\subsection{The NMHV Form Factors for Gluons} \la{formSec}

In sections \ref{matchHex} and \ref{matchHep} we extracted the gluonic measure and pentagon transitions by flattening the hexagon and heptagon bosonic Wilson loops, respectively. 
In this procedure, the gluonic excitation, which controlled the leading OPE behaviour, was produced by the geometric deformation of the Wilson loop in the near-collinear limit. This is however not the only way of creating a gluonic excitation along the flux tube and, accordingly, the MHV amplitudes are not the only ones whose leading OPE behaviours are captured by this gluonic exchange. Instead, just like for scalars, we can insert a gluonic excitation more directly by flattening a polygon WL that is already dressed with a field strength insertion at a cusp. Such an insertion can easily be engineered by considering suitable components of the super Wilson loop~\cite{superloopskinner,superloopsimon} or, equivalently, by looking at appropriate components of N$^k$MHV amplitudes.
In this penultimate section we would like to illustrate how a simple modification of our previous MHV expressions can accommodate for these N$^k$MHV amplitudes as well.

Our main character here is the hexagon NMHV component $\cW_\text{hex}^{(1111)}$, with four $\eta$'s on the bottom edge (numbered $1$ in the text). The motivation for considering this specific component comes from its (tree-level) operatorial definition. This one reveals that it contains the insertion of a gauge field at the bottom cusp which we want to investigate here\footnote{At tree level, $\cW_\text{hex-tree}^{(1111)}=\cR_\text{hex-tree}^{(1111)}=\frac{\langle 2,3,4,6\rangle ^3}{\langle 1,2,3,4\rangle  \langle
   3,4,6,1\rangle  \langle 4,6,1,2\rangle  \langle 6,1,2,3\rangle
   }+\frac{\langle 2,4,5,6\rangle ^3}{\langle 1,2,4,5\rangle  \langle
   4,5,6,1\rangle  \langle 5,6,1,2\rangle  \langle 6,1,2,4\rangle }$ reduces to (\ref{hexagonTree3}) upon the use of the hexagon momentum twistors (\ref{hexagontwistors}).}
\beq\la{hexagonTree3}
\def\svgwidth{13cm}
\begingroup%
  \makeatletter%
  \providecommand\color[2][]{%
    \errmessage{(Inkscape) Color is used for the text in Inkscape, but the package 'color.sty' is not loaded}%
    \renewcommand\color[2][]{}%
  }%
  \providecommand\transparent[1]{%
    \errmessage{(Inkscape) Transparency is used (non-zero) for the text in Inkscape, but the package 'transparent.sty' is not loaded}%
    \renewcommand\transparent[1]{}%
  }%
  \providecommand\rotatebox[2]{#2}%
  \ifx\svgwidth\undefined%
    \setlength{\unitlength}{1199.44003906bp}%
    \ifx\svgscale\undefined%
      \relax%
    \else%
      \setlength{\unitlength}{\unitlength * \real{\svgscale}}%
    \fi%
  \else%
    \setlength{\unitlength}{\svgwidth}%
  \fi%
  \global\let\svgwidth\undefined%
  \global\let\svgscale\undefined%
  \makeatother%
  \begin{picture}(1,0.19136512)%
    \put(0,0){\includegraphics[width=\unitlength]{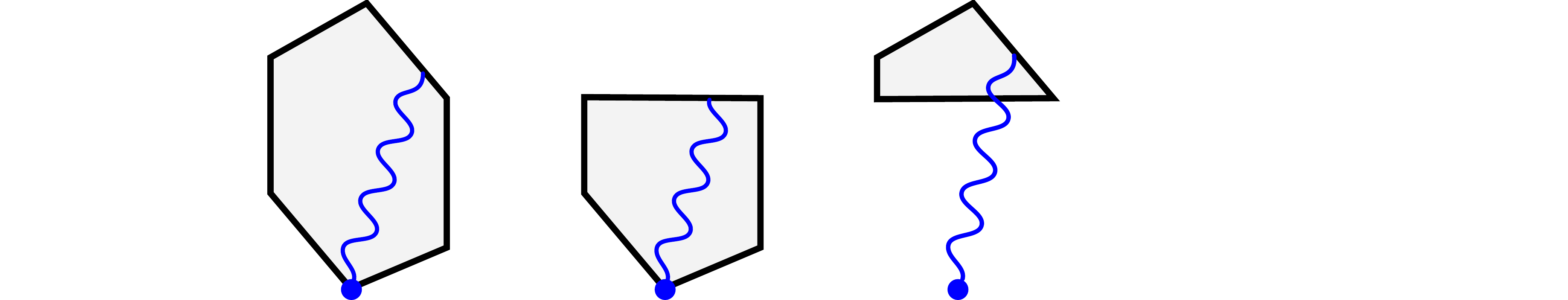}}%
    \put(-0.00086649,0.08361178){\color[rgb]{0,0,0}\makebox(0,0)[lb]{\smash{$\displaystyle \mathcal{W}^{(1111)}_\text{tree} = $}}}%
    \put(0.31261317,0.08361178){\color[rgb]{0,0,0}\makebox(0,0)[lb]{\smash{$\displaystyle =$}}}%
    \put(0.51270654,0.08361178){\color[rgb]{0,0,0}\makebox(0,0)[lb]{\smash{$\displaystyle +$}}}%
    \put(0.71279991,0.08361178){\color[rgb]{0,0,0}\makebox(0,0)[lb]{\smash{$\displaystyle =1-{e^{-\tau+i\phi}\over2\cosh\sigma}
+\cO(e^{-2\tau})$}}}%
    \put(0.24541075,0.03541713){\color[rgb]{0,0,0}\makebox(0,0)[lb]{\smash{$\,_1$}}}%
    \put(0.24219717,0.14577962){\color[rgb]{0,0,0}\makebox(0,0)[lb]{\smash{$\,_3$}}}%
    \put(0.19536754,0.15765202){\color[rgb]{0,0,0}\makebox(0,0)[lb]{\smash{$\,_4$}}}%
    \put(0.17409492,0.10608976){\color[rgb]{0,0,0}\makebox(0,0)[lb]{\smash{$\,_5$}}}%
    \put(0.19428175,0.04949287){\color[rgb]{0,0,0}\makebox(0,0)[lb]{\smash{$\,_6$}}}%
    \put(0.26008426,0.08477349){\color[rgb]{0,0,0}\makebox(0,0)[lb]{\smash{$\,_2$}}}%
  \end{picture}%
\endgroup%

\eeq
This is actually not the only component that leads to such an insertion.
We could also distribute the Grassmann variables more democratically over a neighbouring edge by looking at components with the superscript $(6111), (6611), (6661),$ or even $(6666)$. These five configurations are all physically inequivalent in that they involve different insertions along different edges of the WL. But, for what matters here, they all contain the sought insertion and they all share the same leading OPE behaviour. More precisely they start differing at twist $2$, i.e, at the level of contributions of order $\mathcal{O}(e^{-2\tau})$ in the near-collinear limit $\tau \rightarrow \infty$. We leave for the future the interesting question of disentangling among these various NMHV components by means of the OPE. Here we simply point out that the degeneracy seems to be lifted by different choice of contours of integration for the fermions that come in pairs and thus kick in at twist $2$.

What distinguishes the NMHV component $(1111)$ from the MHV one is that  {the first transition from the vacuum at the bottom of the hexagon is charged}. We already met this situation before when considering scalars and we can handle it similarly by introducing a new creation amplitude that takes the effect of the insertion into account. We denote the corresponding  {charged transition by $P_*(0|\psi)$}. 
With an $F$ insertion at the bottom, the lightest states which can be excited are the vacuum and the gluon excitations. They correspond to twist-$0$ and twist-$1$ contributions, respectively. 
The two processes are sketched in (\ref{hexagonTree3}) that illustrates the decomposition of our (renormalized) super-WL component at tree level.} 

The first term in~(\ref{hexagonTree3}), with no dependence on the kinematics, stands for the case where we just have the vacuum in the middle square, $\psi=\textrm{vac} =0$, so $P_*(0|\psi)=P_*(0|0)$. It corresponds to an $\overline{\textrm{MHV}}$ pentagon which is the same as the MHV one and therefore is independent of the coupling. Stricly speaking, this vacuum form factor $P_{*}(0|0)$ carries helicity weight since it comes from an NMHV amplitude. Here we chosen to work in the normalization where $P_{*}(0|0) =1$.

The second piece in~(\ref{hexagonTree3}) is interpreted as originating from the exchange of a gauge field excitation with positive $U(1)$ charge, i.e., what we call an $F$. Its contribution is of the type
\beq\label{treelevelh}
-{1\over2\cosh\sigma}=\int {du\over2\pi}P_*(0|u)\mu(u)P(-u|0)e^{2iu\sigma}\, ,
\eeq
which, upon Fourier transformation and comparison with the MHV result~(\ref{gluonpert}), gives a direct access to the creation amplitude $P_{*}(0|u)$ for an $F$ carrying the rapidity $u$. 
This new creation amplitude, or rather the ratio $P_{*}(0|u)/P(0|u)$, can be extracted by comparing the NMHV and MHV hexagon integrands in the OPE limit. 
The result is conveniently written in terms of the form factor
\beq\la{PstaroverP}
h(u)\equiv {P_*(0|u)\over P(0|u)}\times {1\over P_*(0|0)}= {u^2+\frac{1}{4} \over g^2}+\cO(g^0)\, ,
\eeq
where the second equality follows from the tree-level result (\ref{treelevelh}). The factor $P_*(0|0)$ was included such that $h(u)$ carries no helicity weight at the end.

The $F$ excitation is not the only twist-one particle that can be produced and similarly to the ratio $h(u)$ in (\ref{PstaroverP}) there is a form factor for the creation of the conjugate $\bar F$ excitation
\beq
\bar h(u)\equiv{\bar P_*(0|u)\over  P(0|u)}{1\over P_*(0|0)} \,.
\eeq
This form factor is a further example of a $U(1)$ violating process. It ought to start at higher loop in perturbation theory, since no tree level diagram can connect this excitation to the decorated pentagon under consideration. This is why no contribution proportional to $e^{-\tau-i\phi}$ shows up in~(\ref{hexagonTree3}).

We already exhausted the information available from the tree level analysis. To proceed further with the determination of the two form factors $h$ and $\bar h$, we need some axioms that we shall now present and motivate.
\begin{description}
\item[Axiom I] $$\vspace{-0.5cm}$$
The first axiom simply follows from the reflexion property of the pentagon,
\beq
h(u)=h(-u)\, .
\eeq
We can immediately check that this property is true at tree level, see~(\ref{PstaroverP}).
\item[Axiom II] $$\vspace{-0.5cm}$$
The second axiom is related to mirror rotation and reads
\beq
h(u^{-\gamma})=\bar h(u)\, .
\eeq
It is inspired by the mirror transformation of the gauge field. As  discussed in the previous section, under a mirror rotation $F\rightarrow \bar{F}$. Our assumption is then that this transformation will swap the two form factors for gluon and anti-gluon.
\item[Axiom III] $$\vspace{-0.5cm}$$
The last axiom reads
\beq
h(u)\bar h(u) =1\, .
\eeq
It has a somewhat different origin. The particular combination displays above appears in the computation of a certain N$^2$MHV hexagon amplitude, as we shall see below. An N$^2$MHV hexagon is the same as $\overline{\textrm{MHV}}$. The condition above guarantees that this is true at any coupling. 
\end{description}
As in the previous examples, we look for the simplest solution to these axioms. In particular we expect the relevant solution to display the minimal amount of singularities that are needed to fulfill the above equations. The most obvious solution in this respect is certainly $h(u) = \bar{h}(u) =1$. However this would not reproduce the tree-level expectation quoted before. A more promising candidate is given by the pair
\beq\label{ansatzh}
h(u) = \frac{x^{+}(u)x^{-}(u)}{g^2}\, , \qquad \bar{h}(u) = \frac{g^2}{x^{+}(u)x^{-}(u)}\, ,
\eeq
where we recall that $x^{\pm}(u) = x(u\pm \ft{i}{2})$ with $x(u) = \ft{1}{2}(u+\sqrt{u^2-4g^2})$ the Zhukowky variable. This expression has the right weak coupling limit, as one can verify with $x(u) = u+O(g^2)$. It also satisfies our three axioms; Axiom II and III for instance follow from the fact that $x^{\pm}(u^{-\gamma}) = x^{\pm}(u^{\gamma})  = g^2/x^{\pm}(u)$ under the mirror path $\gamma$ of a gauge field, while Axiom I is obvious.

Using our from factors~(\ref{ansatzh}) we can now write down a very precise expression for the leading OPE contribution to the NMHV ratio function $\cR_\text{hex}^{(1111)} = \cW_\text{hex}^{(1111)}/\cW_\text{hex}$. It reads
\beqa\label{ansatzR1111}
\cR_\text{hex}^{(1111)} = 1 &+& e^{i\phi-\tau}\int \frac{du}{2\pi}\mu(u)(h(u)-1)e^{ip(u)\sigma-\gamma(u)\tau} \notag \\
&+&  e^{-i\phi-\tau}\int \frac{du}{2\pi}\mu(u)(\bar{h}(u)-1)e^{ip(u)\sigma-\gamma(u)\tau}  + \ldots ,
\eeqa
where dots stand for higher twist corrections and with $\gamma(u) = E(u)-1$ the anomalous energy of a gauge field excitation. Notice that the form factors appear subtracted by $-1$ in the two integrands. This comes from the fact that we divided by the MHV contribution (here truncated to its leading OPE behaviour $\sim e^{-\tau}$) to get to the ratio function. We also specialized to the normalization used throughout this paper for the MHV creation/annihilation amplitudes, that is $P(0|u) = P(u|0) = 1$ and the charge conjugate relations.

We see that the form factors play their expected role and break the symmetry between positively and negatively charged excitations. This is immediate at weak coupling since to the leading order 
\beq\label{WCh}
\mu(u)h(u) = -\pi \,{\rm sech}(\pi u)\, , \,\,\,\,\,\, \mu(u) = -\frac{\pi g^2 \,{\rm sech}(\pi u)}{u^2+\ft{1}{4} }\, , \,\,\,\, \,\,\mu(u)\bar{h}(u) = -\frac{\pi g^4 \,{\rm sech}(\pi u)}{(u^2+\ft{1}{4})^2}\, ,
\eeq 
hence delaying the contribution of the negatively charged excitation. By construction the expression~(\ref{ansatzR1111}) reduces properly to the tree-level result~(\ref{hexagonTree3}). What is less trivial is that~(\ref{ansatzR1111}) with the form factors~(\ref{ansatzh}) works perfectly at higher loops. We verified it explicitly by comparing with the one- and two-loop expressions for the NMHV hexagon ratio function~\cite{Dixon:2011nj}.\footnote{We are very grateful to L.~Dixon for sharing with us the collinear OPE expansion of all relevant functions in \cite{Dixon:2011nj}.}
This confirms in particular that $\bar{h}(u)\mu(u)$ only starts at two loops, as predicted by~(\ref{WCh}).

These checks also serve the purpose of providing evidence for the existence of a well defined super loop for all possible components. This one was previously questioned in \cite{Belitsky:2011zm} where it was shown that there are some components for which unwanted anomalous contributions survive in traditional dimensional regularization.
In \cite{AndreiSimon,Andrei} it was argued that these anomalies can be cured by a more careful operatorial definition and regularization of the super loop. Our analysis here supports this picture, since it successfully combines higher-loop scattering amplitude data with the OPE interpretation of the super Wilson loop. 

Another interesting application of our form factors~(\ref{ansatzh}) comes from considering the N$^2$MHV hexagon ratio function, and in particular its component
\beq
\cR_\text{hex}^{(1111)(4444)} = \cW_\text{hex}^{(1111)(4444)}/\cW_\text{hex}\, .
\eeq
At leading twist it can be thought as an hexagon Wilson loop with an $F$ insertion at both the bottom and top cusps,
\beq\label{N2MHV}
\def\svgwidth{13cm}
\begingroup%
  \makeatletter%
  \providecommand\color[2][]{%
    \errmessage{(Inkscape) Color is used for the text in Inkscape, but the package 'color.sty' is not loaded}%
    \renewcommand\color[2][]{}%
  }%
  \providecommand\transparent[1]{%
    \errmessage{(Inkscape) Transparency is used (non-zero) for the text in Inkscape, but the package 'transparent.sty' is not loaded}%
    \renewcommand\transparent[1]{}%
  }%
  \providecommand\rotatebox[2]{#2}%
  \ifx\svgwidth\undefined%
    \setlength{\unitlength}{1571.64111328bp}%
    \ifx\svgscale\undefined%
      \relax%
    \else%
      \setlength{\unitlength}{\unitlength * \real{\svgscale}}%
    \fi%
  \else%
    \setlength{\unitlength}{\svgwidth}%
  \fi%
  \global\let\svgwidth\undefined%
  \global\let\svgscale\undefined%
  \makeatother%
  \begin{picture}(1,0.292009)%
    \put(0,0){\includegraphics[width=\unitlength]{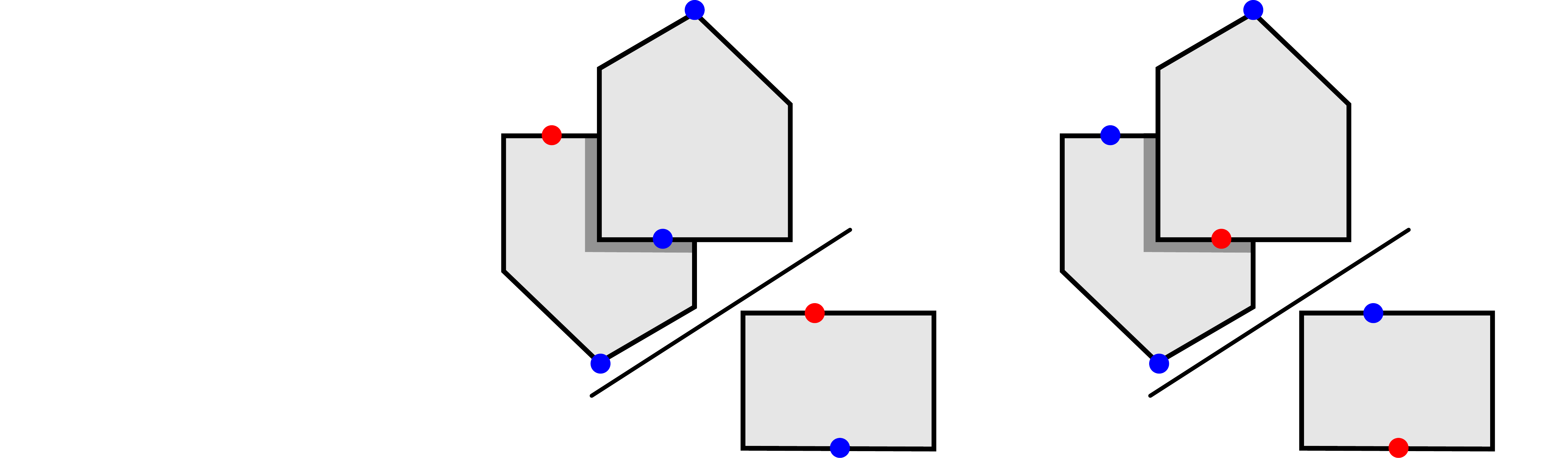}}%
    \put(0.35387367,0.10020863){\color[rgb]{0,0,0}\makebox(0,0)[lb]{\smash{$P_*(u)$}}}%
    \put(0.49130962,0.04523425){\color[rgb]{0,0,0}\makebox(0,0)[lb]{\smash{$1/\mu(u)$}}}%
    \put(0.40070369,0.22237393){\color[rgb]{0,0,0}\makebox(0,0)[lb]{\smash{$\bar P_*(u)$}}}%
    \put(0.92171864,0.14099312){\color[rgb]{0,0,0}\makebox(0,0)[lb]{\smash{$\displaystyle +\dots$}}}%
    \put(-0.0006293,0.14099312){\color[rgb]{0,0,0}\makebox(0,0)[lb]{\smash{$\displaystyle \cW_\text{hex}^{(1111)(4444)}=1+\ e^{i\phi}$}}}%
    \put(0.71018909,0.10020863){\color[rgb]{0,0,0}\makebox(0,0)[lb]{\smash{$\bar P_*(u)$}}}%
    \put(0.84762505,0.04523425){\color[rgb]{0,0,0}\makebox(0,0)[lb]{\smash{$1/\mu(u)$}}}%
    \put(0.75701912,0.22237393){\color[rgb]{0,0,0}\makebox(0,0)[lb]{\smash{$P_*(u)$}}}%
    \put(0.57558367,0.14099312){\color[rgb]{0,0,0}\makebox(0,0)[lb]{\smash{$\displaystyle +\ e^{-i\phi}$}}}%
  \end{picture}%
\endgroup%

\eeq
This component involves two form factors for both creation and annihilation of the gauge field excitation at the top and bottom. It leads to
\beqa\label{ansatzR11114444}
\cR_\text{hex}^{(1111)(4444)} = 1 &+& e^{i\phi-\tau}\int \frac{du}{2\pi}\mu(u)(h(u)\bar{h}(u)-1)e^{ip(u)\sigma-\gamma(u)\tau} \notag \\
&+&  e^{-i\phi-\tau}\int \frac{du}{2\pi}\mu(u)(h(u)\bar{h}(u)-1)e^{ip(u)\sigma-\gamma(u)\tau}  + \ldots .
\eeqa
We see that due to the specific combination of the form factors implied by this component the symmetry between positively and negatively charged excitations has been restored. This does not stop here and we know actually that this particular ratio function should be identically $1$ (in our normalization). This is because N$^2$MHV hexagons describe $\overline{\text{MHV}}$ amplitudes which are the same as MHV ones. This property will be properly observed if and only if $h(u)\bar{h}(u)=1$. This final remark makes clear why we enforced our Axiom III for the form factors.

\section{Discussion}

In this paper, following~\cite{short}, we presented a new non-perturbative formulation of scattering amplitudes or equivalently null polygonal Wilson loops in planar $\mathcal{N}=4$ SYM theory. In this approach, scattering amplitudes and Wilson loops are given by sums over flux tube states, similar to partition functions. The fundamental building blocks in these sums are the so called \textit{pentagon transitions}. They encode the dynamical information about the creation, propagation and annihilation of the excitations living on the flux-tube spanned by the Wilson loop. These transitions allow us to fully characterize the most elementary Wilson loops with edge-insertions and to glue them together into the more complicated ones that are dual to scattering amplitudes. In this paper, we have shown, using a minimal set of axioms, how the expression for the transitions of a single scalar or gauge field excitations can be found at finite coupling.
Once expanded in perturbation theory, our conjectures were found to match with all available perturbative data. 

The OPE approach armed with the pentagon bootstrap axioms nicely combines features of the celebrated bootstrap programs for CFT \cite{Polyakov} and integrable QFT \cite{Zamolodchikov}. Both programs are based on pushing elementary ideas such as crossing and unitarity to their ultimate consequences. In the context of correlation functions, the bootstrap program was recently revived in \cite{boostrapCFT}, see also \cite{moreBoot} and references therein for latest developments. We believe there is room for fruitful interplay between the Wilson loops bootstrap and the more conventional CFT bootstrap for correlation functions. 

It is quite remarkable, at the end, that all the fundamental ingredients of the underlying \textit{integrable} flux tube dynamics appear nicely combined in the form of our conjectures. Of course, this was already hinted in the pioneering OPE paper~\cite{OPEpaper} where the details of the spectrum of GKP excitations and of their dispersion relations~\cite{BenDispPaper,Straps} were shown to be embraced by the WLs. This relationship between 2D and 4D data has now become more stringent. Take for example the leading behaviour off the collinear limit of an heptagon Wilson loop. It reads (here for gluons)
\beq
\mathcal{W}_\text{heptagon}=\dots+e^{\pm i\phi_1 \pm i\phi_2-\tau_1-\tau_2} \int \frac{du}{2\pi}\int \frac{dv}{2\pi} \,\verb"integrand" + \dots
\eeq
where, up to a very simple function of $u$ and $v$ (with no obvious flux tube interpretation), 
\beq
\verb"integrand" \propto  \({x^{+}(u)x^{-}(u)}{}\)^{\eta_1} \({x^{+}(v)x^{-}(v)}{}\)^{\eta_2} \sqrt{\frac{S(u,v)}{S(u^\gamma,v)}} e^{-\gamma(u)\tau_1+i p(u)\sigma_1-\gamma(v)\tau_2-i p(v)\sigma_2}\, .  \la{integrandCon}
\eeq
In this expression the famous Zhukowsky variables $x^+$ and $x^-$ appear with a nice geometrical interpretation: they are form factors for the production of gauge fields. They appear raised to the powers $\eta_i$ which can be $-1,0$ or $1$ depending on which kind of scattering component we are considering. For MHV amplitudes $\eta_i=0$ for example. The (anomalous) energy $\gamma$ and the momentum $p$ of the fundamental excitations appears directly multiplying the space-time cross-ratios, as proclaimed by~\cite{OPEpaper}. The scattering of excitations factorizes on the flux tube, thanks to Integrability, and is therefore uniquely determined by the 2 body S-matrix $S(u,v)$. This object nicely shows up in our pentagon transitions. It comes along with the `mirror' S-matrix $S(u^{\gamma},v)$ that is a physical object on its own; it yields for instance the Casimir energy of the Integrable system in finite volume through the so-called L\"uscher formula \cite{Luscher,Bajnok-Janik}. Altogether, we see that all the physical observables that determine the flux tube dynamics beautifully govern the behaviour of the four dimensional scattering amplitudes. 

In the OPE construction of scattering amplitudes we need to sum over all possible states and there are infinitely many of them. If we restrict the sum to a finite subset, then the more states we include the better is the approximation for the amplitude. This expansion is quite distinct from other approaches such as perturbation theory. This feature is in practice very fortunate since it means that we can learn a lot about the OPE approach by analyzing perturbative data. For instance, already the tree level amplitude (\ref{TreeLevelScalar}) contains transitions involving any number of particles.
Conversely we can easily generate infinitely many predictions for perturbative amplitudes from the OPE. In this paper we illustrated this important interplay on several examples.

Apart from scalars and gauge fields, which were the focus of our analysis, the remaining twist-one excitations in the theory are fermions. We have a conjecture for their pentagon transitions that matches successfully against data, as well as for the other twist-one excitations. 
This proposal deserves however a separate discussion and will be reported elsewhere. The main reason is that 
there is strong evidence that crossing symmetry does not exist for fermions (see for example~\cite{MoreDispPaper} and appendix \ref{FmirrorApp}). It means in practice that we cannot easily move a fermion from one edge of the Wilson loop to another or at least not just through a simple analytical continuation.
This obstruction makes the construction of the pentagon transitions for fermions a bit harder and our conjecture for them is based to a large extent on the experience acquired here with the gluonic and scalar transitions.
We hope that a more rigorous understanding of fermions transitions will be achieved in the near future. 

As we move to higher twist we encounter bound states and multi-particle states. The bound states can be treated similarly to single-particle states. Their S-matrices and dispersion relations are known at any coupling and they obey nice fusion relations. The latter relations connect bound states of different sizes to one another and also relate them to their fundamental twist-one constituents. It follows then that the bound-state transitions satisfy the very same kind of axioms as those for the twist-one particles considered in this paper. The bound-state family comprise the two towers $D_z^m F_{z-}$ and $D_{\bar z}^m F_{\bar z-}$ which played an important role in previous OPE analysis~\cite{OPEpaper,Straps}. They also play a prominent role at strong coupling and they will be studied in greater details in a forthcoming publication~\cite{to appear}.  

The multi-particle transitions are also equipped with their own bootstrap equations. These can often be solved explicitely in terms of single-particle transitions, as exemplified in \cite{short} with gauge fields. More generally we believe that their solution can always be written as a product of two contributions: the \textit{dynamical part} and the \textit{matrix structure}, which transform respectively as scalar and tensor under the R-symmetry group $SU(4)$. For example, for a $2\to2$ transition among scalars we found that
\beq
P({\bf u}|{\bf v})_{i_1i_2}^{j_1j_2}= P_\text{dyn}({\bf u}|{\bf v}) \[ \pi_1({\bf u}|{\bf v})  \delta_{i_1}^{j_1}\delta_{i_2}^{j_2}+\pi_2({\bf u}|{\bf v}) \delta_{i_1}^{j_2}\delta_{i_2}^{j_1}+\pi_3({\bf u}|{\bf v})\delta_{i_1i_2}\delta^{j_1j_2} \] \,, \la{conj2}
\eeq
where ${\bf u}=\{u_1,u_2\}$ and ${\bf v}=\{v_1,v_2\}$. The incoming $SO(6)$ indices $i_1,i_2$ run from $1$ to $6$ and indicate which pair of scalars we insert at the bottom while the outgoing indices $j_1,j_2$ parametrize the state at the top. This transition fulfills several important requirements, which almost uniquely specify it.
For instance, suppose we take a bottom particle and move it to the top through a sequence of mirror transformations while at the same time we take a top particle and move it to the bottom. We should end up with the very same object up to a relabelling of R-symmetry indices and rapidities. More precisely, since for scalars a mirror transformation is simply a shift of rapidity, we should observe that $P(u_1,u_2-3i|v_1-2i,v_2)_{i_1i_2}^{j_1j_2}=P(v_1,u_1|v_2,u_2)_{j_1i_1}^{j_2i_2}$. Further constraints come from the Watson's equations that tell us that exchanging two particles can be compensated by the action of the S-matrix. All these equations have their counterparts in the bootstrap program for form factors in integrable models~\cite{Watson} which provides us with valuable strategy for solving them (see in particular the analysis of matrix structures for form factors in models with $O(N)$ symmetry~\cite{Babujian}). At the end of the day we found that the most natural solution to all these equations reads
\beq
P_\text{dyn}({\bf u}|{\bf v}) = \frac{P(u_1|v_1)P(u_1|v_2)P(u_2|v_1)P(u_2|v_2)}{P(u_2|u_1)P(v_1|v_2)} \,,  \nn
\eeq
while 
\beqa
\pi_1({\bf u}|{\bf v})+\pi_2({\bf u}|{\bf v}) &=&1 \,, \qquad\qquad\qquad \pi_2({\bf u}|{\bf v}) = \frac{(u_1-v_1)(u_2-v_2+i)}{(u_1-u_2-i)(v_1-v_2+i)} \,, \nn \\
\pi_2({\bf u}|{\bf v}) +\pi_3({\bf u}|{\bf v}) &=& \frac{(u_1-v_1)(u_2-v_2+i)(u_1-v_1-i)(u_2-v_2+2i)}{(u_1-u_2-i)(u_1-u_2-2i)(v_1-v_2+i)(v_1-v_2+2i)}  \,. \la{sol2}
\eeqa
We expect (and partially checked) a similar pattern for a larger number of scalars and also for a broader class of excitations. We believe for instance that the factorization of the dynamical part (into a product of single particle transitions) is \textit{universal} and should be a consequence of the Integrability of the theory.
This part should also capture all the non-trivial dependence on the coupling (which is hidden inside the single-particle transitions). The matrix part is then coupling independent and governed by a bunch of functions $\pi_i$, whose number grows fast with the number of particles. Our guess for them is that they are all rational functions of the differences of rapidities. This is clearly the case for the two-particle solution in (\ref{sol2}). Our algorithm for producing all these functions $\pi_i$ for transitions involving higher number of scalars also supports this assumption. It would be very interesting to explore these matrix structures on their own, to investigate whether they admit a factorization of sort, whether they are subject to some kind of simple graphical relation or to some fusion-hierarchy equations which will make their construction easier and their analysis more transparent.

There is a also a very nice story about data extraction for multi particle transitions. We can check  for example a particular component of the conjecture (\ref{conj2}) by first using the mirror transformation to distribute the excitations on different edges of the pentagon and then apply the flattening procedure of section \ref{summarySection} to each of the edges 
\beq \la{last}
\def\svgwidth{11cm}
\begingroup%
  \makeatletter%
  \providecommand\color[2][]{%
    \errmessage{(Inkscape) Color is used for the text in Inkscape, but the package 'color.sty' is not loaded}%
    \renewcommand\color[2][]{}%
  }%
  \providecommand\transparent[1]{%
    \errmessage{(Inkscape) Transparency is used (non-zero) for the text in Inkscape, but the package 'transparent.sty' is not loaded}%
    \renewcommand\transparent[1]{}%
  }%
  \providecommand\rotatebox[2]{#2}%
  \ifx\svgwidth\undefined%
    \setlength{\unitlength}{672.38549805bp}%
    \ifx\svgscale\undefined%
      \relax%
    \else%
      \setlength{\unitlength}{\unitlength * \real{\svgscale}}%
    \fi%
  \else%
    \setlength{\unitlength}{\svgwidth}%
  \fi%
  \global\let\svgwidth\undefined%
  \global\let\svgscale\undefined%
  \makeatother%
  \begin{picture}(1,0.48714498)%
    \put(0,0){\includegraphics[width=\unitlength]{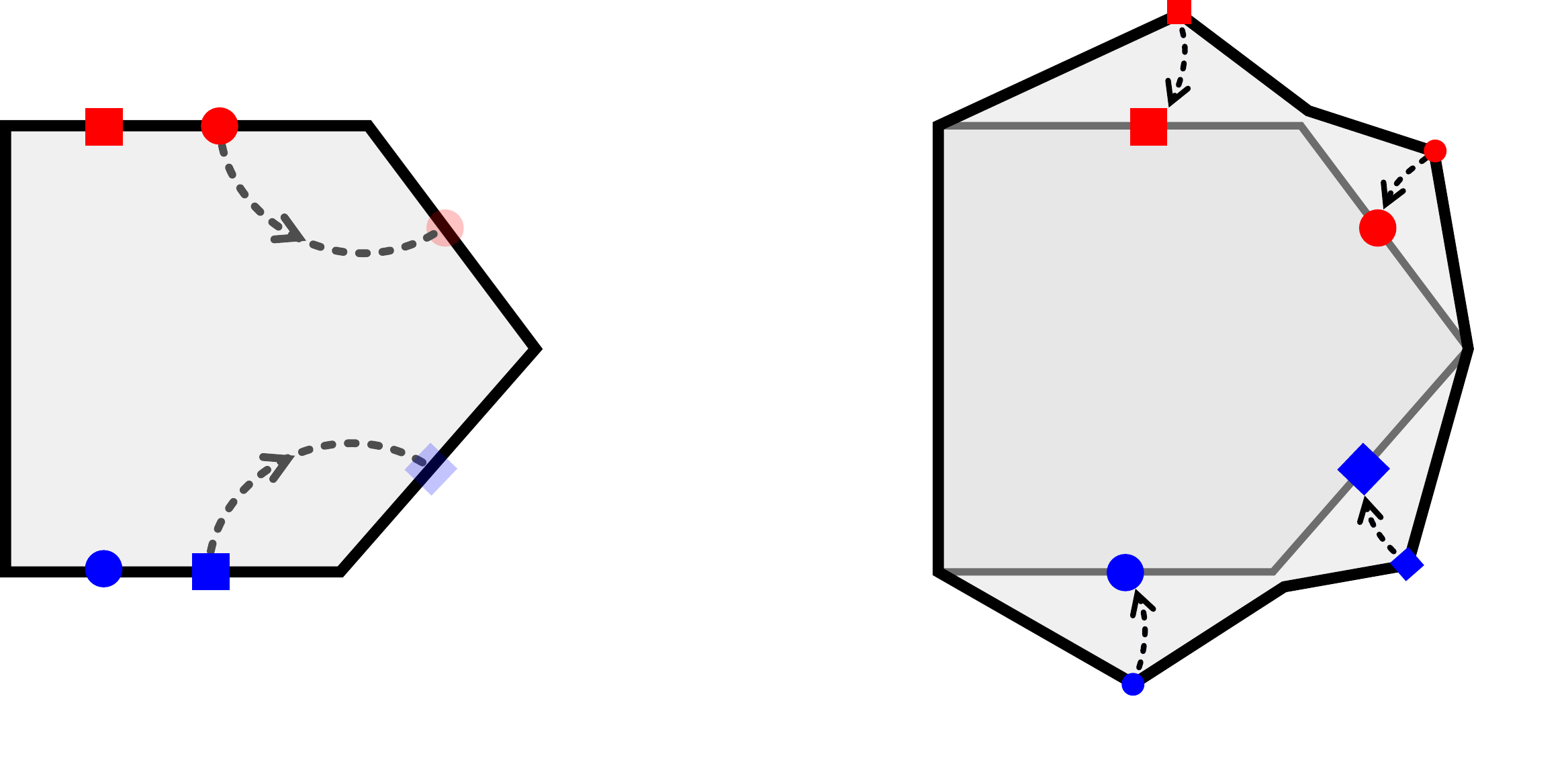}}%
    \put(0.1759874,0.36727622){\color[rgb]{0,0,0}\makebox(0,0)[lb]{\smash{$v_2^{\gamma}$}}}%
    \put(0.06176721,0.3696558){\color[rgb]{0,0,0}\makebox(0,0)[lb]{\smash{$v_1$}}}%
    \put(0.16170987,0.1459746){\color[rgb]{0,0,0}\makebox(0,0)[lb]{\smash{$u_2^{-\gamma}$}}}%
    \put(0.06176721,0.14835418){\color[rgb]{0,0,0}\makebox(0,0)[lb]{\smash{$u_1$}}}%
    \put(0.04986927,0.07458698){\color[rgb]{0,0,0}\makebox(0,0)[lb]{\smash{$Z$}}}%
    \put(0.11173854,0.07458698){\color[rgb]{0,0,0}\makebox(0,0)[lb]{\smash{$X$}}}%
    \put(0.44250118,0.26257438){\color[rgb]{0,0,0}\makebox(0,0)[lb]{\smash{$=$}}}%
    \put(0.04748968,0.43152507){\color[rgb]{0,0,0}\makebox(0,0)[lb]{\smash{$\bar X$}}}%
    \put(0.12125689,0.43152507){\color[rgb]{0,0,0}\makebox(0,0)[lb]{\smash{$\bar Z$}}}%
    \put(0.70663537,0.1459746){\color[rgb]{0,0,0}\makebox(0,0)[lb]{\smash{$u_1$}}}%
    \put(0.83037391,0.21260304){\color[rgb]{0,0,0}\makebox(0,0)[lb]{\smash{$u_2$}}}%
    \put(0.83751267,0.31730488){\color[rgb]{0,0,0}\makebox(0,0)[lb]{\smash{$v_2$}}}%
    \put(0.7185333,0.36489663){\color[rgb]{0,0,0}\makebox(0,0)[lb]{\smash{$v_1$}}}%
    \put(0.03797133,0.00319932){\color[rgb]{0,0,0}\makebox(0,0)[lb]{\smash{$\text{Integrability}$}}}%
    \put(0.65666403,0.00319932){\color[rgb]{0,0,0}\makebox(0,0)[lb]{\smash{$\text{Amplitude}$}}}%
    \put(0.63286816,0.26257438){\color[rgb]{0,0,0}\makebox(0,0)[lb]{\smash{$W_\text{nonagon}^{{\rm N}^2{\rm MHV}}$}}}%
  \end{picture}%
\endgroup%

\eeq
This is quite a nice trick actually, for at least two reasons. First, it probes the mirror transformation, which we recall is non-perturbative in the 't Hooft coupling and thus always constitutes a very non-trivial check of the finite coupling structure. Second, it is very useful at the technical level: since we end up with a single scalar at each edge after the mirror transformations, we do not need to use the two-particle wave function (which is not known yet in the relevant channel).

For gluons, because their scattering is reflectionless, everything is much simpler. There is no matrix part and their multi-particle transitions were already reported in \cite{short}. For fermions, again the lack of ability of moving them around using mirror transformations amputates the bootstrap from one of its legs. This is where a better understanding of the algebraic properties of the matrix structures might help remedying to this unpleasant situation.  

In fact, and probably related to the above subtleties, multi-particle states involving fermions behave in an interesting and non-trivial fashion. Consider for example a neutral state made out of a fermion $\psi$ and an anti-fermion $\bar \psi$. An important contribution, that dominates in perturbation theory, comes from the region where the momentum of one of the excitations vanishes. When this happens, the fermion excitation becomes a generator of a super-symmetry broken by the Wilson loop \cite{AldayMaldacena}. That generator can act  on the vacuum or it can act on the other fermion, thus transforming it into another excitation; for example a $\bar \psi$ can become an $F_{+-}$ . At leading order in perturbation theory the action on the vacuum is suppressed and we only produce an $F_{+-}$. Hence, at leading order in perturbation theory the state $F_{+-}$ can be treated as a single particle excitation but as we go to finite coupling it is more appropriate to think of it as a pair of two fermionic excitations. This is the origin of a symmetry enhancement at weak coupling~\cite{BenDispPaper} and this is compatible with the analysis of several OPE studies \cite{Bootstrapping,Straps,SuperOPE,heptagonPaper}, where $F_{+-}$ (and several other states dubbed descendants there) were treated as single-particles.
This shows that the fermions are pivotal for the OPE since already at tree level infinitely many of them are needed to recover the amplitude in full kinematics. There are also central at strong coupling where they relate to the mass 2 boson of the GKP string.

One additional point of data for scattering amplitudes is their value at strong coupling \cite{AMApril,AGM,AMSV}. There the amplitude is captured by a classical string saddle-point \cite{AM}. As previously sketched~\cite{short} and analyzed more thoroughly in~\cite{toappear}, our pentagon decomposition and transitions are perfectly consistent with the string saddle-point predictions.
Hopefully the OPE approach will allow us to study perturbative corrections to scattering amplitudes at strong coupling, for which almost nothing is known to date.  

Our final comment concerns the radius of convergency of the OPE. The OPE is an expansion around the (multi) collinear limit. As such, by including more and more excitations in the decomposition, one can obtain an arbitrarily precise description of an amplitude at finite coupling. This is assuming that the conformal cross ratios lie inside the radius of convergency of the OPE. What eventually controls the size of this radius is the mass gap in the spectrum of flux-tube excitations. This one is known to become exponentially small at strong coupling~\cite{AldayMaldacena} since scalar excitations become massless. Though this phenomenon might be of no concern at the semiclassical level at strong coupling, it will certainly plays an important role at subleading orders. This immediately raises the fascinating question of the re-summation of the scalar contributions at strong coupling and more generally of the complete OPE at finite coupling.
It is not yet clear if any one of these re-summations is possible. An encouragement comes from strong coupling where the OPE indeed re-sums into a very natural object from the Integrability point of view: the Thermodynamic Bethe Ansatz Yang-Yang Functional. Is this a strong coupling accident or are four dimensional scattering amplitudes given by a Yang-Yang functional at any coupling?

\subsection*{Acknowledgements} 
We thank F.~Cachazo, J.~Caetano, L.~Dixon, D.~Gaiotto, J.~Henn, G.~ Korchemsky, J.~Maldacena, J.~M.~Maillet, J.~Toledo and T.~Wang for discussions. We are very grateful to J.~Bourjaily, S.~Caron-Huot, L.~Dixon, S.~He,  M.~von~Hippel and J.~Pennington for discussions and for providing us with invaluable data used to check several conjectures. We are grateful to Andrei Belitsky for critical correspondence and valuable comments. Research at the Perimeter Institute is supported in part by the Government of Canada through NSERC and by the Province of Ontario through MRI. The research of A.S. has been supported in part by the Province of Ontario through ERA grant ER 06-02-293 and by the U.S. Department of Energy grant DE-SC0009988.

\appendix

\section{Geometry of the Decomposition}\la{geomeryappendix}

In this appendix we study the geometry of null polygons. We consider squares, pentagons, hexagons and heptagons. 
Squares and pentagons are the fundamental building blocks involved in the decomposition of higher $n$--gons while hexagons and heptagons are the simplest polygons one can use for extracting measures and transitions. For each one of these geometries we shall provide an explicit representation of the associated twistors, which could be directly plugged into \texttt{mathematica} if needed. More general expressions, i.e., not referring to a specific representation, could of course be inferred from the ones presented below.

\subsection*{Momentum Twistors}
We represent $\mathbb{R}^{1,3}$ as a lightcone in $\mathbb{R}^{2,4}$. That is, any point $x\in \mathbb{R}^{1,3}$ is associated with a null ray in $\mathbb{R}^{2,4}$, $\{X\in \mathbb{R}^{2,4},\ X^2=0,\ X\simeq tX\}$.
The map to the usual Poincare coordinates is given by
\beq
x_\mu={X_\mu\over X_{-1}+X_4}\ .
\eeq
The conformal group is then realized linearly in the form of the $SO(2,4)$ rotations of the embedding coordinates $X$. A convenient way of representing a null polygon is found by using momentum twistors \cite{Hodges:2009hk}.
These are $\mathbb{R}^{2,4}$ spinors that are defined up to rescaling,  $ Z\equiv ( Z_1, Z_2, Z_3, Z_4)\simeq t Z$. The conformal group acts on these spinors by multiplication on the right  with a group element $\in SL(4)$. To any pair of momentum twistors $ Z,\tilde Z$ we can associate a null ray in $\mathbb{R}^{2,4}$ and therefore a point $x\in \mathbb{R}^{1,3}$ as
\beq \label{PointDef}
X_{ab}=X_M\Gamma^M_{ab}= Z_{[a}\tilde Z_{b]}\ ,\qquad \begin{array}{l}M=-1,0,1,2,3,4\\ a,b=1,2,3,4\end{array}\ ,
\eeq
where $\Gamma_{M}$ are sigma matrices in $\mathbb{R}^{2,4}$. {Group theoretically, the representation~(\ref{PointDef}) is nothing but the isomorphism between the vector and the antisymmetric representation of $SO(2,4)$.}
A polygon with $N$ null sides can be given as a sequence of $N$ twistors $ Z_i$, such that the intersection
 of the sides $i$ and $i+1$ is the point $X_{i } =  Z_i \wedge  Z_{i+1}$. The distance between two cusps of the polygon reads
\beq\la{distance}
(x_i-x_j)^2={\<Z_i, Z_{i+1}, Z_j, Z_{j+1}\>\over
( Z_i\cdot\Gamma^+\cdot Z_{i+1}^T)( Z_j\cdot\Gamma^+\cdot Z_{j+1}^T)}\ ,
\eeq
where $\Gamma^+=\frac{i}{\sqrt{2}}\(\Gamma^{-1}+\Gamma^4\)$.\footnote{For implementation into \texttt{mathematica}, one can use $\Gamma^+={i\over2}\left(
\begin{array}{cccc}
 0 & -\sqrt{2} & 1 & 1 \\
 \sqrt{2} & 0 & -1 & 1 \\
 -1 & 1 & 0 & -\sqrt{2} \\
 -1 & -1 & \sqrt{2} & 0
\end{array}
\right)$ as a particular represention. Note, however, that most of the time we do not need to compute distances, since these are not conformally invariant quantities. In the latter cases, only $SL(4)$ invariant products $\<Z_i,Z_j,Z_k,Z_l\>$ (or cross-ratios) can appear and these are of course independent of any specific representation for the {sigma} matrices. An example of when we do need to compute distances is for checking whether the non-neighboring cusps of a given polygon are space-like separated from one another. 
\la{sigmapfootnote} }
Here, $\<Z_i,Z_j,Z_k,Z_l\>=\det\[\(Z_i,Z_j,Z_k,Z_l\)\]$ is the $SL(4)$ invariant product of four twistors. Note that we automatically have $X_{i-1 } \cdot X_{i }=0$,
which
is the condition that the $i$-th side should be null. 

\subsection*{The Square}
A square is specified by four twistors. For example, these can be
\beq\la{squaretwistors}
\def\svgwidth{8cm}
\begingroup%
  \makeatletter%
  \providecommand\color[2][]{%
    \errmessage{(Inkscape) Color is used for the text in Inkscape, but the package 'color.sty' is not loaded}%
    \renewcommand\color[2][]{}%
  }%
  \providecommand\transparent[1]{%
    \errmessage{(Inkscape) Transparency is used (non-zero) for the text in Inkscape, but the package 'transparent.sty' is not loaded}%
    \renewcommand\transparent[1]{}%
  }%
  \providecommand\rotatebox[2]{#2}%
  \ifx\svgwidth\undefined%
    \setlength{\unitlength}{590.14272461bp}%
    \ifx\svgscale\undefined%
      \relax%
    \else%
      \setlength{\unitlength}{\unitlength * \real{\svgscale}}%
    \fi%
  \else%
    \setlength{\unitlength}{\svgwidth}%
  \fi%
  \global\let\svgwidth\undefined%
  \global\let\svgscale\undefined%
  \makeatother%
  \begin{picture}(1,0.32037459)%
    \put(0,0){\includegraphics[width=\unitlength]{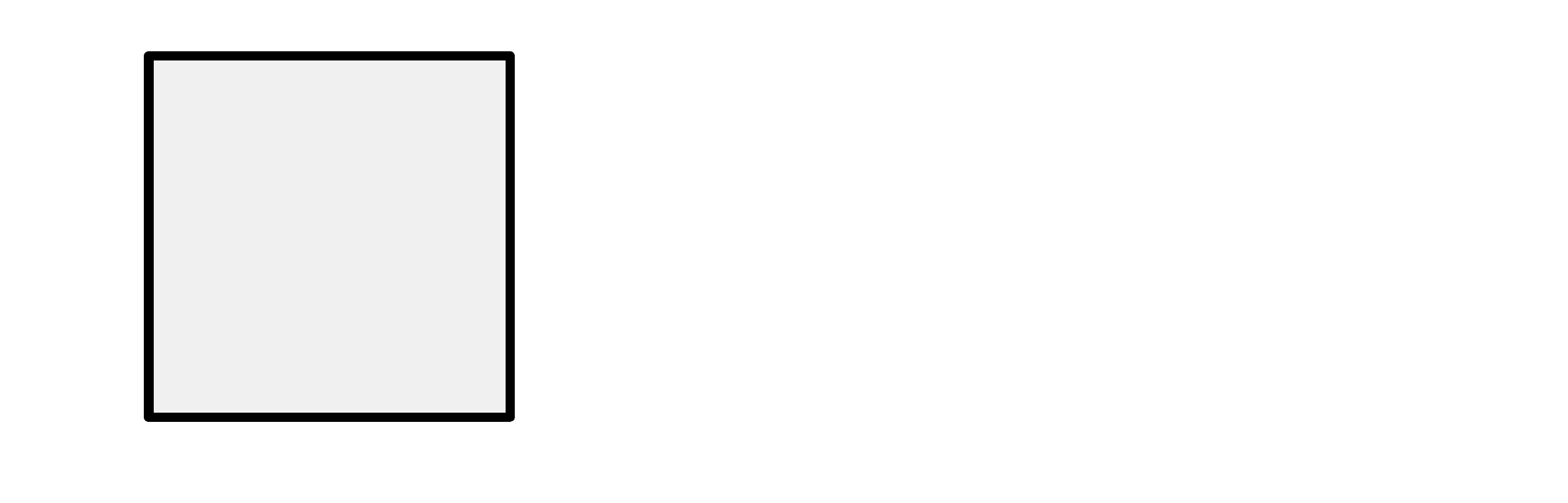}}%
    \put(0.34821662,0.21671951){\color[rgb]{0,0,0}\rotatebox{-90}{\makebox(0,0)[lb]{\smash{$\text{right}$}}}}%
    \put(0.07273043,0.12884013){\color[rgb]{0,0,0}\rotatebox{90}{\makebox(0,0)[lb]{\smash{$\text{left}$}}}}%
    \put(0.17148336,0.30437789){\color[rgb]{0,0,0}\makebox(0,0)[lb]{\smash{$\text{top}$}}}%
    \put(0.17283896,0.00343374){\color[rgb]{0,0,0}\makebox(0,0)[lb]{\smash{$\text{bot}$}}}%
    \put(0.61747733,0.16068384){\color[rgb]{0,0,0}\makebox(0,0)[lb]{\smash{$\begin{array}{lll}
{\cal Z}_\text{left}&=&(0,1,0,0) \\ 
{\cal Z}_\text{right}&=&(1,0,0,0) \\ 
{\cal Z}_\text{top}&=&(0,0,0,1) \\
{\cal Z}_\text{bot}&=&(0,0,1,0)
\end{array}$}}}%
  \end{picture}%
\endgroup%

\eeq
These twistors parametrize a null square in Euclidian kinematics. That is,  any two non-neighboring cusps of the polygon are space-like separated. The square preserves three (commuting) conformal symmetries \cite{OPEpaper} parametrized by $\tau,\sigma$ and $\phi$ and discussed in section \ref{Intropart2}. For the choice (\ref{squaretwistors}) the $SL(4)$ group element implementing these symmetries reads\footnote{All the twistors in this appendix are written in the (2,2) signature where $\phi$ is purely imaginary. To go to (1,3) signature one needs to take $\phi$ real and apply a conformal transformation to the sigma matrix $\Gamma^+$ appearing in footnote \ref{sigmapfootnote}.}
\beq\la{squaresymm}
M_1=\left(
\begin{array}{cccc}
 e^{\sigma -\frac{i\phi }{2}} & 0 & 0 & 0 \\
 0 & e^{-\sigma-\frac{i\phi }{2}} & 0 & 0 \\
 0 & 0 & e^{\tau +\frac{i\phi }{2}} & 0 \\
 0 & 0 & 0 & e^{-\tau +\frac{i\phi }{2}}
\end{array}
\right)\, .
\eeq
\subsection*{The Pentagon}
A pentagon is specified by five twistors. For example, these can be
\beq\la{pentagontwistors}
\def\svgwidth{12cm}
\begingroup%
  \makeatletter%
  \providecommand\color[2][]{%
    \errmessage{(Inkscape) Color is used for the text in Inkscape, but the package 'color.sty' is not loaded}%
    \renewcommand\color[2][]{}%
  }%
  \providecommand\transparent[1]{%
    \errmessage{(Inkscape) Transparency is used (non-zero) for the text in Inkscape, but the package 'transparent.sty' is not loaded}%
    \renewcommand\transparent[1]{}%
  }%
  \providecommand\rotatebox[2]{#2}%
  \ifx\svgwidth\undefined%
    \setlength{\unitlength}{831.39355469bp}%
    \ifx\svgscale\undefined%
      \relax%
    \else%
      \setlength{\unitlength}{\unitlength * \real{\svgscale}}%
    \fi%
  \else%
    \setlength{\unitlength}{\svgwidth}%
  \fi%
  \global\let\svgwidth\undefined%
  \global\let\svgscale\undefined%
  \makeatother%
  \begin{picture}(1,0.41408395)%
    \put(0,0){\includegraphics[width=\unitlength]{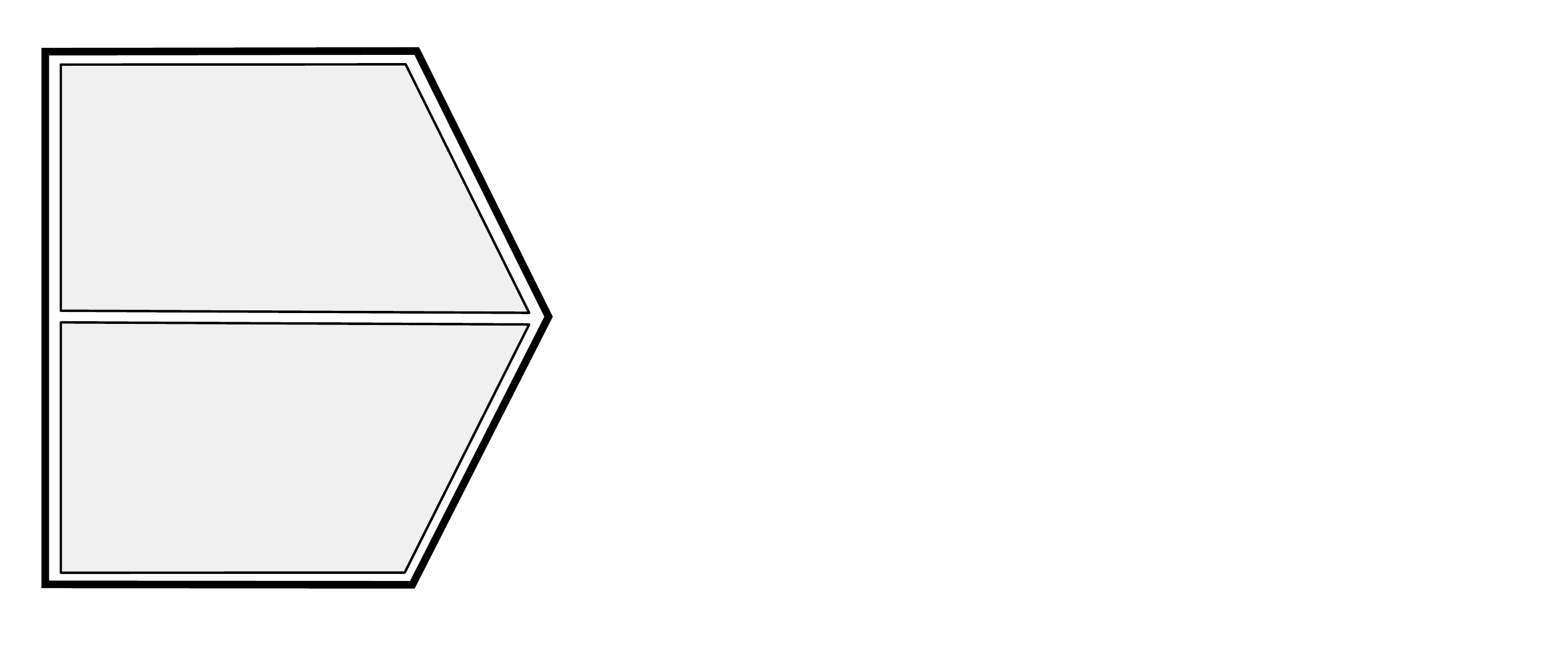}}%
    \put(0.1054536,0.00243734){\color[rgb]{0,0,0}\makebox(0,0)[lb]{\smash{$\text{bottom}$}}}%
    \put(0.11122704,0.40272911){\color[rgb]{0,0,0}\makebox(0,0)[lb]{\smash{$\text{top}$}}}%
    \put(0.01135484,0.19360029){\color[rgb]{0,0,0}\rotatebox{90}{\makebox(0,0)[lb]{\smash{$\text{left}$}}}}%
    \put(0.30929842,0.05086179){\color[rgb]{0,0,0}\rotatebox{63.08162251}{\makebox(0,0)[lb]{\smash{$\text{right-bot}$}}}}%
    \put(0.2997762,0.36788063){\color[rgb]{0,0,0}\rotatebox{-63.73334936}{\makebox(0,0)[lb]{\smash{$\text{right-top}$}}}}%
    \put(0.13047184,0.23145042){\color[rgb]{0,0,0}\makebox(0,0)[lb]{\smash{$\text{middle}$}}}%
    \put(0.58474113,0.20409976){\color[rgb]{0,0,0}\makebox(0,0)[lb]{\smash{$\begin{array}{lll}
Z_\text{left}&=&(\ \ 0,\ \ 1,\ \ 0,\ \ 0) \\ 
Z_\text{top}&=&(\ \ 0,\ \ 1,-1,\ \, 1) \\
Z_\text{right top}&=&(-1,\ \ 0,\ \ 0,\ \, 1) \\ 
Z_\text{right bot}&=&(\ \ 1,\ \ 0,\ \ 0,\ \ 0) \\ 
Z_\text{bot}&=&(\ \ 0,\ \ 0,\ \ 1,\ \ 0)
\end{array}$}}}%
  \end{picture}%
\endgroup%

\eeq
As for the square, these twistors parametrize a pentagon in Euclidian kinematics. In (1,3) signature it is not possible to have a real configuration of five null separated cusps such that the distance between any two non-consecutive cusps is spacelike.\footnote{Note also that for polygons with an odd number of edges we can not have all the cusps being of the in-out type.} This is however possible in (2,2) signature which is the signature in which the pentagon parametrized by the twistors (\ref{pentagontwistors}) is living. The discrete geometrical symmetries of the pentagon transition, namely cyclicity and reflection, are only manifest in Euclidian kinematics:
\beq\la{PentagonInR22}
\def\svgwidth{10cm}
\begingroup%
  \makeatletter%
  \providecommand\color[2][]{%
    \errmessage{(Inkscape) Color is used for the text in Inkscape, but the package 'color.sty' is not loaded}%
    \renewcommand\color[2][]{}%
  }%
  \providecommand\transparent[1]{%
    \errmessage{(Inkscape) Transparency is used (non-zero) for the text in Inkscape, but the package 'transparent.sty' is not loaded}%
    \renewcommand\transparent[1]{}%
  }%
  \providecommand\rotatebox[2]{#2}%
  \ifx\svgwidth\undefined%
    \setlength{\unitlength}{930.39794922bp}%
    \ifx\svgscale\undefined%
      \relax%
    \else%
      \setlength{\unitlength}{\unitlength * \real{\svgscale}}%
    \fi%
  \else%
    \setlength{\unitlength}{\svgwidth}%
  \fi%
  \global\let\svgwidth\undefined%
  \global\let\svgscale\undefined%
  \makeatother%
  \begin{picture}(1,0.3315801)%
    \put(0,0){\includegraphics[width=\unitlength]{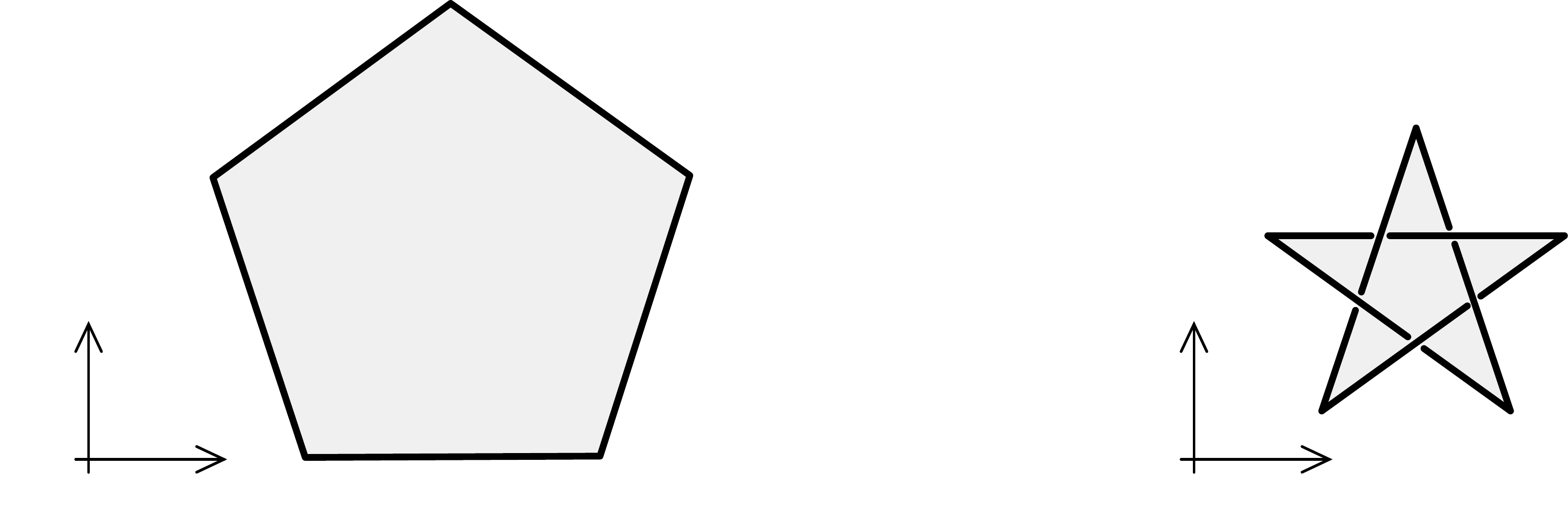}}%
    \put(0.10217506,0.00543144){\color[rgb]{0,0,0}\makebox(0,0)[lb]{\smash{$x_1$}}}%
    \put(0.0015277,0.09367892){\color[rgb]{0,0,0}\makebox(0,0)[lb]{\smash{$x_2$}}}%
    \put(0.8072497,0.00543144){\color[rgb]{0,0,0}\makebox(0,0)[lb]{\smash{$t_1$}}}%
    \put(0.71004173,0.09367892){\color[rgb]{0,0,0}\makebox(0,0)[lb]{\smash{$t_2$}}}%
  \end{picture}%
\endgroup%

\eeq
Note also that null polygons in (1,3) and (2,2) signature are related by analytic continuation in the external data, {i.e.}, in the conformal cross ratios $\{\tau_i,\sigma_i,\phi_i\}$. Such analytic continuation leaves the {transitions} in the decomposition (\ref{decompositionIntro}) untouched. 

Any pentagon  can be  divided into a top and a bottom squares using a null line that starts from the right cusp and ends on the left edge, see (\ref{pentagontwistors}). These two squares share three of their twistors with the pentagon,
\beq
\text{Top square}=\left\{Z_\text{left},Z_\text{top},Z_\text{right-top},Z_\text{middle}\right\}\,,\quad \text{Bot square}=\left\{Z_\text{left},Z_\text{middle},Z_\text{right-bot},Z_\text{bot}\right\}\, ,\nn
\eeq
where $Z_\text{middle}$ is the twistor associated to the dividing line. It is given by
\beq
Z_\text{mid}={\<Z_\text{bot},Z_\text{left},Z_\text{top},Z_\text{right-bot}\>Z_\text{right-top}-\<Z_\text{bot},Z_\text{left},Z_\text{top},Z_\text{right-top}\>Z_\text{right-bot}\over\<Z_\text{bot},Z_\text{top},Z_\text{right-top},Z_\text{right-bot}\>}={\cal Z}_\text{top}\, .
\eeq

\subsection*{The Hexagon}
An hexagon is specified by six twistors. We think of it as a sequence of three squares or, equivalently, as two pentagons overlapping on a middle square. We coordinatize all conformally inequivalent hexagons by acting with the symmetries of the middle square on the cusps located to its bottom. The corresponding six twistors are
\beq\la{hexagontwistors}
\def\svgwidth{12cm}
\begingroup%
  \makeatletter%
  \providecommand\color[2][]{%
    \errmessage{(Inkscape) Color is used for the text in Inkscape, but the package 'color.sty' is not loaded}%
    \renewcommand\color[2][]{}%
  }%
  \providecommand\transparent[1]{%
    \errmessage{(Inkscape) Transparency is used (non-zero) for the text in Inkscape, but the package 'transparent.sty' is not loaded}%
    \renewcommand\transparent[1]{}%
  }%
  \providecommand\rotatebox[2]{#2}%
  \ifx\svgwidth\undefined%
    \setlength{\unitlength}{1008.71386719bp}%
    \ifx\svgscale\undefined%
      \relax%
    \else%
      \setlength{\unitlength}{\unitlength * \real{\svgscale}}%
    \fi%
  \else%
    \setlength{\unitlength}{\svgwidth}%
  \fi%
  \global\let\svgwidth\undefined%
  \global\let\svgscale\undefined%
  \makeatother%
  \begin{picture}(1,0.36606194)%
    \put(0,0){\includegraphics[width=\unitlength]{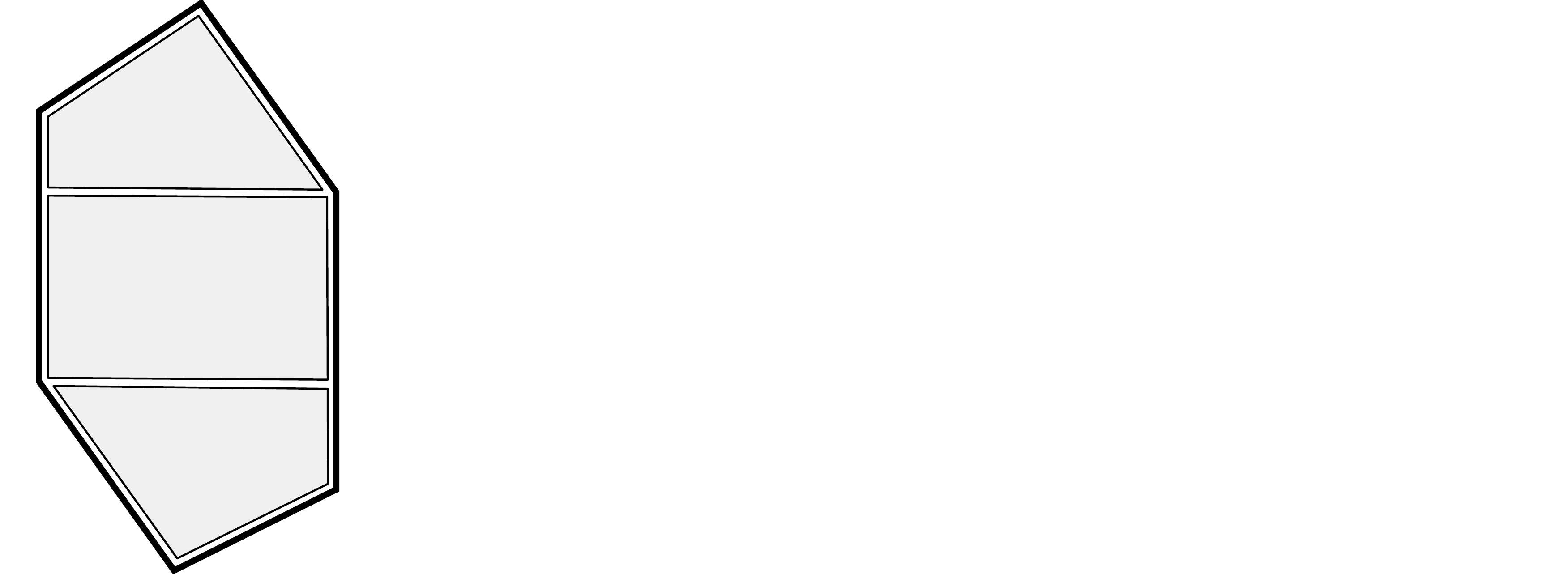}}%
    \put(0.1748869,0.00626581){\color[rgb]{0,0,0}\makebox(0,0)[lb]{\smash{$1$}}}%
    \put(0.22405842,0.14267714){\color[rgb]{0,0,0}\makebox(0,0)[lb]{\smash{$2$}}}%
    \put(0.17964543,0.30846626){\color[rgb]{0,0,0}\makebox(0,0)[lb]{\smash{$3$}}}%
    \put(0.04989162,0.33339832){\color[rgb]{0,0,0}\makebox(0,0)[lb]{\smash{$4$}}}%
    \put(-0.00102249,0.20179021){\color[rgb]{0,0,0}\makebox(0,0)[lb]{\smash{$5$}}}%
    \put(0.03869901,0.04261387){\color[rgb]{0,0,0}\makebox(0,0)[lb]{\smash{$6$}}}%
    \put(0.41413755,0.18647784){\color[rgb]{0,0,0}\makebox(0,0)[lb]{\smash{$\(\!\! \begin{array}{c}
Z_1 \\ Z_2 \\ Z_3 \\ Z_4 \\ Z_5 \\ Z_6
\end{array}\!\!\) =
\left(
\begin{array}{cccc}
 e^{\sigma-\frac{i\phi }{2}} & 0 & e^{\tau+\frac{i\phi }{2}} & e^{-\tau+\frac{i\phi }{2}}\\
 1 & 0 & 0 & 0 \\
 -1 & 0 & 0 & 1 \\
 0 & 1 & -1 & 1 \\
 0 & 1 & 0 & 0 \\
 0 & e^{-\sigma-\frac{i\phi }{2}} & e^{\tau+\frac{i\phi }{2}} & 0 
\end{array}
\right)$}}}%
  \end{picture}%
\endgroup%

\eeq
Here, $Z_6=(0,1,1,0)\cdot M_1(\tau,\sigma,\phi)$ and $Z_1=(1,0,1,1)\cdot M_1(\tau,\sigma,\phi)$ are the two bottom twistors on which we acted with the symmetries of the middle square. The middle square is given by (\ref{squaretwistors}) and one of the two pentagons in (\ref{hexagontwistors}) coincides with (\ref{pentagontwistors}).

{In the OPE parametrization, the usual three cross-ratios for the hexagon read
\beqa
u_3&=&\frac{(x_2-x_6)^2(x_3-x_5)^2}{(x_2-x_5)^2(x_6-x_3)^2}=\frac{\langle 2,3,6,1\rangle  \langle 3,4,5,6\rangle }{\langle
   2,3,5,6\rangle  \langle 6,1,3,4\rangle } = \frac{1}{1+e^{2 \sigma }+ 2\,e^{\sigma -\tau } \cos (\phi )+e^{-2 \tau }}  \,, \nn \\
u_2&=&\frac{(x_1-x_5)^2(x_2-x_4)^2}{(x_1-x_4)^2(x_5-x_2)^2}= \frac{\langle 1,2,5,6\rangle  \langle 2,3,4,5\rangle }{\langle
   1,2,4,5\rangle  \langle 5,6,2,3\rangle } = \frac{1}{2} e^{-\tau } \text{sech}(\tau ) \,, \nn \\
u_1&=&\frac{(x_6-x_4)^2(x_1-x_3)^2}{(x_6-x_3)^2(x_4-x_1)^2} =\frac{\langle 1,2,3,4\rangle  \langle 6,1,4,5\rangle }{\langle
   4,5,1,2\rangle  \langle 6,1,3,4\rangle } = e^{2 \sigma+2\tau} u_2 \,u_3 \,.\la{theus}
\eeqa
}

Notice that the hexagon has four of its cusps ($X_{1,2,4,5}$) lying in the plane of the middle square. The two cusps which stand out of this plane are the top and bottom ones, denoted $X_3$ and $X_6$ respectively. In the parametrization (\ref{hexagontwistors}) the coordinates $(\tau, \sigma, \phi)$ move the bottom cusp while holding both the middle square and top cusp fixed. One can always shift the origin of the $(\tau, \sigma, \phi)$ by a constant, see discussion below (\ref{pentagontransition}), but a natural choice is provided by the hexagon geometry. The origin in the $\phi$ direction is fixed such that at $\phi=0$ the top and bottom cusps are pointing in opposite directions in the two-dimensional plane transverse to the middle square. This implies that the scalar propagator in (\ref{hexagonTree2}), i.e, $1/\<6,1,3,4\>$, is minimal at $\phi=0$. The origin in the $\tau$ direction is chosen such that at $\tau=i\pi/2$ the points $X_1$ and $X_3$ are null separated. 
Finally, the origin in the $\sigma$ direction is chosen such that at $\sigma=i\pi/2$ and large $\tau$ the points $X_6$ and $X_3$ become null separated.

\subsection*{The Heptagon}

Finally, an heptagon is specified by seven twistors. We think of it as a sequence of four squares or three pentagons overlapping on two middle squares. We coordinatize all conformally inequivalent heptagons by acting with the symmetries of the two middle squares on the cusps to their bottom. The corresponding twistors (which are such that (\ref{hexagontwistors}) is one of the hexagons in the heptagon decomposition) are\footnote{{Here, for cosmetic reasons, instead of acting with the symmetries of the top square on the cusps on its bottom, we acted with the inverse conformal transformation on the cusps on its top.}}
\beq\la{Heptagon2}
\def\svgwidth{14.5cm}
\begingroup%
  \makeatletter%
  \providecommand\color[2][]{%
    \errmessage{(Inkscape) Color is used for the text in Inkscape, but the package 'color.sty' is not loaded}%
    \renewcommand\color[2][]{}%
  }%
  \providecommand\transparent[1]{%
    \errmessage{(Inkscape) Transparency is used (non-zero) for the text in Inkscape, but the package 'transparent.sty' is not loaded}%
    \renewcommand\transparent[1]{}%
  }%
  \providecommand\rotatebox[2]{#2}%
  \ifx\svgwidth\undefined%
    \setlength{\unitlength}{754.18999023bp}%
    \ifx\svgscale\undefined%
      \relax%
    \else%
      \setlength{\unitlength}{\unitlength * \real{\svgscale}}%
    \fi%
  \else%
    \setlength{\unitlength}{\svgwidth}%
  \fi%
  \global\let\svgwidth\undefined%
  \global\let\svgscale\undefined%
  \makeatother%
  \begin{picture}(1,0.54553102)%
    \put(0,0){\includegraphics[width=\unitlength]{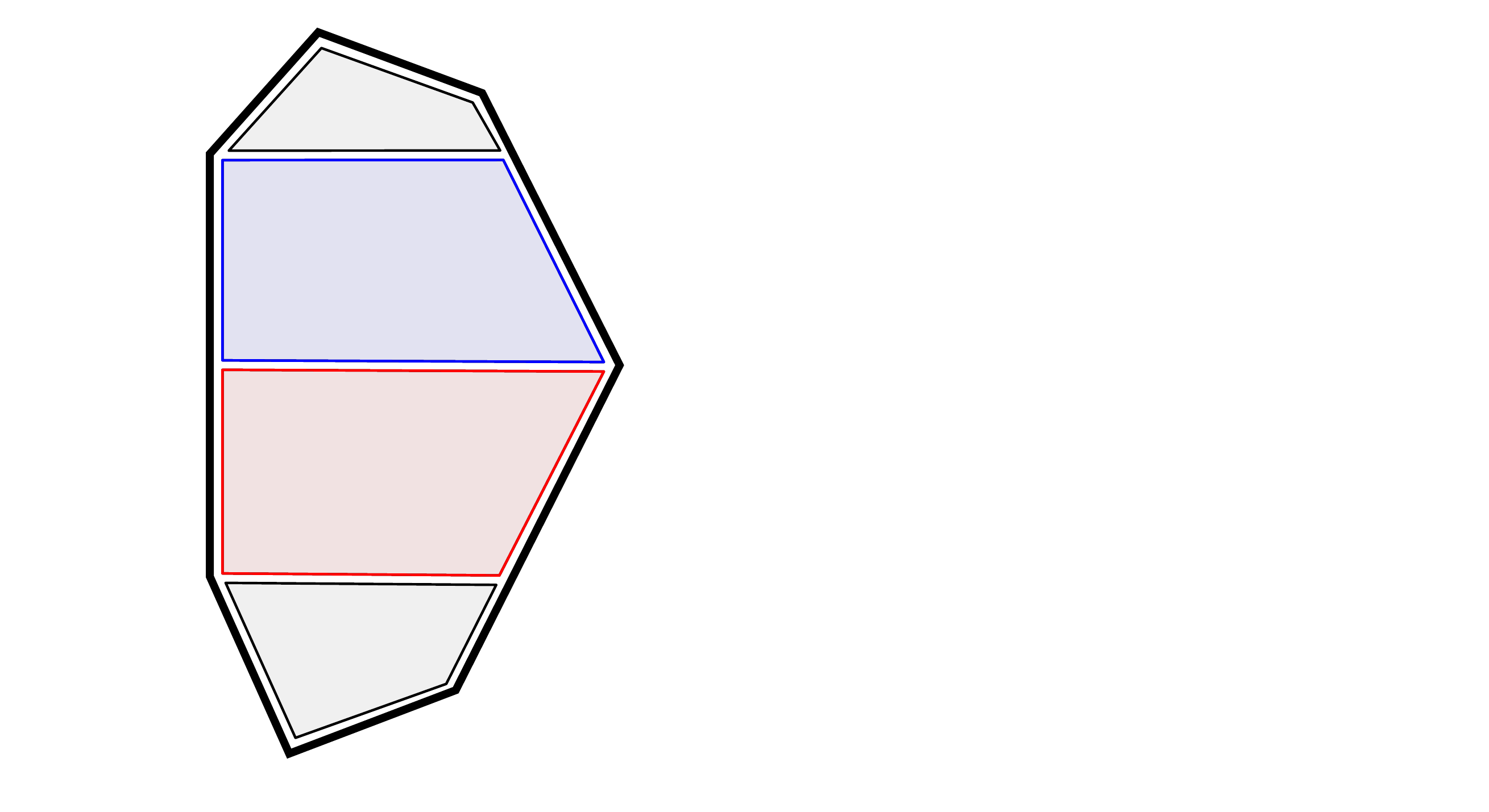}}%
    \put(0.2664173,0.03028257){\color[rgb]{0,0,0}\makebox(0,0)[lb]{\smash{$Z_1$}}}%
    \put(0.37036989,0.16817879){\color[rgb]{0,0,0}\makebox(0,0)[lb]{\smash{$Z_2=Z_\text{right bot}$}}}%
    \put(0.38160478,0.39451997){\color[rgb]{0,0,0}\makebox(0,0)[lb]{\smash{$Z_3=Z_\text{right top}$}}}%
    \put(0.27950473,0.51179901){\color[rgb]{0,0,0}\makebox(0,0)[lb]{\smash{$Z_4$}}}%
    \put(0.12583174,0.49037502){\color[rgb]{0,0,0}\makebox(0,0)[lb]{\smash{$Z_5$}}}%
    \put(-0.00136754,0.29666289){\color[rgb]{0,0,0}\makebox(0,0)[lb]{\smash{$Z_6=Z_\text{left}$}}}%
    \put(0.1103838,0.08227375){\color[rgb]{0,0,0}\makebox(0,0)[lb]{\smash{$Z_7$}}}%
    \put(0.21586029,0.4566405){\color[rgb]{0,0,0}\makebox(0,0)[lb]{\smash{$Z_\text{top}$}}}%
    \put(0.21586029,0.31662273){\color[rgb]{0,0,0}\makebox(0,0)[lb]{\smash{$Z_\text{middle}$}}}%
    \put(0.21586029,0.17448348){\color[rgb]{0,0,0}\makebox(0,0)[lb]{\smash{$Z_\text{bot}$}}}%
    \put(0.61894188,0.27843607){\color[rgb]{0,0,0}\makebox(0,0)[lb]{\smash{$\begin{array}{lll}
Z_4&=&(-1,\ \ 1,-1,\ \ 3).{\color{blue} M_2^{-1}} \\
Z_5&=&(\ \ 0,\ \ 2,-1,\ \ 1).{\color{blue} M_2^{-1}}\\ \\
\color{blue} Z_\text{left}&=&(\ \ 0,\ \ 1,\ \ 0,\ \ 0) \\ 
\color{blue} Z_\text{right top}&=&(-1,\ \ 0,\ \ 0,\ \, 1) \\ 
\color{blue} Z_\text{top}&=&(\ \ 0,\ \ 1,-1,\ \, 1) \\
\color{blue} Z_\text{middle}&=&(\ \ 0,\ \ 0,\ \ 0,\ \ 1)\\ \\
\color{red} Z_\text{left}&=&(\ \ 0,\ \ 1,\ \ 0,\ \ 0) \\ 
\color{red}Z_\text{right bot}&=&(\ \ 1,\ \ 0,\ \ 0,\ \ 0) \\ 
\color{red}Z_\text{middle}&=&(\ \ 0,\ \ 0,\ \ 0,\ \ 1) \\
\color{red}Z_\text{bot}&=&(\ \ 0,\ \ 0,\ \ 1,\ \ 0)\\ \\
Z_7&=&(\ \ 0,\ \ 1,\ \ 1,\ \ 0).{\color{red} M_1} \\
Z_1&=&(\ \ 1,\ \ 0,\ \ 1,\ \ 1).{\color{red} M_1}
\end{array}$}}}%
  \end{picture}%
\endgroup%

\eeq
The matrix $M_1$ realizes the symmetries of the bottom middle (red) square and is given by (\ref{squaresymm}). The matrix $M_2$ implements the symmetries of the top middle (blue) square in (\ref{Heptagon2}). It is easy to see that 
\beq
M_2=\left(
\begin{array}{cccc}
 e^{-\sigma -\frac{i \phi }{2}} & 0 & 0 & -e^{-\sigma -\frac{i \phi
   }{2}}+e^{\tau +\frac{i \phi }{2}} \\
 0 & e^{\sigma -\frac{i \phi }{2}} & 0 & 0 \\
 0 & e^{\sigma -\frac{i \phi }{2}}-e^{\frac{i \phi }{2}-\tau } & e^{\frac{i \phi
   }{2}-\tau } & -e^{\frac{i \phi }{2}-\tau }+e^{\tau +\frac{i \phi }{2}} \\
 0 & 0 & 0 & e^{\tau +\frac{i \phi }{2}}
\end{array}
\right)
\eeq
leaves the twistors (marked in blue in (\ref{Heptagon2})) of this square invariant.

\subsection*{General n-gons}
For a general n-edges polygon, the three conformal cross ratios of the $j$'th middle square $(\tau_j, \sigma_j, \phi_j)$ can be expressed in terms of the momentum twistors as
\beqa
e^{2\tau_{2j}}&\equiv&{\< {-j}, j+1,  {j+2}, {j+3}\>\< {-j-1}, {-j}, {-j+1}, {j+2}\>\over\< {-j-1}, {-j}, {j+2}, {j+3}\>\< {-j}, {-j+1}, j+1, {j+2}\>}\, ,\nn\\
e^{\sigma_{2j}+\tau_{2j}-i\phi_{2j}}&\equiv&{\< {-j-1}, {-j}, {j+1}, {j+2}\>\< {j}, {j+1}, {j+2}, {j+3}\>\over\<  {-j-1}, {j+1}, {j+2}, {j+3}\>\< {-j}, {j}, {j+1}, {j+2}\>}\, ,\la{ccr}\\
e^{\sigma_{2j}+\tau_{2j}+i\phi_{2j}}&\equiv&{\< {-j-2}, {-j-1}, {-j}, {-j+1}\>\< {-j-1}, {-j}, j+1, {j+2}\>\over\< {-j-2}, {-j-1}, {-j}, {j+2}\>\< {-j-1}, {-j}, {-j+1}, {j+1}\>}\, ,\nn\\
e^{2\tau_{2j+1}}&\equiv&{\< {-j-1}, {j+1}, {j+2}, {j+3}\>\< {-j-2}, {-j-1}, {-j}, {j+2}\>\over\< {-j-2}, {-j-1}, {j+2}, {j+3}\>\< {-j-1}, {-j}, j, {j+2}\>}\, ,\nn\\
e^{\sigma_{2j+1}+\tau_{2j+1}-i\phi_{2j+1}}&\equiv&{\< {-j-2}, {-j-1}, {-j}, {-j+1}\>\< {-j-1}, {-j}, {j+2}, {j+3}\>\over\< {-j-2}, {-j-1}, {-j}, {j+3}\>\< {-j-1}, {-j}, {-j+1}, {j+2}\>}\, ,\nn\\	
e^{\sigma_{2j+1}+\tau_{2j+1}+i\phi_{2j+1}}&\equiv&{\< {j+1}, {j+2}, {j+3}, {j+4}\>\< {-j-1}, {-j}, {j+2}, {j+3}\>\over\<  {-j-1}, {j+2}, {j+3}, {j+4}\>\< {-j}, {j+1}, {j+2}, {j+3}\>}\, ,\nn	
\eeqa
where in the convention of (\ref{OPEvsOPE}) the edge 1 is the very bottom one, the edge 2 is the next on its right, ..... Here, $\<i,j,k,l\>$ is a shorthand for $\<Z_i,Z_j,Z_k,Z_l\>$ and our definitions here are equivalent to the ones in figure 2 of \cite{short}.
Note that we have different expressions for even and odd $j$'s because the tessellation oscillates as indicated in (\ref{OPEvsOPE}). Flipping the signs of all $\phi_i$ is a symmetry of the problem and thus describes {conformally} equivalent polygons.

\section{Flipping and Gluing}\la{FlippingAp}

In the decomposition of the Wilson loop (\ref{Whex}) and (\ref{Whept}), some rapidities are flipped and some are not. In this appendix we comment on this feature which relates to the way pentagons are glued together in the decomposition.

Our convention for the sign and ordering of the rapidities entering the pentagon transitions is as follows 
\beq
\def\svgwidth{16cm} 
\begingroup%
  \makeatletter%
  \providecommand\color[2][]{%
    \errmessage{(Inkscape) Color is used for the text in Inkscape, but the package 'color.sty' is not loaded}%
    \renewcommand\color[2][]{}%
  }%
  \providecommand\transparent[1]{%
    \errmessage{(Inkscape) Transparency is used (non-zero) for the text in Inkscape, but the package 'transparent.sty' is not loaded}%
    \renewcommand\transparent[1]{}%
  }%
  \providecommand\rotatebox[2]{#2}%
  \ifx\svgwidth\undefined%
    \setlength{\unitlength}{1439.446875bp}%
    \ifx\svgscale\undefined%
      \relax%
    \else%
      \setlength{\unitlength}{\unitlength * \real{\svgscale}}%
    \fi%
  \else%
    \setlength{\unitlength}{\svgwidth}%
  \fi%
  \global\let\svgwidth\undefined%
  \global\let\svgscale\undefined%
  \makeatother%
  \begin{picture}(1,0.19270533)%
    \put(0,0){\includegraphics[width=\unitlength]{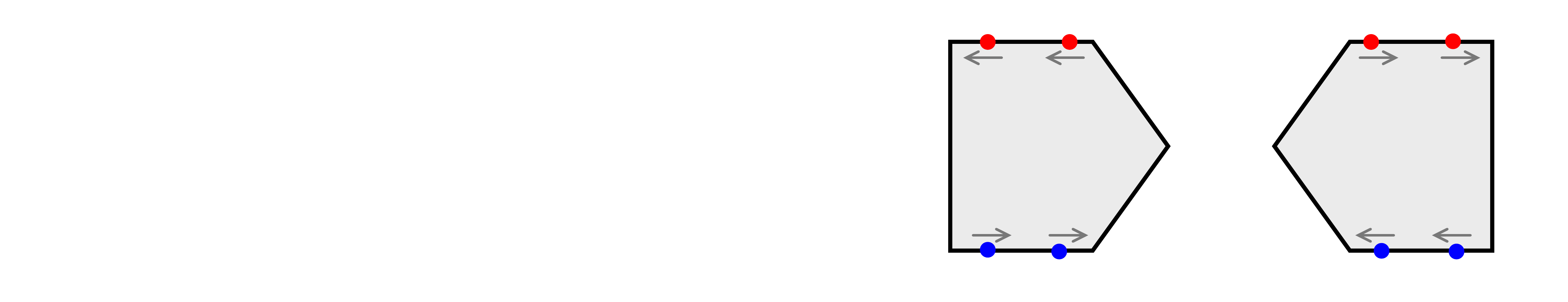}}%
    \put(-0.00093845,0.09474843){\color[rgb]{0,0,0}\makebox(0,0)[lb]{\smash{$e^{\Blue{-}i[p(u_1)+p(u_2)]\sigma_\text{Bot}}\, P(u_1,u_2|v_1,v_2)\, e^{\Blue{+}i[p(v_1)+p(v_2)]\sigma_\text{Top}}\ =$}}}%
    \put(0.76602282,0.09474843){\color[rgb]{0,0,0}\makebox(0,0)[lb]{\smash{$=$}}}%
    \put(0.6139458,0.00262592){\color[rgb]{0,0,0}\makebox(0,0)[lb]{\smash{$\text{Bottom}$}}}%
    \put(0.62728426,0.18047201){\color[rgb]{0,0,0}\makebox(0,0)[lb]{\smash{$\text{Top}$}}}%
    \put(0.61839196,0.05264513){\color[rgb]{0,0,0}\makebox(0,0)[lb]{\smash{$u_1$}}}%
    \put(0.66618809,0.05264513){\color[rgb]{0,0,0}\makebox(0,0)[lb]{\smash{$u_2$}}}%
    \put(0.62394965,0.13712202){\color[rgb]{0,0,0}\makebox(0,0)[lb]{\smash{$v_1$}}}%
    \put(0.67619194,0.13712202){\color[rgb]{0,0,0}\makebox(0,0)[lb]{\smash{$v_2$}}}%
    \put(0.87404571,0.05264513){\color[rgb]{0,0,0}\makebox(0,0)[lb]{\smash{$u_2$}}}%
    \put(0.92184185,0.05264513){\color[rgb]{0,0,0}\makebox(0,0)[lb]{\smash{$u_1$}}}%
    \put(0.86404187,0.13712202){\color[rgb]{0,0,0}\makebox(0,0)[lb]{\smash{$v_2$}}}%
    \put(0.91628416,0.13712202){\color[rgb]{0,0,0}\makebox(0,0)[lb]{\smash{$v_1$}}}%
    \put(0.86404187,0.00262592){\color[rgb]{0,0,0}\makebox(0,0)[lb]{\smash{$\text{Bottom}$}}}%
    \put(0.8796034,0.18047201){\color[rgb]{0,0,0}\makebox(0,0)[lb]{\smash{$\text{Top}$}}}%
  \end{picture}%
\endgroup%
\la{convention}
\eeq
That is, we measure the momentum flow in the bottom edge w.r.t. the direction pointing toward the middle cusp, while in the top edge this is the other way around. At both the top and the bottom, the particles are ordered according to their distance to the middle cusp. This convention is quite natural from the transition/integrability point of view, since they amount to treating the bottom excitations as incoming and the top excitations as outgoing. Within this integrability friendly convention, the bootstrap equations that determine the pentagon transitions become quite nicer, see (\ref{funrel}) in particular. 

However, from a more geometrical perspective, the above convention does not treat bottom and top in the same way. For example, the pentagon is symmetric under a reflection that interchanges its bottom and top. In our convention, this translates into 
\beq
P(u_1,\dots,u_N|v_1,\dots,v_M)=P(-v_1,\dots,-v_M|-u_1,\dots,-u_M)\, ,
\eeq
whose content is identical to the first pentagon axiom in \cite{short}. We could, of course, adopt a different convention (where all momenta flow towards the middle cusp for example) to get rid of the minus signs in this expression. The price to pay is that this would introduce some unpleasant minus signs in the fundamental relation (\ref{funrel}). 

We now explain why the integrability friendly convention (\ref{convention}) leads to the flipping of some rapidities, i.e., to the bar notation, in (\ref{Whex}) and (\ref{Whept}). Consider a square that arises from the overlap of two consecutive pentagons in the tessellation~(\ref{OPEvsOPE}). Moving the excitations at the bottom of this square in a given direction is conformally equivalent to moving the excitations at the top in the opposite direction. This is how the coordinate $\sigma$ parameterizes the overall position of the excitations:
\beq
\qquad\def\svgwidth{3cm} 
\begingroup%
  \makeatletter%
  \providecommand\color[2][]{%
    \errmessage{(Inkscape) Color is used for the text in Inkscape, but the package 'color.sty' is not loaded}%
    \renewcommand\color[2][]{}%
  }%
  \providecommand\transparent[1]{%
    \errmessage{(Inkscape) Transparency is used (non-zero) for the text in Inkscape, but the package 'transparent.sty' is not loaded}%
    \renewcommand\transparent[1]{}%
  }%
  \providecommand\rotatebox[2]{#2}%
  \ifx\svgwidth\undefined%
    \setlength{\unitlength}{185.29887695bp}%
    \ifx\svgscale\undefined%
      \relax%
    \else%
      \setlength{\unitlength}{\unitlength * \real{\svgscale}}%
    \fi%
  \else%
    \setlength{\unitlength}{\svgwidth}%
  \fi%
  \global\let\svgwidth\undefined%
  \global\let\svgscale\undefined%
  \makeatother%
  \begin{picture}(1,1.70520006)%
    \put(0,0){\includegraphics[width=\unitlength]{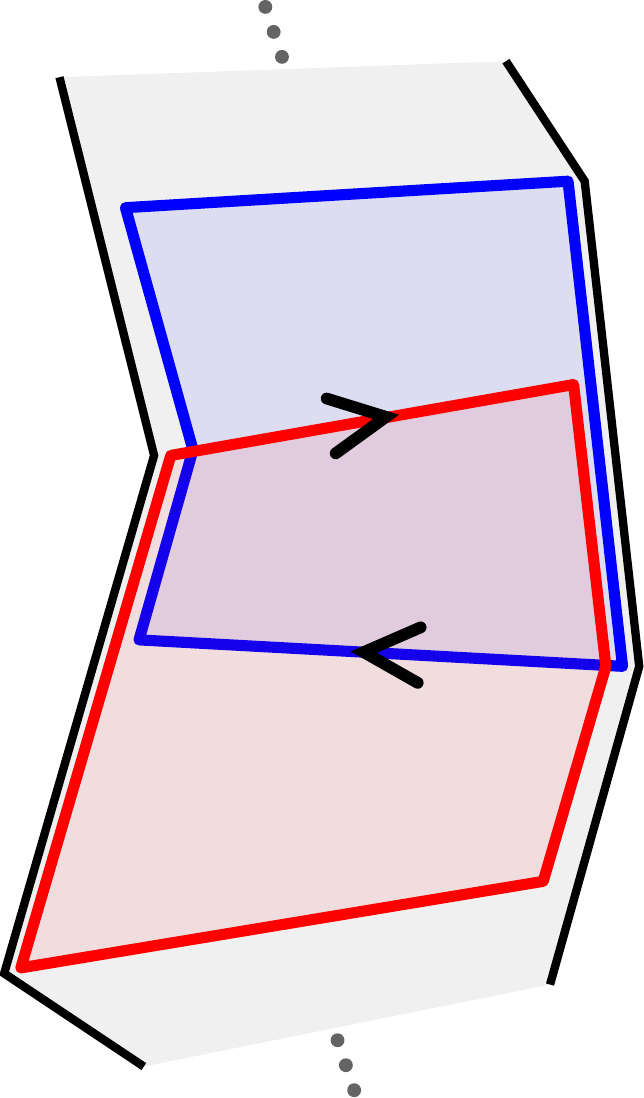}}%
    \put(0.55341042,0.83032047){\color[rgb]{0,0,0}\makebox(0,0)[lb]{\smash{$\sigma$}}}%
  \end{picture}%
\endgroup%
\nn
\eeq
An excitation moving to the right at the bottom will of course keep moving in the same direction when it reaches the top. However, due to the way $\sigma$ acts, it will be viewed at the top with the reversed momentum. Similarly, if we have more than than one excitation in the eigenstate, they are viewed from the top as if their momenta were flipped. 
For illustration, consider a sequence of two transitions involving two particles (the generalization to any number of particles is obvious):
\beq
\qquad\def\svgwidth{11cm} 
\begingroup%
  \makeatletter%
  \providecommand\color[2][]{%
    \errmessage{(Inkscape) Color is used for the text in Inkscape, but the package 'color.sty' is not loaded}%
    \renewcommand\color[2][]{}%
  }%
  \providecommand\transparent[1]{%
    \errmessage{(Inkscape) Transparency is used (non-zero) for the text in Inkscape, but the package 'transparent.sty' is not loaded}%
    \renewcommand\transparent[1]{}%
  }%
  \providecommand\rotatebox[2]{#2}%
  \ifx\svgwidth\undefined%
    \setlength{\unitlength}{585.29887695bp}%
    \ifx\svgscale\undefined%
      \relax%
    \else%
      \setlength{\unitlength}{\unitlength * \real{\svgscale}}%
    \fi%
  \else%
    \setlength{\unitlength}{\svgwidth}%
  \fi%
  \global\let\svgwidth\undefined%
  \global\let\svgscale\undefined%
  \makeatother%
  \begin{picture}(1,0.53984668)%
    \put(0,0){\includegraphics[width=\unitlength]{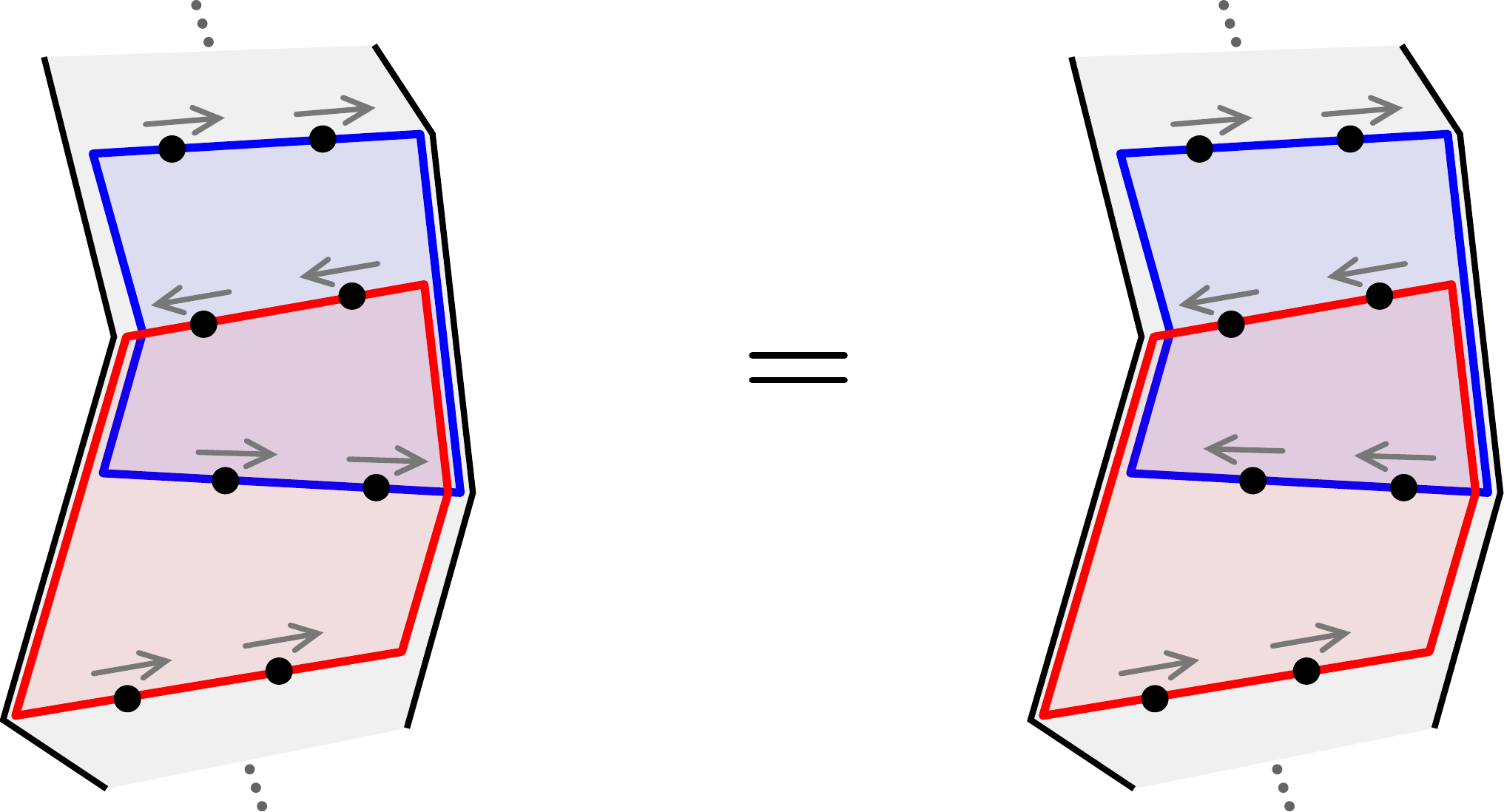}}%
    \put(0.15261082,0.12589836){\color[rgb]{0,0,0}\makebox(0,0)[lb]{\smash{$u_2$}}}%
    \put(0.05384259,0.10818589){\color[rgb]{0,0,0}\makebox(0,0)[lb]{\smash{$u_1$}}}%
    \put(0.20728374,0.37466015){\color[rgb]{0,0,0}\makebox(0,0)[lb]{\smash{$v_2$}}}%
    \put(0.11124916,0.35694768){\color[rgb]{0,0,0}\makebox(0,0)[lb]{\smash{$v_1$}}}%
    \put(0.22945666,0.24891243){\color[rgb]{0,0,0}\makebox(0,0)[lb]{\smash{$v_2$}}}%
    \put(0.12856219,0.25398037){\color[rgb]{0,0,0}\makebox(0,0)[lb]{\smash{$v_1$}}}%
    \put(0.17994728,0.47853871){\color[rgb]{0,0,0}\makebox(0,0)[lb]{\smash{$w_2$}}}%
    \put(0.08391269,0.47176082){\color[rgb]{0,0,0}\makebox(0,0)[lb]{\smash{$w_1$}}}%
    \put(0.83602226,0.12589836){\color[rgb]{0,0,0}\makebox(0,0)[lb]{\smash{$u_2$}}}%
    \put(0.73725412,0.10818589){\color[rgb]{0,0,0}\makebox(0,0)[lb]{\smash{$u_1$}}}%
    \put(0.89069519,0.37466015){\color[rgb]{0,0,0}\makebox(0,0)[lb]{\smash{$v_2$}}}%
    \put(0.79466068,0.35694768){\color[rgb]{0,0,0}\makebox(0,0)[lb]{\smash{$v_1$}}}%
    \put(0.89099902,0.24891243){\color[rgb]{0,0,0}\makebox(0,0)[lb]{\smash{$-v_2$}}}%
    \put(0.7928382,0.25398037){\color[rgb]{0,0,0}\makebox(0,0)[lb]{\smash{$-v_1$}}}%
    \put(0.86335873,0.47853871){\color[rgb]{0,0,0}\makebox(0,0)[lb]{\smash{$w_2$}}}%
    \put(0.76732422,0.47176082){\color[rgb]{0,0,0}\makebox(0,0)[lb]{\smash{$w_1$}}}%
    \put(0.43721376,0.24891243){\color[rgb]{0,0,0}\makebox(0,0)[lb]{\smash{$\text{convention}$}}}%
  \end{picture}%
\endgroup%
\nn
\eeq
The dependence on $\sigma$ in the middle square only enters through the phase factor $e^{i\sigma[p(v_1)+p(v_2)]}$. On the right, we have flipped the sign of the rapidities and the direction of the arrows of all the bottom excitations of the top transition, such as to fit  this transition with our convention~(\ref{convention}). This way we see that the phase factor is multiplied by the bottom transition $P(\dots|v_1,v_2)=P(\dots|{\bf v})$ and by the top transition $P(-v_2,-v_1|\dots)=P(\bar{\bf v}|\dots)$, hence explaining the notation used in~(\ref{Whex}) and (\ref{Whept}).

\section{Gluing Pentagon Transitions at Born Level}\la{gluingappendix}

In this paper we focused on single-particle transitions and measures, see for instance~(\ref{mu-P-BL}) for a scalar at tree level. The OPE decomposition for polygons with more than seven edges includes more than two middle squares. Such polygons involve therefore several transitions between general states which are combined as in (\ref{decompositionIntro}). We will now illustrate how two such transitions are nicely glued together in the case of the leading OPE contribution of an octagon NMHV component.
 
An octagon has three middle squares and therefore nine OPE parameters (or conformal cross ratios) $\{\tau_i,\sigma_i,\phi_i\}$ with $i=1,2,3$. Consider the scalar NMHV component ${\cal W}_\text{oct}^{(8145)}$ at tree level
\beq \la{OctagonNMHV}
\def\svgwidth{5cm}
\begingroup%
  \makeatletter%
  \providecommand\color[2][]{%
    \errmessage{(Inkscape) Color is used for the text in Inkscape, but the package 'color.sty' is not loaded}%
    \renewcommand\color[2][]{}%
  }%
  \providecommand\transparent[1]{%
    \errmessage{(Inkscape) Transparency is used (non-zero) for the text in Inkscape, but the package 'transparent.sty' is not loaded}%
    \renewcommand\transparent[1]{}%
  }%
  \providecommand\rotatebox[2]{#2}%
  \ifx\svgwidth\undefined%
    \setlength{\unitlength}{365.63137207bp}%
    \ifx\svgscale\undefined%
      \relax%
    \else%
      \setlength{\unitlength}{\unitlength * \real{\svgscale}}%
    \fi%
  \else%
    \setlength{\unitlength}{\svgwidth}%
  \fi%
  \global\let\svgwidth\undefined%
  \global\let\svgscale\undefined%
  \makeatother%
  \begin{picture}(1,1.04993977)%
    \put(0,0){\includegraphics[width=\unitlength]{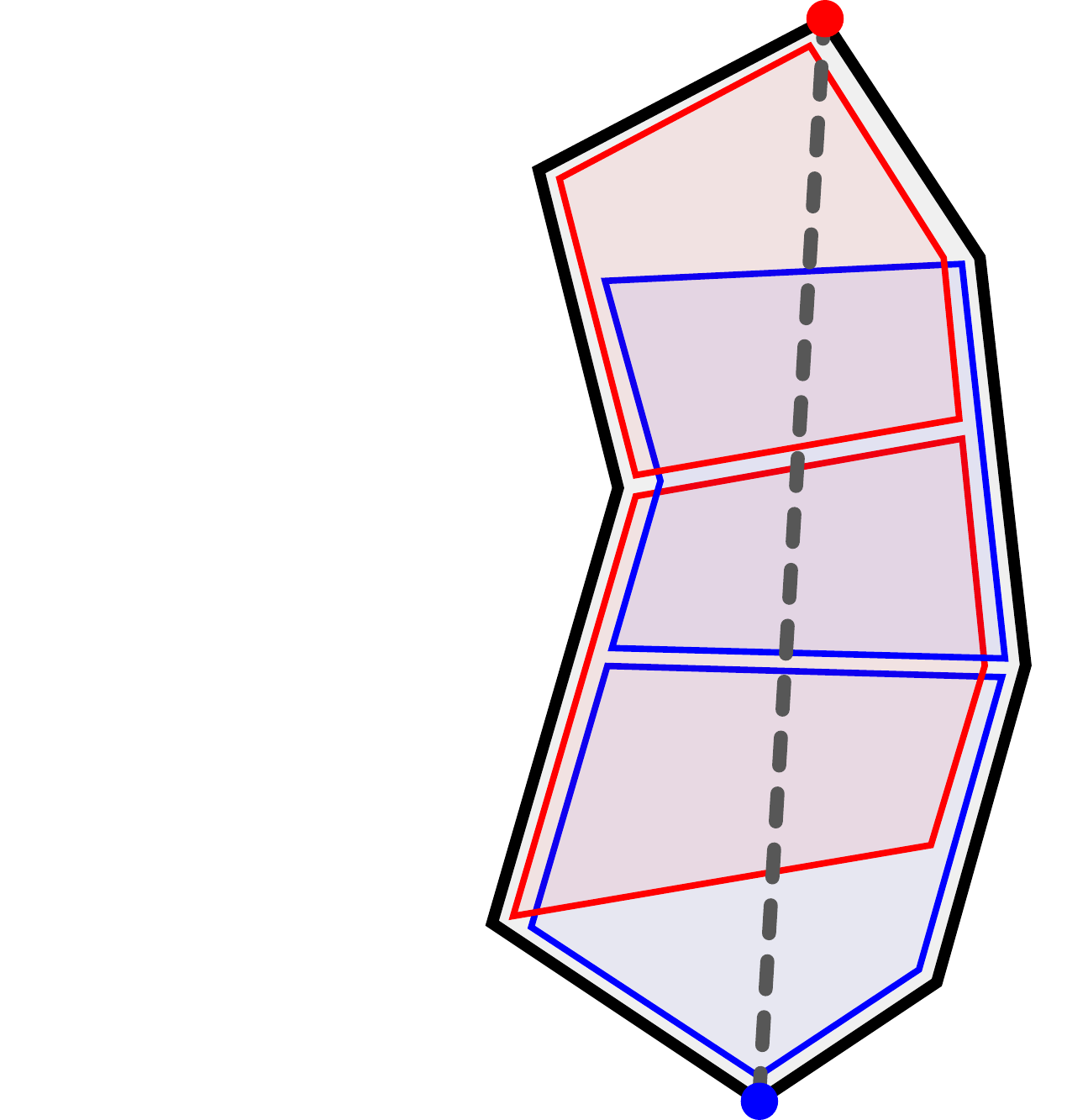}}%
    \put(-0.00257239,0.54538828){\color[rgb]{0,0,0}\makebox(0,0)[lb]{\smash{$\displaystyle {\cal W}^{(8145)}_\text{tree}=$}}}%
    \put(0.81129991,0.00865378){\color[rgb]{0.50196078,0.50196078,0.50196078}\makebox(0,0)[lb]{\smash{$1$}}}%
    \put(0.94695567,0.23182937){\color[rgb]{0.50196078,0.50196078,0.50196078}\makebox(0,0)[lb]{\smash{$2$}}}%
    \put(0.9530835,0.61952668){\color[rgb]{0.50196078,0.50196078,0.50196078}\makebox(0,0)[lb]{\smash{$3$}}}%
    \put(0.86532381,0.92036265){\color[rgb]{0.50196078,0.50196078,0.50196078}\makebox(0,0)[lb]{\smash{$4$}}}%
    \put(0.58767507,0.98024528){\color[rgb]{0.50196078,0.50196078,0.50196078}\makebox(0,0)[lb]{\smash{$5$}}}%
    \put(0.45922405,0.71741052){\color[rgb]{0.50196078,0.50196078,0.50196078}\makebox(0,0)[lb]{\smash{$6$}}}%
    \put(0.53514073,0.03334031){\color[rgb]{0.50196078,0.50196078,0.50196078}\makebox(0,0)[lb]{\smash{$8$}}}%
    \put(0.45545489,0.39398627){\color[rgb]{0.50196078,0.50196078,0.50196078}\makebox(0,0)[lb]{\smash{$7$}}}%
  \end{picture}%
\endgroup%

\eeq
The leading contribution at large $\tau_1$, $\tau_2$ and $\tau_3$ comes from a single scalar propagating through each of the three channels. We find
\beq\la{octagonNMHV}
{\cal W}_\text{oct}^{(8145)}={1\over\<8,1,4,5\>}={e^{-\tau_1-\tau_2-\tau_3}\over e^{\sigma_1+\sigma_2+\sigma_3}+e^{\sigma_2+\sigma_3-\sigma_1}+e^{\sigma_1+\sigma_3-\sigma_2}+e^{\sigma_1+\sigma_2-\sigma_3}+e^{\sigma_2-\sigma_1-\sigma_3}}+\cO(e^{-2\tau_i})
\eeq 
On the other hand, we expect this contribution to be equal to
\beq\la{octagondecomp}
{\cal W}_\text{oct}^{(8145)}=\int{du\,dv\,dw\over g^2(2\pi)^3}\,e^{-\tau_1-\tau_2-\tau_3+2iu\sigma_1+2iv\sigma_2+2iw\sigma_3}\mu(u)\,P(-u|v)\mu(v)P(-v|w)\mu(w)
+\cO(e^{-2\tau_i})\, ,
\eeq
where the overall factor $1/g^2$ comes from the form factors for creation/annihilation of a scalar, $P_*(0|u) = P_*(-w|0) = 1/g$, and with the contour $\mathbb{R}+i0^+$ for each one of the three integrals. By plugging the tree level results (\ref{mu-P-BL}) obtained from the heptagon and hexagon one can easily check that, indeed, (\ref{octagondecomp}) is the Fourier transform of~(\ref{octagonNMHV}). 

\section{The Mirror of a Gluon}\la{FmirrorApp}
In this appendix we determine the transformation property of the gauge field excitation $F$ (or equivalently $\bar F$) under analytic continuation to the mirror sheet. The most natural expectation is that under this transformation $F\to F$. However, as stated in section~\ref{IntG}, this is not the case and instead under the mirror rotation $F\to\bar F$. To derive this relation, we start with the more general transformation rule
 \beq\la{Fmirror}
|\psi(u^{-\gamma})\>_{\textrm{edge I}} =|\varphi(u)\>_\text{edge II}
\eeq
where $|\psi(u^{-\gamma})\>_{\textrm{edge I}} $ and $|\varphi(u)\>_\text{edge II}$ are arbitrary excitations inserted on the edge I and II, with the latter standing on the right side of the former.  Equation (\ref{Fmirror}) must be consistent with the symmetries of the square, that is, with the symmetries of the background on which the excitations are living. In other words, both sides on (\ref{Fmirror}) must transform in the same way under a conformal transformation that leave the flux invariant. We recall that the square have three conformal symmetries, conjugate to the flux energy (twist), momentum (conformal spin) and U(1) charge. A convenient description of any null polygon, which has the virtue of linearizing the action of the conformal generators, is found by using momentum twistors \cite{Hodges:2009hk}. This representation, in the specific form relevant to the OPE, is presented in detail in appendix \ref{geomeryappendix}. In particular, the square and its conformal symmetries are given by
\beq
M=\left(
\begin{array}{cccc}
 e^{\sigma-\frac{i\phi }{2}} & 0 & 0 & 0 \\
 0 & e^{-\sigma-\frac{i\phi }{2}} & 0 & 0 \\
 0 & 0 & e^{\tau+\frac{i\phi }{2}} & 0 \\
 0 & 0 & 0 & e^{-\tau+\frac{i\phi }{2}}
\end{array}
\right)\quad\text{and}\quad\(\!\! \begin{array}{l}
{\cal Z}_\text{bot}\\
{\cal Z}_\text{right}\\
{\cal Z}_\text{top}\\
{\cal Z}_\text{left}
\end{array}\!\!\)=
\left(
\begin{array}{cccc}
 0 & 0 & 1 & 0 \\
 1 & 0 & 0 & 0 \\
 0 & 0 & 0 & 1 \\
 0 & 1 & 0 & 0
\end{array}
\right)
\eeq
where the conformal transformation $M(\tau,\sigma,\phi)$ acts on the four momentum twistors of the square by multiplication ${\cal Z}_i\to{\cal Z}_i\cdot M$.

The mirror rotation $u\to u^{-\gamma}$ in (\ref{Fmirror}) maps an insertion on the bottom edge to an insertion on the right neighbouring edge. If we now cyclicly relabel the edges of the square, so that the right edge becomes the bottom edge, etc., we do not change its symmetries but simply relabel them. This particular cyclic rotation of the twistors is obtained by acting with the matrix
\beq
{\mathbb R}=\left(
\begin{array}{cccc}
 0 & 0 & 0 & 1 \\
 0 & 0 & 1 & 0 \\
 1 & 0 & 0 & 0 \\
 0 & 1 & 0 & 0 \\
\end{array}
\right)\qquad\text{such that}\qquad\(\!\! \begin{array}{l}
{\cal Z}_\text{right}\\
{\cal Z}_\text{top}\\
{\cal Z}_\text{left}\\
{\cal Z}_\text{bot}
\end{array}\!\!\)=\(\!\! \begin{array}{l}
{\cal Z}_\text{bot}\\
{\cal Z}_\text{right}\\
{\cal Z}_\text{top}\\
{\cal Z}_\text{left}
\end{array}\!\!\)\cdot{\mathbb R}\, .
\eeq
The symmetry generators acts on the cyclic rotated twistors as
\beq\la{gentransform}
{\mathbb R^{-1}}\cdot M(\tau,\sigma,\phi)\cdot{\mathbb R} =M(-\sigma,\tau,-\phi)\equiv M(\tilde\tau,\tilde\sigma,\tilde\phi)\, .
\eeq
In other words, the space coordinate $\sigma$ becomes minus the time coordinate when we the square is viewed from the right and, similarly, the time coordinate $\tau$ becomes the space coordinate. This, of course, could be more simply derived by drawing a picture of a square, labelled with time and space direction, and performing a direct $90^{\circ}$ rotation. What is crucial for the analysis here, and is somewhat less intuitive, is that the $U(1)$ angle $\phi$ has its sign reversed $\phi\rightarrow -\phi$ in the process.

Now let $-E(u)$, $ip(u)$ and $im$ be the charge conjugate to $\partial_{\tau}$, $\partial_{\sigma}$ and $\partial_{\phi}$, respectively. By acting with these generators on both sides of (\ref{Fmirror}) and demanding for consistency that both sides transform in the same way, we conclude that
\beq\label{mirror-transf-charge}
e^{-E_\varphi(u)\tilde\tau+ip_\varphi(u)\tilde\sigma+im_\varphi\tilde\phi}\overset{\Blue{!}}{=}e^{-E_\psi(u^{-\gamma})\tau+ip_\psi(u^{-\gamma})\sigma+im_\psi\phi}\overset{(\ref{gentransform})}{=}
e^{-E_\psi(u^{-\gamma})\tilde\sigma-ip_\psi(u^{-\gamma})\tilde\tau-im_\psi\tilde\phi}\, ,
\eeq
and therefore
\beq
E_\psi(u^{-\gamma})=-ip_\varphi(u)\ ,\quad p_\psi(u^{-\gamma})=-iE_\varphi(u)\, ,\quad\text{and}\quad m_\psi=-m_\varphi\, .
\eeq
Finally, we know that for a gauge field, i.e, $\psi = F$ or $\bar{F}$, we have~\cite{MoreDispPaper} that $E_\psi(u^{-\gamma})=-ip_\psi(u)$ and $p_\psi(u^{-\gamma})=-iE_\psi(u)$. We therefore conclude that $\varphi$ has the same dispersion relation as $\psi$ but carries opposite $U(1)$ charge. In other words, if we start with the gluonic excitation $\psi=\Blue{F}$ then its mirror excitation is $\varphi=\Red{\bar F}$, and reciprocally. 

We close this appendix with a comment on the mirror rotation of the twist one fermions. For them the construction above lends support to the thesis according to which there cannot exist an analytic continuation to the mirror sheet, which would map a fermion back to itself. This is in line with the difficulties encountered in~\cite{MoreDispPaper} to make sense of such a transformation at finite coupling. The argument goes as follows. Suppose there is an analytic continuation $\gamma$ that transforms a twist one fermion on one edge into some excitation on the neighbouring edge. We thus start on the left hand side of (\ref{Fmirror}) with a twist one fermion on edge I. Among the two kinds of twist one fermions we choose the one with $U(1)$ charge equal to $+1/2$ for definiteness (the other possible choice being $-1/2$). By applying~(\ref{mirror-transf-charge}), we are led to conclude that the state on the right hand side of (\ref{Fmirror}), i.e., $|\varphi(u)\>_\text{edge II}$, has the same R-charge as the fermion, since $SU(4)$ rotations should commute with the mirror transformation, but carries opposite $U(1)$ charge, equal to $-1/2$. 
The problem is that in ${\cal N}=4$ SYM the $U(1)$ charge of the twist one fermion is correlated with its R-charge, in such a way that we cannot flip the former without changing the latter. As a result, the state $|\varphi(u)\>_\text{edge II}$ cannot be a twist one fermion on edge II. This obstruction raises many interesting questions regarding the crossing properties of fermions, which we hope to be able to address in the future.

\section{Expanding the Ans\"atze at Weak Coupling}\la{ansatz-App}

In this section we explain how to generate the expressions for the scalar and gluon pentagon transitions order by order in perturbation theory. The starting points are the general formulae for the S-matrix of scalar and gluon. For the scalar they were extracted in~\cite{toappearAdam,Peng} and are recalled below for the reader's convenience. The formulae for gluon were obtained by following a similar approach. We shall give the general expressions in terms of certain functions $f_{1,2,3,4}(u, v)$, satisfying
\beq\label{f-gen-prop}
f_{1}(u, v) = f_{2}(v, u)\, , \qquad f_{3, 4}(u, v) = f_{3,4}(v, u)\, .
\eeq
These are the main dynamical quantities that parameterize the S-matrix on top of the GKP background. We shall then explain how to compute these functions order by order in perturbation.

\subsection*{Formulae for Scalar}

The S-matrix for scalar (and its mirror-rotated version) can be parameterized as
\beq
\begin{aligned}
S(u, v) &= \frac{\Gamma(\frac{1}{2}-iu)\Gamma(\frac{1}{2}+iv)\Gamma(iu-iv)}{\Gamma(\frac{1}{2}+iu)\Gamma(\frac{1}{2}-iv)\Gamma(iv-iu)}F(u, v)\, ,\\
S(u^{\gamma}, v) &= \frac{\pi g^2\sinh{(\pi(u-v))}}{(u-v+i)\cosh{(\pi u)}\cosh{(\pi v)}}G(u, v)\, .\\
\end{aligned} \la{smat1}
\eeq 
In this way of writing the leading order expressions are factored out explicitly and all the higher-loop corrections are absorbed into the functions $F(u, v)$ and $G(u, v)$. As a consequence the latter functions start with $1$ when $g\rightarrow 0$. They read explicitly
\beq\label{FGscalar}
\begin{aligned}
&\log{F} = 2i\int\limits_{0}^{\infty}\frac{dt}{t}(J_{0}(2gt)-1)\frac{e^{t/2}(\sin{(ut)}-\sin{(vt)})}{e^{t}-1} -2if_{1}+2if_{2}\, ,\\
&\log{G} = 2\int\limits_{0}^{\infty}\frac{dt}{t}(J_{0}(2gt)-1)\frac{e^{t/2}(\cos{(ut)}+\cos{(vt)})-J_{0}(2gt)-1}{e^{t}-1} +2f_{3}-2f_{4}\, ,\\
\end{aligned}
\eeq
where $J_{0}(z) = 1+O(z^2)$ is the zero-th Bessel function and we note that $F(u, v) = 1/F(v, u),$ $G(u, v) = G(v, u)$.

The scalar pentagon transition and measure are related to the S-matrix by our ansatz (\ref{conj-scalar}-\ref{mu2}). After combining it with~(\ref{FGscalar}), one immediately obtains the following representation for the scalar pentagon transition
\beq
\begin{aligned}
&P(u|v) = \frac{\Gamma(iu-iv)}{g^2\Gamma(\frac{1}{2}+iu)\Gamma(\frac{1}{2}-iv)}\times \\
&\exp{\bigg[\int\limits_{0}^{\infty}\frac{dt}{t}(J_{0}(2gt)-1)\frac{J_{0}(2gt)+1-e^{t/2}(e^{-iut}+e^{ivt})}{e^{t}-1} -if_1 + if_2-f_3+f_4\bigg]}\, .
\end{aligned}
\eeq
For the measure we have 
\beq
\begin{aligned}
&\mu(u) =\frac{\pi g^2}{\cosh{(\pi u)}}\times\\
&\exp{\bigg[\int\limits_{0}^{\infty}\frac{dt}{t}(J_{0}(2gt)-1)\frac{2e^{t/2}\cos(ut)-J_{0}(2gt)-1}{e^{t}-1} + f_3(u, u)-f_4(u, u)\bigg]}\, ,
\end{aligned}
\eeq
where we used that $f_1(u, u) - f_{2}(u, u) = 0$, see~(\ref{f-gen-prop}).

\subsection*{Formulae for Gluon}

Similarly, we can write
\beq
\begin{aligned}
S(u, v) &= \frac{\Gamma(\frac{3}{2}-iu)\Gamma(\frac{3}{2}+iv)\Gamma(iu-iv)}{\Gamma(\frac{3}{2}+iu)\Gamma(\frac{3}{2}-iv)\Gamma(iv-iu)}F(u, v)\, ,\\
S(u^{\gamma}, v) &= \frac{\pi g^2\sinh{(\pi(u-v))}}{(u-v-i)\cosh{(\pi u)}\cosh{(\pi v)}}G(u, v)\, ,\\
\end{aligned} \la{smat2}
\eeq
for the gluons, where now
\beq\la{gluonFG}
\begin{aligned}
&\log{F} = 2i\int\limits_{0}^{\infty}\frac{dt}{t}(J_{0}(2gt)-1)\frac{e^{-t/2}(\sin{(ut)}-\sin{(vt)})}{e^{t}-1} -2if_{1}+2if_{2}\, ,\\
&\log{G} = 2\int\limits_{0}^{\infty}\frac{dt}{t}(J_{0}(2gt)-1)\frac{e^{t/2}(\cos{(ut)}+\cos{(vt)})-J_{0}(2gt)-1}{e^{t}-1} +2f_{3}-2f_{4}\, .\\
\end{aligned}
\eeq
Our ansatz for the gluon transitions and measure in terms of the S-matrix (\ref{smat2}) are given in (\ref{finalP}), (\ref{measureEq}). Using (\ref{gluonFG}) we obtain 
\beq
\begin{aligned}
&P(u|v) = -\frac{\Gamma(iu-iv)}{g^2\Gamma(\frac{3}{2}+iu)\Gamma(\frac{3}{2}-iv)}\sqrt{\left(x^{+}y^{-}-g^2\right)\left(x^{-}y^{+}-g^2\right)\left(x^{+}y^{+}-g^2\right)\left(x^{-}y^{-}-g^2\right)}\times \\
&\exp{\bigg[\int\limits_{0}^{\infty}\frac{dt}{t}(J_{0}(2gt)-1)\frac{J_{0}(2gt)+1-e^{-t/2}(e^{-iut}+e^{ivt})}{e^{t}-1} -if_1 + if_2-f_3+f_4\bigg]}\, . 
\end{aligned} \la{transitionPG}
\eeq
for the transition, where we recall that $x^{\pm} = x(u\pm \ft{i}{2})$, $y^{\pm} = x(v\pm \ft{i}{2})$ and $x(u) = \ft{1}{2}(u+\sqrt{u^2-4g^2})$. For the measure we have
\beq\la{measureansatz}
\begin{aligned}
&\mu(u) =-\frac{\pi g^2}{\cosh{(\pi u)}}\frac{(u^2+\ft{1}{4})}{(x^{+}x^{-}-g^2)\sqrt{(x^+x^{+}-g^2)(x^-x^{-}-g^2)}}\times\\
&\exp{\bigg[\int\limits_{0}^{\infty}\frac{dt}{t}(J_{0}(2gt)-1)\frac{2e^{-t/2}\cos(ut)-J_{0}(2gt)-1}{e^{t}-1} +f_3(u, u)-f_4(u, u)\bigg]}\, .
\end{aligned}
\eeq
We now explain how to compute the functions $f_i$.

\subsection*{Computing the $f$-functions}

A particular way of presenting the $f$-functions, which is most convenient for expanding them in perturbation theory, makes use of two auxiliary vectors $\kappa_i(u)$, $\tilde \kappa_j(u)$ (that depend on the coupling and also on the rapidity $u$) and of a matrix $K_{ij}$ (which only depends on the coupling). The indices take values over all positive integers. In perturbation theory we can effectively truncate the range of the indices as explained below. The matrix elements are given by 
\beq
K_{ij}=2j(-1)^{j(i+1)} \int\limits_{0}^\infty \frac{dt}{t} \frac{J_i(2gt)J_j(2gt)}{e^t-1} \, , \la{Kdef}
\eeq
where $J_i$ is the $i$-th Bessel function. This same matrix $K$ is useful for computing the $f$-functions for the scalar as well as for the gluon excitations. It corresponds to the kernel of the Beisert-Eden-Staudacher equation~\cite{BES}, when written in the manner of~\cite{Benna}, and it is universal, in that it is the same for all the excitations of the GKP background~\cite{BenDispPaper}.
The vectors $\kappa,\tilde\kappa$, on the other hand, depend on which case we are interested in, but, actually, in a very minimal way. In fact, we can introduce a parameter $\eta$ such that $\eta=0$ for scalars and $\eta=1$ for gluons and describe $f_i$ for both cases at once. We have
\beqa
\kappa_{j}(u) &\equiv&  -\int\limits_{0}^\infty \frac{dt}{t} \frac{J_j(2gt)(J_0(2gt)-\cos(ut)\[e^{t/2}\]^{(-1)^{\eta \times j}})}{e^t-1} \nn  \\
\tilde\kappa_{j}(u) &\equiv&  -\int\limits_{0}^\infty \frac{dt}{t} (-1)^{j+1}\frac{J_j(2gt) \sin(u t) \[e^{t/2}\]^{(-1)^{\eta \times(j+1)}}}{e^t-1}  \nn
\eeqa
Next we construct the inverse of the identity matrix plus the matrix $K$, 
\beq
\mathbb{M} \equiv (1+K)^{-1}=1-K+K^2-K^3+\dots \la{Mdef}
\eeq
together with a trivial diagonal matrix $\mathbb{Q}$ with entries $\mathbb{Q}_{ij}=\delta_{ij}(-1)^{i+1}i$. The functions $f_i$ are then given by 
\beqa\label{fkAMk}
&&f_1(u,v)=2 \, \tilde \kappa(u) \cdot \mathbb{Q} \cdot  \mathbb{M} \cdot  \kappa(v) \, , \qquad f_2(u,v)=2 \, \tilde \kappa(v) \cdot \mathbb{Q} \cdot  \mathbb{M} \cdot  \kappa(v)\\
&&f_3(u,v)=2 \, \tilde \kappa(u) \cdot \mathbb{Q} \cdot  \mathbb{M} \cdot  \tilde\kappa(v) \, , \qquad f_4(u,v)=2 \,  \kappa(v) \cdot \mathbb{Q} \cdot  \mathbb{M} \cdot  \kappa(v)
\eeqa
\subsection*{Perturbation theory}
In perturbation theory $K_{ij}=O(g^{i+j})$ so that we can truncate the range of the indices to go over $i,j=1,2,\dots,\Lambda-1$ and get accurate results to order $g^{\Lambda}$. The inverse $\mathbb{M}$ is also trivial to compute exactly in perturbation theory since we can truncate (\ref{Mdef}). For example, to get results up to order $g^4$ we simply need to use 
\beq
\mathbb{Q}\cdot \mathbb{M}=\left(
\begin{array}{ccc}
 \frac{11 \pi ^4 g^4}{45}-\frac{\pi ^2 g^2}{3}+1 & -4 g^3 \zeta (3) & -\frac{1}{15} g^4 \pi ^4 \\
 -4 g^3 \zeta (3) & \frac{2 g^4 \pi ^4}{15}-2 & 0 \\
 -\frac{1}{15} g^4 \pi ^4 & 0 & 3 \\
\end{array}
\right) \la{QMexample}
\eeq
and compute the first three components of the vectors $\kappa$ and $\tilde\kappa$. For scalars the first component of $\kappa$ reads
\beqa
\kappa_1(u)&=&\frac{g}{2}\[\psi ^{(0)}\left(\tfrac{1}{2}-iu\right)+\psi ^{(0)}\left(\tfrac{1}{2}+i u\right)-2\psi^{(0)}(1) \]\\
&-&{g^3\over4}\[\psi ^{(2)}\left(\tfrac{1}{2}-i u\right)+\psi ^{(2)}\left(\tfrac{1}{2}+iu\right)+12 \zeta (3)\] + O(g^5)\, , \nn
\eeqa
for instance. Note that to compute any component of the vectors $\kappa,\tilde \kappa$ or any element of the matrix $K$ in perturbation theory all we need to do is to Taylor expand the Bessel functions at small~$g$. This generates powers of $t$ in the integrands of the relevant integrals, which always boil down to instances of
\beq
\int_{0}^\infty dt \frac{e^{iut}}{e^t-1} t^n = (-1)^{n+1} \psi^{(n)}(1-iu) \,, \qquad \psi^{(n-1)}(x) \equiv \frac{d^{n}}{dx^{n}} \log \Gamma(x) \,. \la{gammaInt}
\eeq
In fact, by inspecting the remaining contributions to the pentagon transitions, it is easy to see that this is all we need for computing the expansion of these objects to any order in perturbation theory, see for example the first terms in the second line in (\ref{transitionPG}). Hence, it is trivial to systematize this expansion to any desired order in perturbation theory. We attach a \verb"mathematica"  notebook where the functions $f_i$ are computed to any desired order. 

\subsection*{A few comments} 
We conclude with a few remarks.
\begin{itemize}
\item The matrix $\mathbb{Q}\cdot \mathbb{M}$ is symmetric. This follows from the fact that $K$ as defined in (\ref{Kdef}) is clearly of the form $K=K_\text{sym}\cdot \mathbb{Q}$ where $K_\text{sym}$ is a symmetric matrix. Hence $$\mathbb{Q}\cdot \mathbb{M}= \mathbb{Q}- \mathbb{Q} \cdot K_\text{sym}\cdot \mathbb{Q}+\mathbb{Q} \cdot K_\text{sym}\cdot \mathbb{Q} \cdot K_\text{sym}\cdot \mathbb{Q}- \dots$$ which is clearly symmetric sincel both $\mathbb{Q}$ and $K_\text{sym}$ are. The illustration given in (\ref{QMexample}) clearly exhibits this property. From this and from~(\ref{fkAMk}) the general properties~(\ref{f-gen-prop}) of the $f$-functions follow. These relations are quite useful in establishing several properties of the pentagon transitions as well as of the flux tube S-matrices. For example, unitarity of the physical S-matrices in (\ref{smat1}) or (\ref{smat2}) readily follows from these symmetry properties for $f_1$ and $f_2$. 
\item The upper left corner of the matrix $\mathbb{Q}\cdot \mathbb{M}$ is a quite physical quantity: it is nothing but the cusp anomalous dimension, up to an overall factor of $4g^2$. For example, by multiplying the upper left corner in the example (\ref{QMexample}) with $4g^2$ we obtain the three-loop cusp anomalous dimension 
\beq
\Gamma_\text{cusp}(g)=4g^2-\frac{4}{3}\pi^2 g^4+ \frac{44}{45} \pi^4 g^6+\mathcal{O}(g^8)\, .
\eeq
\item The energy and momentum can also be read from the quantities considered here, as already presented in~\cite{BenDispPaper}. We have
\beq\la{EnergyMomenta}
E(u)=1+ 4 g \( \mathbb{Q}\cdot \mathbb{M} \cdot \kappa(u) \)_1  \, , \qquad p(u) =2u- 4 g \( \mathbb{Q}\cdot \mathbb{M} \cdot \tilde\kappa(u) \)_1 \, .
\eeq
These expressions are valid at any coupling but they are particularly trivial to evaluate at weak coupling as explained above. 
\end{itemize}

\section{OPE in Position Space for the Scalar NMHV Hexagon} \la{FnlAppendix}
All the expressions in this appendix can be found in the compaying notebook \verb"Fnl.nb", where use is made of the notations of the \verb"HPL" package \cite{HPLpackage}.

We have ($H=H(-x)$ and $\bar H=H(+x)$)
\begin{eqnarray*}
F_0^{(0)}&=&1\\ 
F_1^{(1)}&=&-2 \bar{H}_0-4 H_1 \\
F_0^{(1)}&=& -2 H_1 \bar{H}_0-2 H_1^2 \\
F_2^{(2)}&=&8 H_1 \bar{H}_0+\bar{H}_0^2+8 H_1^2+\frac{\pi ^2}{3}\\
F_1^{(2)}&=&4 \bar{H}_0 H_{0,1}+12 H_1^2 \bar{H}_0+2 H_1 \bar{H}_0^2+\frac{4}{3} \pi ^2 \bar{H}_0+8 H_1^3+2 \pi ^2
   H_1-4 \zeta (3) \\
F_0^{(2)}&=& 4 H_1 \bar{H}_0 H_{0,1}+\frac{1}{2} \bar{H}_0^2 H_{0,1}-4 \bar{H}_0 H_{0,1,1}+2 \zeta (3) \bar{H}_0+4
   H_1^3 \bar{H}_0+H_1^2 \bar{H}_0^2+\frac{4}{3} \pi ^2 H_1 \bar{H}_0\\&&+\frac{1}{12} \pi ^2
   \bar{H}_0^2+\frac{1}{6} \pi ^2 H_{0,1}+4 H_{0,0,0,1}+2 H_{0,1,0,1}-4 H_1 \zeta (3)+2 H_1^4+\pi ^2
   H_1^2+\frac{\pi ^4}{36}
   \end{eqnarray*}
For the three-loop predictions we have
{\footnotesize
\begin{eqnarray}
-F_3^{(3)}&=&8 \bar{H}_0 H_{0,1}+16 H_1^2 \bar{H}_0+4 H_1 \bar{H}_0^2+\frac{2 \bar{H}_0^3}{9}+\frac{2}{9} \pi ^2 \bar{H}_0-\frac{40}{3} H_{0,0,1}+\frac{32 H_1^3}{3}+\frac{4 \pi ^2 H_1}{3}-8 \zeta (3)\nn \\
-F_2^{(3)}&=&40 H_1 \bar{H}_0 H_{0,1}+4 \bar{H}_0^2 H_{0,1}+4 \bar{H}_0 H_{0,0,1}-40 \bar{H}_0 H_{0,1,1}-4 \zeta (3) \bar{H}_0+32 H_1^3 \bar{H}_0+10 H_1^2 \bar{H}_0^2\nn \\ 
&+&\frac{2}{3} H_1 \bar{H}_0^3+\frac{26}{3} \pi ^2 H_1 \bar{H}_0+\frac{4}{3} \pi ^2 \bar{H}_0^2-40 H_1 H_{0,0,1}+\frac{4}{3} \pi ^2 H_{0,1}-8 H_{0,0,0,1}+40 H_{0,0,1,1}\nn\\
&+&16 H_{0,1,0,1}-40 H_1 \zeta (3)+16 H_1^4+\frac{26}{3} \pi ^2 H_1^2+\frac{7 \pi ^4}{15}\nn \\
-F_1^{(3)}&=&48 H_1^2 \bar{H}_0 H_{0,1}+10 H_1 \bar{H}_0^2 H_{0,1}+8 H_1 \bar{H}_0 H_{0,0,1}-96 H_1 \bar{H}_0 H_{0,1,1}+\frac{1}{3} \bar{H}_0^3 H_{0,1}+5 \pi ^2 \bar{H}_0 H_{0,1}\nn\\
   &-&10 \bar{H}_0^2 H_{0,1,1}+4 \bar{H}_0 H_{0,0,0,1}-8 \bar{H}_0 H_{0,0,1,1}+20 \bar{H}_0 H_{0,1,0,1}+96 \bar{H}_0 H_{0,1,1,1}+2
   \zeta (3) \bar{H}_0^2\nn\\
   &+&20 H_1^4 \bar{H}_0+8 H_1^3 \bar{H}_0^2+\frac{2}{3} H_1^2 \bar{H}_0^3+\frac{38}{3} \pi ^2 H_1^2 \bar{H}_0+3 \pi ^2 H_1 \bar{H}_0^2+\frac{1}{18} \pi ^2
   \bar{H}_0^3+\frac{113}{90} \pi ^4 \bar{H}_0-20 \zeta (3) H_{0,1}\nn\\ 
   &-&40 H_1^2 H_{0,0,1}+\frac{10}{3} \pi^2 H_1 H_{0,1}+80 H_1 H_{0,0,1,1}+40 H_1 H_{0,1,0,1}-\frac{10}{3} \pi ^2 H_{0,0,1}-\frac{10}{3} \pi ^2 H_{0,1,1}\nn\\ &+&20 H_{0,0,0,0,1}-20 H_{0,0,1,0,1}-80 H_{0,0,1,1,1}-20 H_{0,1,0,0,1}-40 H_{0,1,0,1,1}-40
   H_{0,1,1,0,1}\nn\\&-&48 H_1^2 \zeta (3)+8 H_1^5+8 \pi ^2 H_1^3+\frac{91 \pi ^4 H_1}{45}-32 \zeta (5)-\frac{8
   \pi ^2 \zeta (3)}{3}\nn
   \eeqa
   and finally
   \beqa 
-F_0^{(3)}  &=&\frac{4 H_1^6}{3}+4 \bar{H}_0 H_1^5+2 \bar{H}_0^2 H_1^4+2 \pi ^2 H_1^4+\frac{2}{9} \bar{H}_0^3
   H_1^3+\frac{38}{9} \pi ^2 \bar{H}_0 H_1^3+16 \bar{H}_0 H_{0,1} H_1^3 \nn\\
   &-&\frac{40}{3} H_{0,0,1} H_1^3-16 \zeta (3) H_1^3+\frac{3}{2} \pi ^2 \bar{H}_0^2 H_1^2+5 \bar{H}_0^2 H_{0,1} H_1^2
   +\frac{5}{3} \pi ^2 H_{0,1} H_1^2+4 \bar{H}_0 H_{0,0,1} H_1^2\nn\\
   &-&48 \bar{H}_0 H_{0,1,1} H_1^2+40 H_{0,0,1,1} H_1^2+20
   H_{0,1,0,1} H_1^2+\frac{91}{90} \pi ^4 H_1^2+\frac{1}{18} \pi ^2 \bar{H}_0^3 H_1+\frac{113}{90} \pi ^4
   \bar{H}_0 H_1\nn\\ 
   &+&\frac{1}{3} \bar{H}_0^3 H_{0,1} H_1+5 \pi ^2 \bar{H}_0 H_{0,1} H_1-\frac{10}{3} \pi ^2
   H_{0,0,1} H_1-10 \bar{H}_0^2 H_{0,1,1} H_1-\frac{10}{3} \pi ^2 H_{0,1,1} H_1\nn\\
   &+&4 \bar{H}_0 H_{0,0,0,1} H_1-8 \bar{H}_0 H_{0,0,1,1} H_1+20 \bar{H}_0 H_{0,1,0,1} H_1
   +96 \bar{H}_0 H_{0,1,1,1} H_1+20 H_{0,0,0,0,1} H_1\nn\\
   &-&20 H_{0,0,1,0,1} H_1 -80 H_{0,0,1,1,1} H_1-20 H_{0,1,0,0,1} H_1-40 H_{0,1,0,1,1} H_1-40 H_{0,1,1,0,1} H_1
   -32 \zeta (5) H_1\nn\\
   &+&2 \bar{H}_0^2 \zeta (3) H_1-20 H_{0,1} \zeta (3) H_1-\frac{8}{3} \pi ^2 \zeta (3) H_1+\frac{7}{60} \pi ^4 \bar{H}_0^2+\frac{2}{3} \pi ^2 \bar{H}_0^2 H_{0,1}+\frac{7}{30} \pi ^4 H_{0,1}\nn\\
   &+&\frac{1}{3} \pi ^2 \bar{H}_0 H_{0,0,1}-\frac{1}{3} \bar{H}_0^3 H_{0,1,1}-5 \pi ^2 \bar{H}_0 H_{0,1,1}-\frac{2}{3} \pi ^2 H_{0,0,0,1}+\frac{10}{3} \pi ^2 H_{0,0,1,1} \nn\\
   &+&2 \bar{H}_0^2 H_{0,1,0,1}+2 \pi ^2 H_{0,1,0,1}+10 \bar{H}_0^2 H_{0,1,1,1}+\frac{10}{3} \pi^2 H_{0,1,1,1}-6 \bar{H}_0 H_{0,0,0,0,1}-4 \bar{H}_0 H_{0,0,0,1,1} \nn\\ 
   &-&2 \bar{H}_0 H_{0,0,1,0,1}+8\bar{H}_0 H_{0,0,1,1,1}-2 \bar{H}_0 H_{0,1,0,0,1}-20 \bar{H}_0 H_{0,1,0,1,1}-20 \bar{H}_0 H_{0,1,1,0,1}-96 \bar{H}_0 H_{0,1,1,1,1}\nn\\ 
   &+&56 H_{0,0,0,0,0,1}-20 H_{0,0,0,0,1,1}+4 H_{0,0,0,1,0,1}+4 H_{0,0,1,0,0,1}+20 H_{0,0,1,0,1,1}+20 H_{0,0,1,1,0,1}\nn\\&+&80 H_{0,0,1,1,1,1}+4 H_{0,1,0,0,0,1} 
   +20 H_{0,1,0,0,1,1}+20 H_{0,1,0,1,0,1}+40 H_{0,1,0,1,1,1}+20 H_{0,1,1,0,0,1}\nn\\&+&40 H_{0,1,1,0,1,1}+40 H_{0,1,1,1,0,1}+16 \bar{H}_0 \zeta (5)-4 \zeta (3)^2+\bar{H}_0^3 \zeta (3)+\frac{4}{3} \pi ^2 \bar{H}_0 \zeta (3)\nn\\
   &+&2 \bar{H}_0 H_{0,1} \zeta (3)-4 H_{0,0,1} \zeta (3)+20 H_{0,1,1} \zeta(3)+\frac{11 \pi ^6}{270} \la{HexPredictions}
\end{eqnarray}}

\section{Description of Attached Notebooks}
Here we briefly describe the few \verb"mathematica" notebooks that the reader can find in attachment to this paper:
\begin{itemize}
\item \verb"Fnl.nb" contains the functions $F_n^{(l)}$ in (\ref{expanded}) up to three loops,
\item \verb"beta.nb" contains the function $\beta$ in (\ref{todo}),
\item \verb"Functionshf.nb" contains the functions $f_i$ in (\ref{fInt}) and the functions $h_i$ in (\ref{Job}),
\item \verb"f1f2f3f4.nb" is a notebook that yields the functions $f_1,f_2,f_3$ and $f_4$ appearing in the finite coupling conjectures for the pentagon transitions to any desired order in perturbation theory. This allows one to straightforwardly expand the pentagon transitions perturbatively at weak coupling. For more details see appendix \ref{ansatz-App}.
\end{itemize}

\end{document}